\documentclass[a4paper,11pt]{article}
\pdfoutput=1 
\usepackage{jheppub} 

\usepackage[T1]{fontenc} 
\usepackage{amsmath,amssymb,graphicx}
\usepackage{subfigure}
\usepackage{multirow}
\usepackage{dcolumn}
\usepackage{color}
\usepackage{bm}
\usepackage{siunitx}
\usepackage{slashed}
\usepackage{natbib}
\usepackage{feynmp-auto}
\usepackage{hyperref}
\usepackage{makecell}
\usepackage{pbox}
\usepackage{fourier} 
\usepackage{array}
\usepackage{colordvi}
\newcommand{\bea}{\begin{eqnarray}}
\newcommand{\eea}{\end{eqnarray}}
\newcommand{\beq}{\begin{eqnarray}}
\newcommand{\eeq}{\end{eqnarray}}

\newcommand{\gwa}[1]{g_{W,a#1}}
\newcommand{\gwb}[1]{g_{W,b#1}}

\newcommand{\gza}[1]{g_{Z,a#1}}
\newcommand{\gzb}[1]{g_{Z,b#1}}

\newcommand{\gab}[1]{g_{A,b#1}}

\newcommand{\gva}[1]{g_{V,a#1}}
\newcommand{\gvb}[1]{g_{V,b#1}}

\newcommand{\gxb}[1]{g_{X,b#1}}

\title{\boldmath Highly Boosted Higgs Bosons and Unitarity \\ in Vector-Boson Fusion
  at Future Hadron Colliders}

\author[a]{Wolfgang Kilian}
\author[b]{Sichun Sun}
\author[c,d]{Qi-Shu Yan}
\author[e]{Xiaoran Zhao}
\author[d]{Zhijie Zhao}

\affiliation[a]{Department of Physics, University of Siegen, 57068 Siegen, Germany}
\affiliation[b]{Department of Physics and INFN, Sapienza University of Rome, Rome I-00185, Italy}
\affiliation[c]{School of Physics Sciences, University of Chinese Academy of Sciences, Beijing 100039, China}
\affiliation[d]{Center for future high energy physics, Institute of High Energy Physics, Chinese Academy of Sciences, Beijing 100039, China}
\affiliation[e]{Centre for Cosmology, Particle Physics and Phenomenology (CP3), Universit catholique de Louvain, Chemin du Cyclotron, 2, B-1348 Louvain-la-Neuve, Belgium}

\emailAdd{kilian@physik.uni-siegen.de}
\emailAdd{sichunssun@gmail.com}
\emailAdd{yanqishu@ucas.ac.cn}
\emailAdd{xiaoran.zhao@uclouvain.be}
\emailAdd{zhao@physik.uni-siegen.de}

\preprint{SI-HEP-2021-05, CP3-21-02}

\abstract{
We study the observability of new interactions which modify
Higgs-pair production
via vector-boson fusion processes at
the LHC and at future proton-proton colliders.
In an effective-Lagrangian approach, we explore in particular the effect of
the operator $h^2 W_{\mu\nu}^a W^{a,\mu\nu}$, which describes the interaction
of the Higgs boson with transverse vector-boson polarization modes. By
tagging highly boosted Higgs bosons in the final state, we 
determine projected bounds for the coefficient of this operator at the LHC
and at a future 27 TeV or 100 TeV collider.  Taking into account unitarity
constraints, 
we estimate the new-physics discovery potential of Higgs pair production in
this channel.
}

\begin{document} 
\maketitle
\flushbottom

\section{Introduction}

In 2012, the Higgs boson was discovered at the LHC~\cite{Aad:2012tfa,Chatrchyan:2012ufa}. 
Current experimental data show that its properties agree with the predictions of Standard Model (SM), 
but further measurements are still necessary to examine the SM and search for new physics.

The SM predicts that the Higgs boson has three kinds of interaction at tree level: 
(1) the Yukawa interaction with fermions;
(2) the interaction with massive vector bosons ($W^\pm$ and $Z$);
(3) the cubic and quartic Higgs self-interactions.
One of the prime targets of the second run and future runs of the LHC is to
measure the Higgs self-couplings~(3).

Higgs pair-production in vector-boson fusion
(VBF) \cite{Jones:1979bq}, $VV\to hh$, is sensitive to the last two kinds of
interaction. VBF-type processes are possible at both electron-positron and
hadron colliders. At a hadron collider, the two incoming vector bosons
$V=W^\pm,Z$ are radiated from two initial quarks. In the final state, in
addition to two Higgs bosons, there are two back-to-back hard jets which
should be tagged in the forward and backward regions of a detector,
respectively. This allows us to use VBF cuts to reject the QCD type backgrounds efficiently.

In this paper, we study double Higgs production in the VBF process in an
effective-field theory (EFT) approach. We follow the conventions given in our
previous paper~\cite{Kilian:2018bhs}, and use the following phenomenological effective Lagrangian:
\begin{align}
\label{eft}  
\mathcal{L}_{EFT} =& \mathcal{L}_{\overline{SM}}+\mathcal{L}_{VVh}+\mathcal{L}_{Vh}, \\
\mathcal{L}_{VVh} = & - \left(g_{W,b1} \frac{h}{v} + g_{W,b2} \frac{h^2}{2 v^2} + g_{W,b3} \frac{h^3}{6 v^3} + \cdots \right)
W^+_{\mu\nu} W^{-\, \mu\nu} \nonumber \\ &
 - \left(g_{A,b1} \frac{h}{2 v}  + g_{A,b2} \frac{h^2}{4v^2} + g_{A,b3} \frac{h^3}{12v^3} + \cdots \right)F_{\mu\nu} F^{ \mu\nu}  \nonumber \\ &
- \left(g_{X,b1} \frac{h}{ v} + g_{X,b2} \frac{h^2}{2 v^2} + g_{X,b3} \frac{h^3}{6 v^3} + \cdots \right)F_{\mu\nu} Z^{ \mu\nu} \nonumber \\ &
- \left(g_{Z,b1} \frac{h}{2 v} + g_{Z,b2} \frac{h^2}{4 v^2} + + g_{Z,b3} \frac{h^3}{12v^2} + \cdots \right)Z_{\mu\nu} Z^{ \mu\nu} \\ 
    \mathcal{L}_{VH}=&\gwa1 \frac{2m_W^2}{v} h W^{+,\mu}W^-_{\mu}
    +\gwa2 \frac{m_W^2}{v^2}h^2 W^{\mu}W_{\mu}
    +\gwa3 \frac{m_W^2}{3v^3} h^3 W^\mu W_\mu \nonumber \\ &
    + \gza1 \frac{m_Z^2}{v} h Z^{\mu} Z_{\mu}
    +\gza2 \frac{m_Z^2}{2 v^2}h^2 Z^{\mu} Z_{\mu}
    +\gza3 \frac{m_Z^2}{6 v^3} h^3 Z^\mu Z_\mu  + \cdots \,.
    \label{lvh}
\end{align}

Dots indicate higher-dimensional interactions which are not relevant for the
VBF Higgs production process that we consider. We only include the
CP-conserving interactions and omit any CP-violating operators. The relations
$\gwa1=\gwa2=\gza1=\gza2=1$ and $\gvb1=\gvb2=\gvb3=\gwa3=\gza3=0$ characterize
the SM reference values at the tree level, where the subscript letter $V$
denotes $W, A, X, Z$, respectively.  The corresponding terms have been removed
from the SM Lagrangian, as indicated by the
overline notation $\overline{SM}$, such that they are not double-counted.

Introducing gauge degrees of freedom in this phenomenological Lagrangian, the
vertices can be rewritten as gauge-invariant operators which are understood as
the low-energy effect of short-range structure or new heavy degrees of
freedom beyond the SM.  Such effects would appear, for instance, in a composite
Higgs model -- for concrete examples, cf.\  \cite{Qi:2019ocx, Agrawal:2019bpm,
  Xu:2019xuo, Li:2019ghf}.   In Sec.\ref{Sec:hhh-EFT}, we relate the
parameterization to the formulation in terms of gauge-invariant operators,
adopting a 
concrete basis and truncating the expansion at dimension six, as commonly done
in the literature.

The single Higgs couplings $\gva1$ can be determined to up to $5-10\%$ via
measuring the decay fractions $h\to WW^*$ and $h\to ZZ^*$ at the LHC.\
Current LHC data exclude deviations from the SM
prediction~\cite{Aaboud:2017vzb,Sirunyan:2017exp,Aaboud:2018jqu,Aad:2019mbh}
of more than about $15\%$.  (This limit, as well as the bounds discussed
below, depends on a universal assumption about the absence of undetected Higgs
decays which we 
will adopt for this paper.)  For the couplings of type $\gvb1$, the bounds
are weaker.  For instance 
($\gvb1\in [0.8, 4.5]$) was reported in Ref.~\cite{Aaboud:2017vzb}.  The
measurement of double Higgs couplings hhVV ($\gva2$ and $\gvb2$) is
challenging at the LHC. Recently, ATLAS reported a search for double Higgs
production in VBF~\cite{Aad:2020kub}, which excludes the ranges
$\gva2<-0.56$ and $\gva2>2.89$.

At the LHC, searches for a double Higgs final state focus on the gluon-gluon
fusion process $gg\to hh$. The Higgs decay channels
$b\bar{b}b\bar{b}$~\cite{Aaboud:2018knk,Sirunyan:2018tki},
$b\bar{b}\gamma\gamma$~\cite{Aaboud:2018ftw,Sirunyan:2018iwt},
$b\bar{b}\tau\tau$~\cite{Aaboud:2018sfw,Sirunyan:2017djm} and
$b\bar{b}VV$~\cite{Aaboud:2018zhh,Sirunyan:2017guj} have been investigated by
both ATLAS and CMS. In addition, result for the channels
$WW\gamma\gamma$~\cite{Aaboud:2018ewm} and $WWWW$~\cite{Aaboud:2018ksn} were
reported by ATLAS.  A combination of these searches can be found in
Ref.~\cite{Sirunyan:2018ayu,Aad:2019uzh}. It is shown that the Higgs
self-coupling $\lambda_3$ can be constrained to $[-5, 12]$ by data, while no
constraint of $\kappa_5$ has been reported. 

As a complementary process,
double Higgs production via VBF at hadron colliders has been extensively
studied in the literature
\cite{Dolan:2013rja,Liu-Sheng:2014gxa,Dolan:2015zja,Bishara:2016kjn,Arganda:2018ftn}. 
The NLO QCD and higher order correction for this process has been calculated in
Ref.~\cite{Baglio:2012np,Frederix:2014hta,Dreyer:2018qbw,Dreyer:2018rfu,Dreyer:2020urf,Dreyer:2020xaj}; they find an enhancement of around
$7\%$ as it is natural for a pure electroweak process.  For the high-luminosity
LHC (HL-LHC), assuming $\mathcal{L}=3$ ab$^{-1}$ at 14 TeV,  
it is expected that the couplings of type $\gva2$ can be constrained to
$20\%$~\cite{Dolan:2015zja},  
while a precision of around $1\%$ should be achievable at a future 100 TeV
hadron collider~\cite{Bishara:2016kjn}.
Furthermore, the hhVV coupling is also accessible via $hVV$ or $hhV$ final
states.  
Ref.~\cite{Englert:2017gdy} argues that a measurement of the $W^\pm W^\pm h$ final state can
constrain the $hhWW$ coupling to $\mathcal{O}(100\%)$ at the HL-LHC, and
to $20\%$ at a 100 TeV collider,  
while the determination of this coupling from the  $hhV$ final state can only yield a weak
bound~\cite{Nordstrom:2018ceg}.

Possible measurements of the $\gva2$ couplings, i.e., the Higgs interacting
with the logitudinal components of massive vector bosons, have thus been
covered in some detail in previous work.  However, without further assumptions
it is not evident that couplings of the Higgs to transversal vector bosons
play a lesser role.  In this work, we will aim at filling this gap and thus
perform a Monte-Carlo study of the sensitivity to couplings of type $\gvb2$,
both for the LHC and for future high-energy hadron colliders, and correlate
this with the determination of~$\gva2$.

The VBF process $VV\to hh$ receives a contribution from the Higgs
cubic self-coupling.  It is well known that at hadron colliders, the Higgs
self-coupling is most accessible in the gluon-gluon fusion
process~\cite{Plehn:1996wb}, whose cross section is one order of
magnitude larger than that of the VBF process.  This fact has received a lot
of
attention~\cite{Baur:2002rb,Li:2013flc,Cao:2015oaa,Cao:2016zob,Baur:2002qd,Ren:2017jbg,Baur:2003gp,Yao:2013ika,Kling:2016lay,Chang:2018uwu,Kim:2018uty,He:2015spf,Papaefstathiou:2012qe,Baur:2003gpa,Dolan:2012rv,Barr:2013tda,deLima:2014dta,Behr:2015oqq,Barger:2013jfa,Barr:2014sga,Papaefstathiou:2015iba,Li:2015yia,Zhao:2016tai,Contino:2016spe,Goncalves:2018yva}.
A measurement of the VBF process will provide additional precision for the Higgs
self-coupling.  However, for simplicity we will assume here that the couplings
which are accessible in gluon-gluon fusion are known, and we fix those at
their SM value.  This allows us to focus on the couplings which are specific
to the VBF class of processes, and lets us more easily estimate the
sensitivity potential for those.

This paper is organized as follows. In
Sec.~\ref{hpsm}, we briefly introduce the mass-drop method used for tagging
highly boosted Higgs bosons, and use it to explore the SM case at the LHC and
at a 100 TeV collider.
In Sec.~\ref{Sec:hhh-EFT}, we investigate the potential for the discovery of new physics via 
multi-Higgs production, in form of the interactions discussed above.
In Sec.~\ref{projgvb2}, we provide projections for bounds on $\gvb2$ and
$\gva2$ for
different collision energies. 
We conclude this paper with a discussion of our findings in Sec.~\ref{Sec:conc}.

\section{ Higgs pair production in the Standard Model }
\label{hpsm}
We will focus on the signal process $pp\to hhjj \to 4b2j$ due to the reason that the decay channel $h\to b\bar{b}$ has the largest branch ratio.
The partonic signal events are generated using WHIZARD~\cite{Kilian:2007gr} with the cuts listed in Table \ref{vbfcuts}. We takes the parton distribution functions from CTEQ6l1~\cite{Pumplin:2002vw}.

In this work, we consider the main backgrounds $pp\to t\bar{t}\to 2b4j$,
$pp\to 2b4j$ (QCD) and $pp\to 4b2j$.  The background events
$pp\to t\bar{t} \to 2b 4j$ are generated by WHIZARD.  We have cross-checked
the results with Madgraph~\cite{Alwall:2014hca}.  Pure-QCD partonic events of type
$pp\to 2b4j$ and $pp\to 4b2j$ are generated using
ALPGEN~\cite{Mangano:2002ea}.

For all event samples, parton shower and hadronization are performed by Pythia 8~\cite{Sjostrand:2007gs}.
Jets are reconstructed by
FastJet~\cite{Cacciari:2011ma} using the anti-$k_t$
algorithm~\cite{Cacciari:2008gp} with a jet radius $R=0.4$ and transverse
momentum cut $P_{t}>20$ GeV.  We do not account for detector effects in
detail but insert values for efficiency and mistagging rates where
appropriate. 

\subsection{Analysis method at the 14 TeV LHC}
\label{Sec:ANA014}


In Table \ref{cut:SM014} (first column) we list the expected number of signal
and background events in the SM
for the LHC with $\sqrt{s}=14$ TeV and luminosity
$\mathcal{L}=3$ ab$^{-1}$.
To suppress the QCD background, we require four b-tagged jets in the final state
($n_b=4$).  We assume the b-tagging efficiency $\epsilon_b=0.7$ and
mistagging rate $\epsilon_{miss}=0.001$.  
The number of events after applying the b-tagging requirement are listed in the 2nd column of Table \ref{cut:SM014}.

To identify two forward jets in the VBF process, 
we first select the jet with highest energy and label it as $j_1$.  If its energy satisfies $E_{j_1} >
500$ GeV, we scan over all other jets and determine the maximal rapidity
difference $\Delta \eta(j_1,j)$ and invariant mass $m(j_1,j)$ with respect to the leading
jet. If the conditions $\Delta \eta(j_1,j)_{max}>3.6$ and $m(j_1,j)_{max}>500$ GeV are met simultaneously, we
identify the most energetic j as $j_2$ and label corresponding pair of jets as the tagging forward jets of a VBF process. 
Otherwise, the event is rejected. The number of events after applying this VBF cut are listed in the 3rd column
of Table \ref{cut:SM014}. 

With these b-jet and forward-jet tagging requirements, the backgrounds $pp\to
t\bar{t}$ and $pp\to 2b4j$ are greatly reduced.  The dominant remaining
background originates from QCD processes $pp\to 4b2j$, where the cross section
after cuts is still five orders of magnitude larger than that of the signal.

To further suppress this huge background, we select those events with massive jets formed by highly boosted Higgs bosons and adopt the mass drop method \cite{Butterworth:2008iy}. 

\begin{center}
\begin{table}
  \begin{center}
  \begin{tabular}{|c|c|c|c|}
  \hline
  Cuts           &  $\sqrt{s} = 14$ TeV    &  $\sqrt{s} = 27$ TeV       &  $\sqrt{s}=100$ TeV \\ 
  \hline
  $P_t(j)$       &  $>20$ GeV  &  $>20$ GeV  &     $> 30$ GeV \\
  \hline
  $\Delta R (j,j)$      &  $>0.8$   &  $>0.8$               &    $> 0.8$ \\
  \hline
  $|\eta(j)|$     &  $<5.0$   &  $<5.0$            &   $<8.0$ \\ 
  \hline \hline
  $\Delta \eta(j_1,j_2)$ & $>3.6$ & $>3.6$  &  $> 4.0$ \\
  \hline
  $m(j_1,j_2)$ & $>500$ GeV  & $>500$ GeV &  $>800$ GeV \\
  \hline
  \end{tabular}
  \end{center}
      \caption{Acceptance cuts used for the calculation of VBF Higgs
        production in $pp$ collision 
        (VBF cuts), for three different collider energies. 
        The $j_1$ and $j_2$ are the tagged forward jets for VBF process. \label{vbfcuts}} 
\end{table}
\end{center}

\begin{center}
\begin{table}
  \begin{center}
  \begin{tabular}{|c|ccc|}
  \hline
  Process           &  $\sigma\times\mathcal{L}$   &  $n_b=4$             &  VBF    \\ 
  \hline
  SM signal         & $993$                       &  $238$               &  $171$          \\ 
  \hline
  $pp\to 4b2j$      & $2.28\times 10^8$           &  $5.47\times 10^7$   &  $1.86\times 10^7$\\ 
  $pp\to 2b4j$ (QCD) & $2.38\times 10^{10}$        &  $1.14\times 10^{4}$ &  $3.85\times 10^4$ \\ 
  $pp\to t\bar{t}\to 2b4j$ & $7.89\times 10^8$   &  $387$               &  $58$\\
  \hline
  \end{tabular}
  \end{center}
  \caption{The cut efficiencies of b-tagging and VBF at 14 TeV LHC are
    demonstrated. The total integrated luminosity is assumed to be
    $\mathcal{L}=3000$ fb$^{-1}$. The b-tagging efficiency is $\epsilon_b=0.7$,
    and the mis-tagging rate of light quarks is $\epsilon_{miss}=0.001$. \label{cut:SM014}} 
\end{table}
\end{center}

\subsubsection{The mass drop method for highly-boosted Higgs boson tagging}
\label{Sec:BDRS}

A significant fraction (of the order of $10\;\%$) of the VBF event sample in the
SM contains highly boosted Higgs bosons.  
A highly boosted Higgs boson has a large transverse momentum ($P_t>200$ GeV)
and can be detected in the central region of the detector. The
decay products of the Higgs boson typically form a fat jet with a large jet
mass, if a large cone parameter is used for the analysis.  A fraction of the
QCD background events also contains massive jets, but the fat jets originating
from Higgs pairs can in principle be distinguished by their characteristic jet
sub-structure.

In recent years, various methods for jet substructure analysis have been
developed:  (1) Jet-grooming methods
aim at removing soft radiation which is unlikely to originate from the
hard process (see, e.g.,\cite{Butterworth:2008iy}). (2) Radiation-constraint
methods 
impose a cut on jet shape to separate the signal from the background (see,
e.g., \cite{Thaler:2010tr}).  (3) Prong-finder methods detect 
a massive 
boosted object as a fat jet with multiple hard cores by exploiting the
recombination history of the jet algorithm.  As a particular prong-finder
algorithm, the mass-drop tagger method is particularly suited for isolating
boosted Higgs 
bosons, decaying to $b\bar{b}$, from the QCD
background~\cite{Butterworth:2008iy}.  A detail review of jet substructure can
be found in Ref.~\cite{Marzani:2019hun} 

In this work, we adopt the mass-drop tagger~\cite{Butterworth:2008iy} as a
means for tagging highly boosted Higgs bosons in the final state of the VBF process. 
The method consists of the following two steps: (1) We first identify jets by
the standard anti-$k_t$ algorithm with the cone parameter $R=0.4$. After
identifying the forward jets associated with the VBF process, we recluster the
remaining 
jets using the Cambridge/Aachen (CA) algorithm~\cite{Dokshitzer:1997in,Wobisch:1998wt} with $R=1.2$.
(2) If there are 2 to 4 CA jets in an event, and its leading or subleading jet has a transverse momentum satisfying $P_t>200$ GeV and a jet mass $m_j>100$ GeV,
we apply the following procedure to tag candidates for highly boosted Higgs bosons.
\begin{enumerate}
  \item Undo the last step of clustering of jet $j$ to get two daughter jets $j_1$ and $j_2$ with $m_{j1}>m_{j2}$.
  \item If the conditions 
  \begin{eqnarray}
    && m_{j1}<\mu  m_{j}
       \qquad\text{and}\qquad
       \frac{\min(P_{t}(j_1),P_{t}(j_2))}{m^2_j}\Delta R^2_{j1,j2}>y_{cut}  \label{muy} 
  \end{eqnarray}
are satified, we identify $j$ as a fat jet associated with a highly boosted Higgs boson.
  \item Otherwise, we redefine $j$ as $j_1$ and repeat the above procedure.
\end{enumerate}
There are two dimensionless parameters $\mu$ and $y_{cut}$ in this method, as given in Eq. (\ref{muy}).
In this work, we fix them as $\mu=0.67$ and $y_{cut}=0.09$.
After two subjets are found, we also apply the filtering method to remove soft
radiation which originates from the underlying event and contaminates CA jets with a larger cone size.

After applying this tagging algorithm, both signal and background events fall
into three classes: 2-boosted Higgs (2BH), 1-boosted Higgs (1BH) and 0-boosted
Higgs (0BH) candidate events. 
The number of events for each class are listed in Table \ref{number:SM014}. 

We observe that the fraction of signal events in the 2BH category
is still small ($2-3\%$).  Nevertheless, the corresponding background is two
orders of magnitude lower than for the 0BH category, where the fraction of
signal events is even less.  The 1BH category falls in between.  The tagger
significantly improves the chance for finding signal events, but by itself it
is clearly not sufficient for a measurement if the rate is SM-like.

\begin{figure}
  \setcounter{subfigure}{0}
  \centering
  \subfigure[2BH]{
  \label{mh_boost}
  \includegraphics[width=0.3\textwidth]{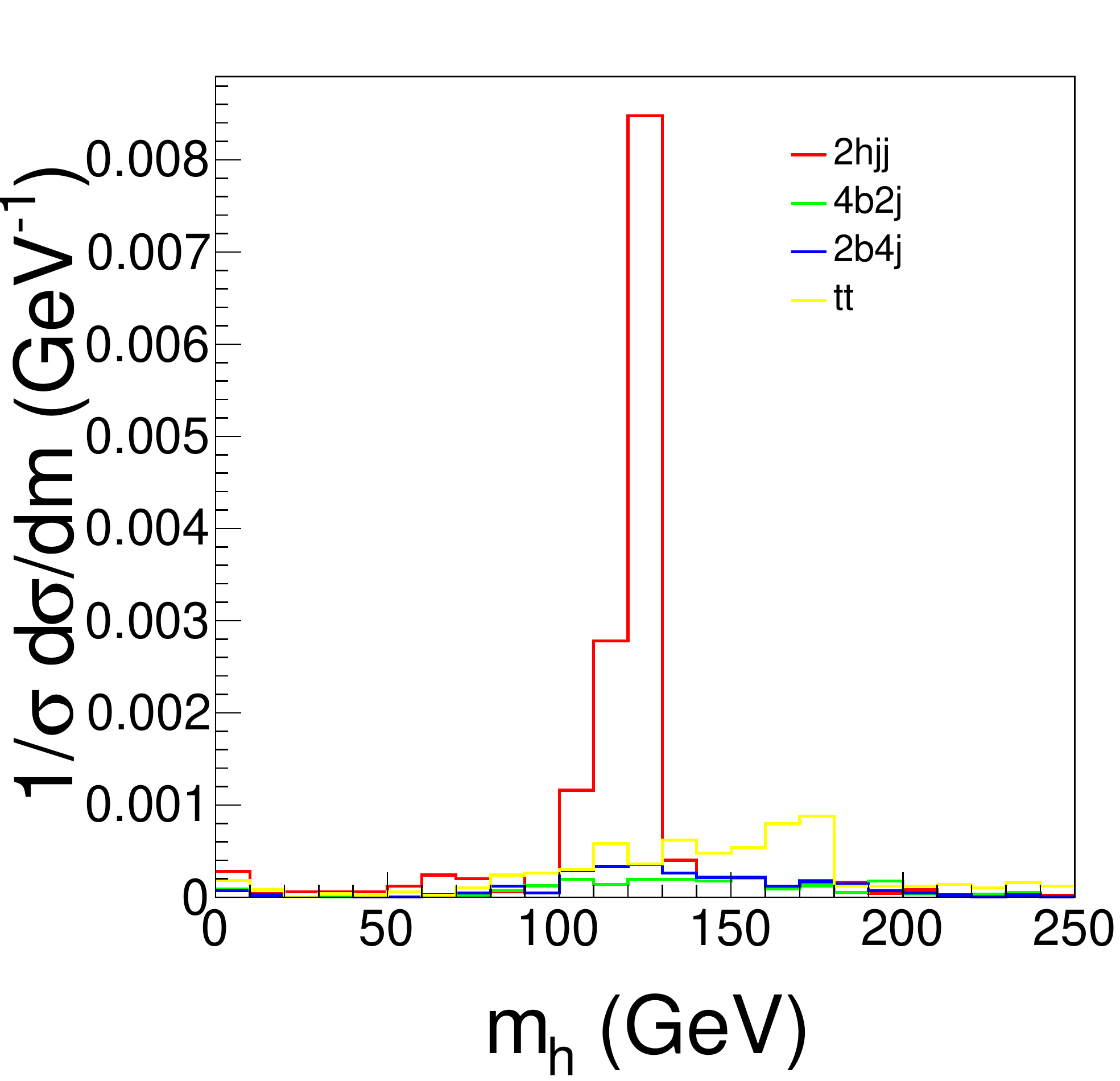}}
  \subfigure[1BH]{
  \label{mh_inter}
  \includegraphics[width=0.3\textwidth]{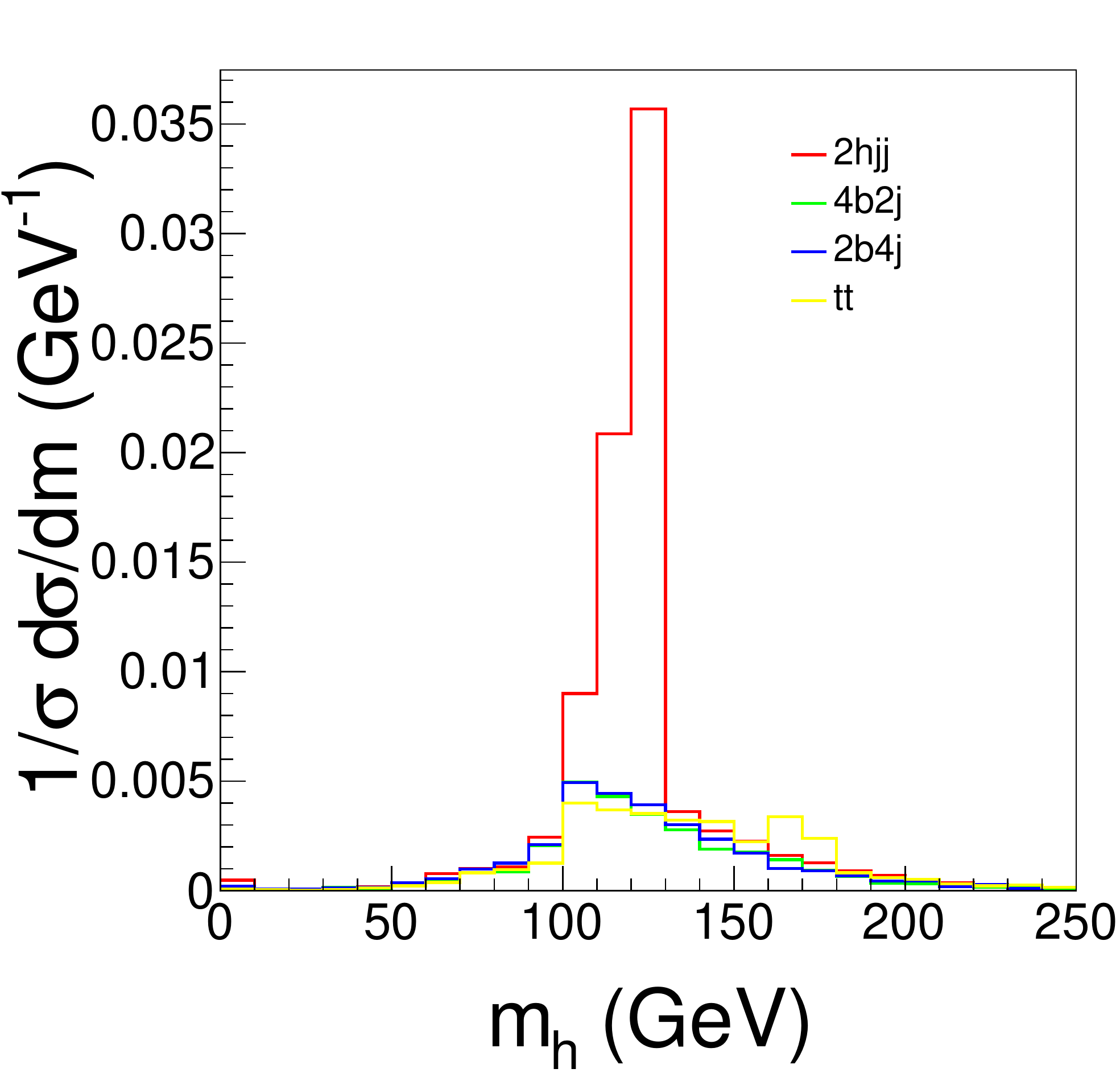}}
  \subfigure[0BH]{
  \label{mh_resolved}
  \includegraphics[width=0.3\textwidth]{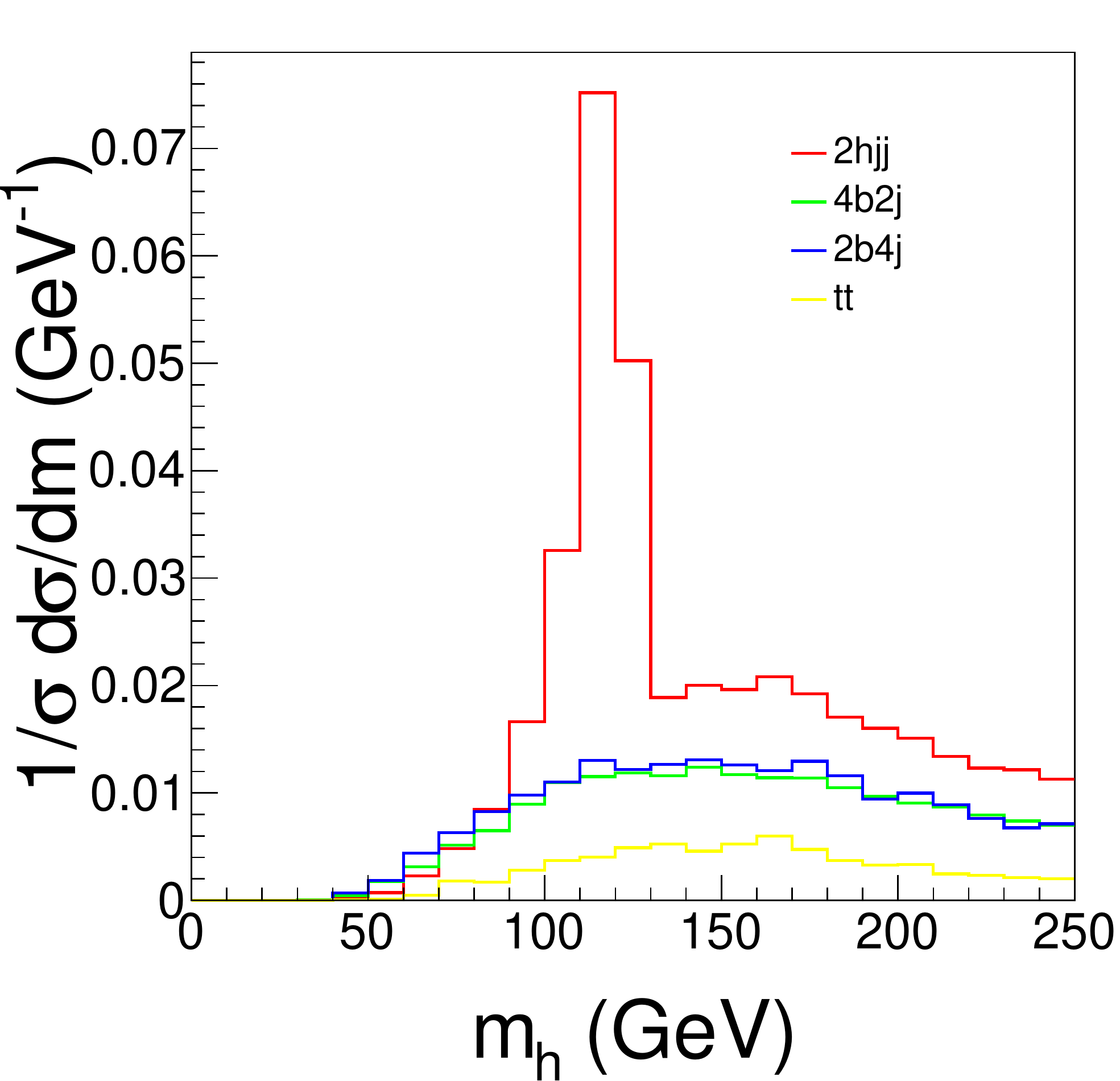}}
  \caption{Distributions of the reconstructed Higgs mass with (a) 2-boosted Higgs events, (b) 1-boosted Higgs events, and (c) 0-boosted Higgs events.} \label{fig:mhrec}
\end{figure}

In order to construct kinematic observables which improve the
signal vs.\ background discrimination, we can reconstruct the mass
peaks of the Higgs bosons for each event category. 
For each of the 2BH events, two jet masses $m_j$ should peak around Higgs the boson mass $m_h$, as shown in Fig. (\ref{fig:mhrec}).
For a 1BH event, the jet mass of the leading fat jet should peak around the
mass of the Higgs boson, while the mass of the second reconstructed Higgs
candidate should coincide with the invariant mass of two b jets.
For a 0BH event, two Higgs boson candidates are reconstructed by using the
$\chi^2$ method.  To this end, we define $\chi^2$ as follows:
\begin{eqnarray}
  \chi^2(m) &=& \frac{|m(j_1,j_2)-m_h|^2}{\sigma^2_j}+\frac{|m(j_3,j_4)-m_h|^2}{\sigma^2_j}\,,
\end{eqnarray}
We assume $m_h=125$ GeV, and $m(j_1,j_2)$ and $m(j_3,j_4)$ are the invariant
mass of two jet pairs in the final state, scanning over each
combination. $\sigma_j=10$ GeV is used to take 
into account the error in jet energy resolution. The pairing which
minimizes $\chi^2$ is selected, and the corresponding
invariant masses determined by the pairs of jets are taken as the
reconstructed masses of Higgs bosons. 

We display the distributions of the reconstructed mass of the leading Higgs
boson in Fig.~\ref{fig:mhrec}.  The shapes of the 2BH and 1BH cases are almost
identical.  We note that the Higgs peak in the 2BH
case is narrower than in the 0BH case, since wrong pairings are rejected
more efficiently. 

\begin{center}
\begin{table}
  \begin{center}
  \begin{tabular}{|c|ccc|}
  \hline
                    &  2-boosted Higgs    &  1-boosted Higgs &  0-boosted Higgs         \\ 
                    & (2BH) & (1BH) & (0BH) \\
  \hline
  SM Signal         &  $4$           &  $21$       &  $146$              \\ 
  \hline
  $pp\to 4b2j$      &  $1.17\times 10^5$           &  $1.56\times 10^6$   &  $1.69\times 10^7$\\ 
  $pp\to 2b4j$ (QCD)&  $28$          &  $349$      &  $3.81\times 10^4$ \\ 
  $pp\to t\bar{t}\to 2b4j$  & $3$    &  $13$       &  $42$\\
  \hline
  \end{tabular}
  \end{center}
  \caption{The numbers of events in the 2BH case, 1BH case and 0BH case at 14 TeV LHC are tabulated. \label{number:SM014}} 
\end{table}
\end{center}

\subsubsection{Multivariate analysis}

After applying the VBF cuts and boosted-Higgs tagging, in order to further
improve the ratio of signal over 
background, we employ the method of boosted decision tree (BDT) which
utilizes the correlation of observables in the signal and can help to further
suppress 
background. We select the following observables as the input to our BDT
analysis: 
\begin{itemize}
   \item $P_t(h_i)$: the transverse momenta of the two reconstructed Higgs
     bosons. 
   \item $m(h_i)$: the invariant masses of the reconstructed Higgs bosons.
   \item $P_t(j_i)$: the transverse momenta of the two forward jets.
   \item $E(j_i)$: the energies of the forward jets.
   \item $\eta(j_i)$: the pseudo-rapidity of the forward jets.
   \item $m(j,j)$: the invariant mass of the forward jets.
   \item $\Delta \eta(j,j)$: the rapidity difference of the forward jets.
   \item $P_t(j^{sub}_i)$: the transverse momenta of the two subjets of each
     highly boosted Higgs boson. 
   \item $m(h,h)$: the invariant mass of the two Higgs boson candidates.
   \item $\chi^2_{min}$ (only for the 0BH case): the minimum value of $\chi^2$.
\end{itemize}

\begin{figure}
  \setcounter{subfigure}{0}
  \centering
  \subfigure[2-boosted Higgs]{
  \label{bdt_boost}
  \includegraphics[width=0.3\textwidth]{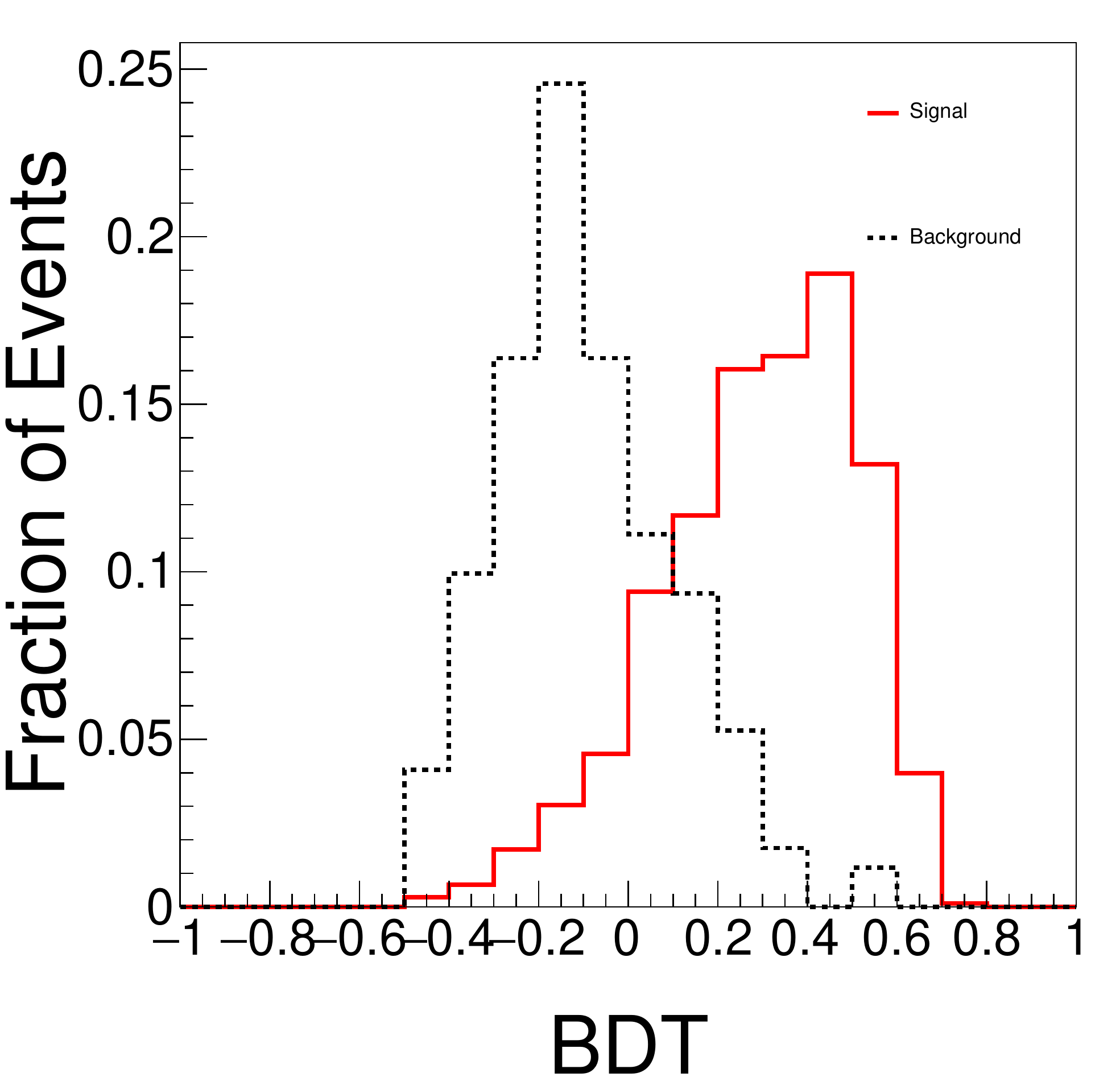}}
  \subfigure[1-boosted Higgs]{
  \label{bdt_inter}
  \includegraphics[width=0.3\textwidth]{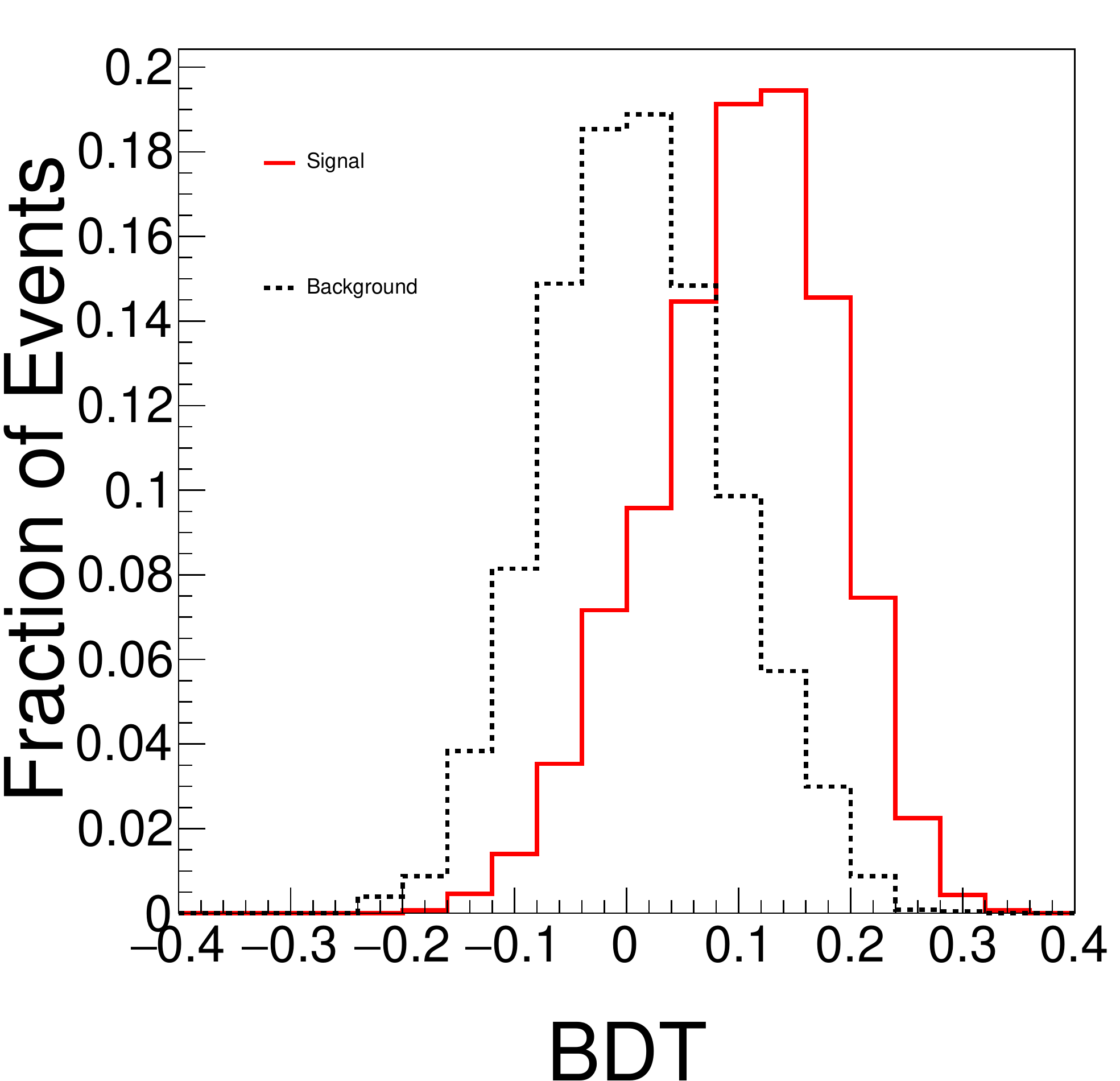}}
  \subfigure[0-boosted Higgs]{
  \label{bdt_resolved}
  \includegraphics[width=0.3\textwidth]{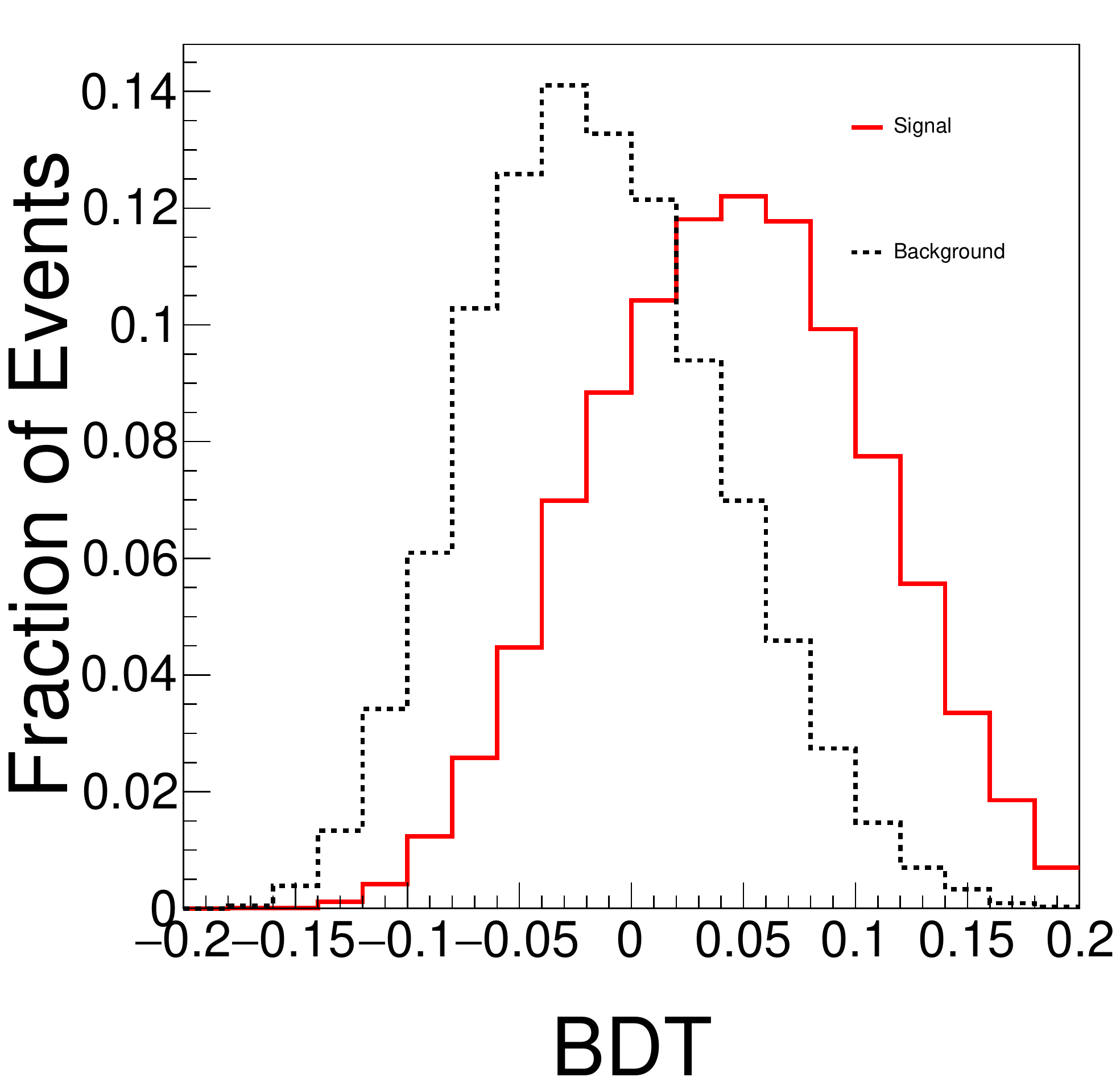}}
  \caption{The response of the discriminants to the SM signal and background at 14 TeV LHC with (a) 2-boosted events, (b) 1-boosted events, and (c) 0-boosted events.} \label{fig:SMBDT014}
\end{figure}

The results of the BDT response are presented in the Fig.~\ref{fig:SMBDT014}.
Obviously, signal and background can separated best in the 2BH case.
For the 1BH and 0BH cases, extracting the signal is challenging even after
exploiting the BDT method.

We can optimize the BDT cut to achieve the maximal significance, which is defined as $S/\sqrt{S+B}$, where $S$ is the event number of the signal and $B$ is the event number of the total background. 
The efficiencies and significances of the optimized BDT cut are listed the
Table \ref{significance:SM014} for all three event classes. 

Comparing the results given in Table \ref{number:SM014} and Table
\ref{significance:SM014}, we conclude that the BDT reduces the background
by one additional order of magnitude, improving on the sequential cut method. 
Nevertheless, the final number of SM signal events is still tiny compared to the
background 
in all cases, and the significance can only reach $0.02$ (2BH, 1BH) or $0.04$
(0BH).  This is still far from the requirement of a discovery at the LHC.

\begin{center}
\begin{table}
  \begin{center}
  \begin{tabular}{|c|ccc|}
  \hline
                    &  2-boosted  Higgs    &  1-boosted Higgs  &  0-boosted Higgs          \\ 
                    & (2BH) & (1BH) & (0BH) \\
  \hline
  SM Signal with BDT cut     &  $3$           &  $13$       &  $90$              \\ 
  \hline
  Background with BDT cut    &  $2.06\times 10^4$    & $3.05\times 10^5$ & $4.42\times 10^6$ \\
  \hline
  $S/B$                      &  $1.40\times 10^{-4}$ & $4.30\times 10^{-5}$ &  $2.04\times 10^{-5}$  \\
  \hline
  $S/\sqrt{S+B}$             &  $0.020$       &  $0.024$    &  $0.043$   \\
  \hline
  \end{tabular}
  \end{center}
  \caption{The significances of the BDT at 14 TeV LHC are demonstrated. \label{significance:SM014}}
\end{table}
\end{center}

\subsection{Analysis of $pp\to hhjj$ in SM at 100 TeV hadron collider}
\label{Sec:ANA100}

We now apply the methods as described above to the VBF process at a future
100 TeV hadron collider.   We assume a high integrated luminosity of
$\mathcal{L}=30$ ab$^{-1}$, and adapt all selection parameters to the
different environment as appropriate.

We require that the most energetic jet has an energy $E_j>800$ GeV. 
The VBF cuts are adjusted as $\Delta \eta_{jj,max}>4.0$ and $m_{jj,max}>800$ GeV.
Table \ref{cut:SM100} shows the number of events after imposing b-tagging and VBF cuts. 
Due to the increased cross section at high energy and the high luminosity of
the collider, the number of signal events is increased by a factor 1000
compared to the high-luminosity LHC.  
After applying b-tagging and VBF cuts, the total number of the signal events is $8.96\times 10^4$.
Of course, the number of background events also increases and reaches $10^9$, so
background reduction is still essential.

\begin{figure}
  \setcounter{subfigure}{0}
  \centering
  \subfigure[2BH]{
  \label{bdt_boost}
  \includegraphics[width=0.3\textwidth]{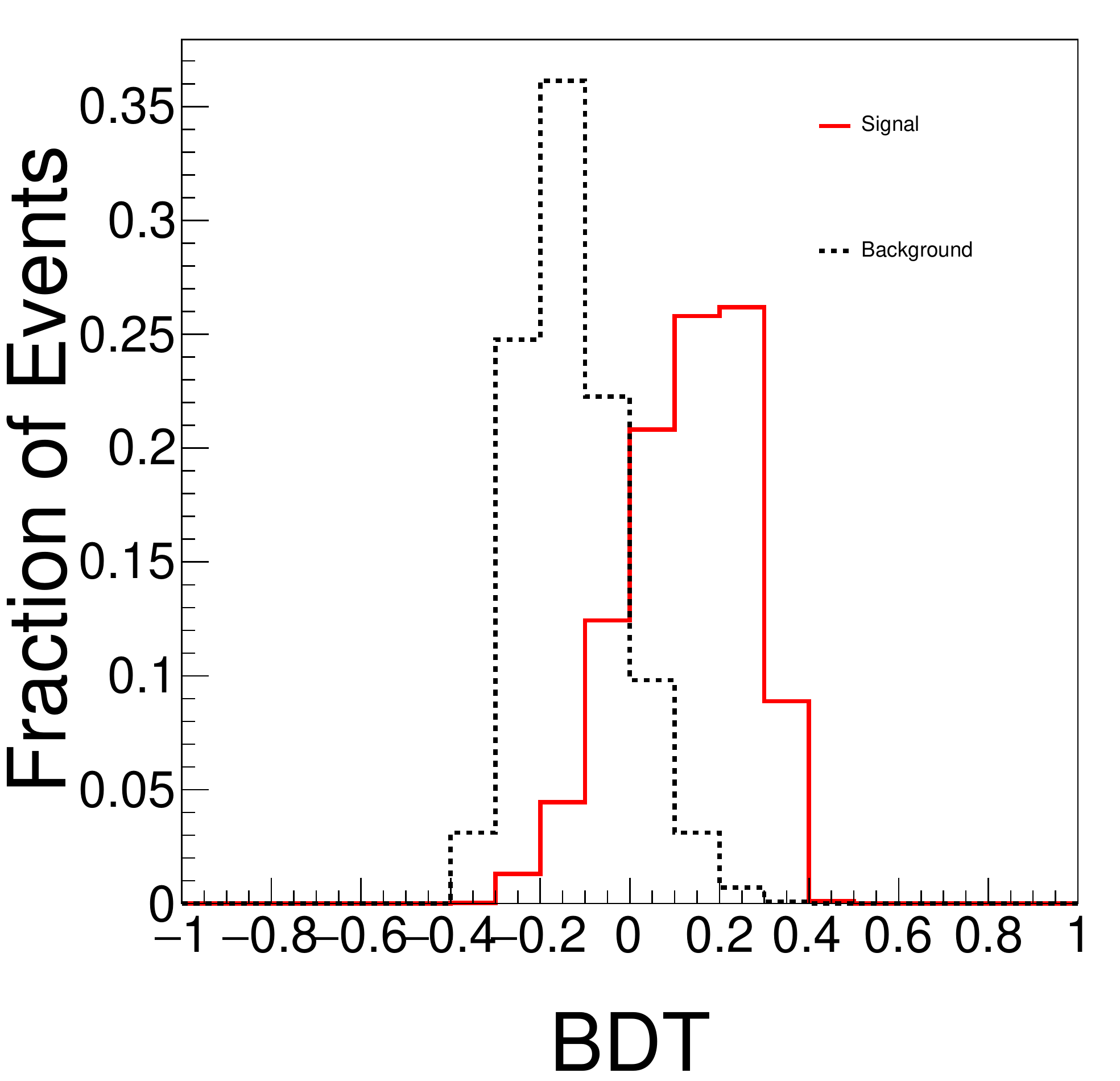}}
  \subfigure[1BH]{
  \label{bdt_inter}
  \includegraphics[width=0.3\textwidth]{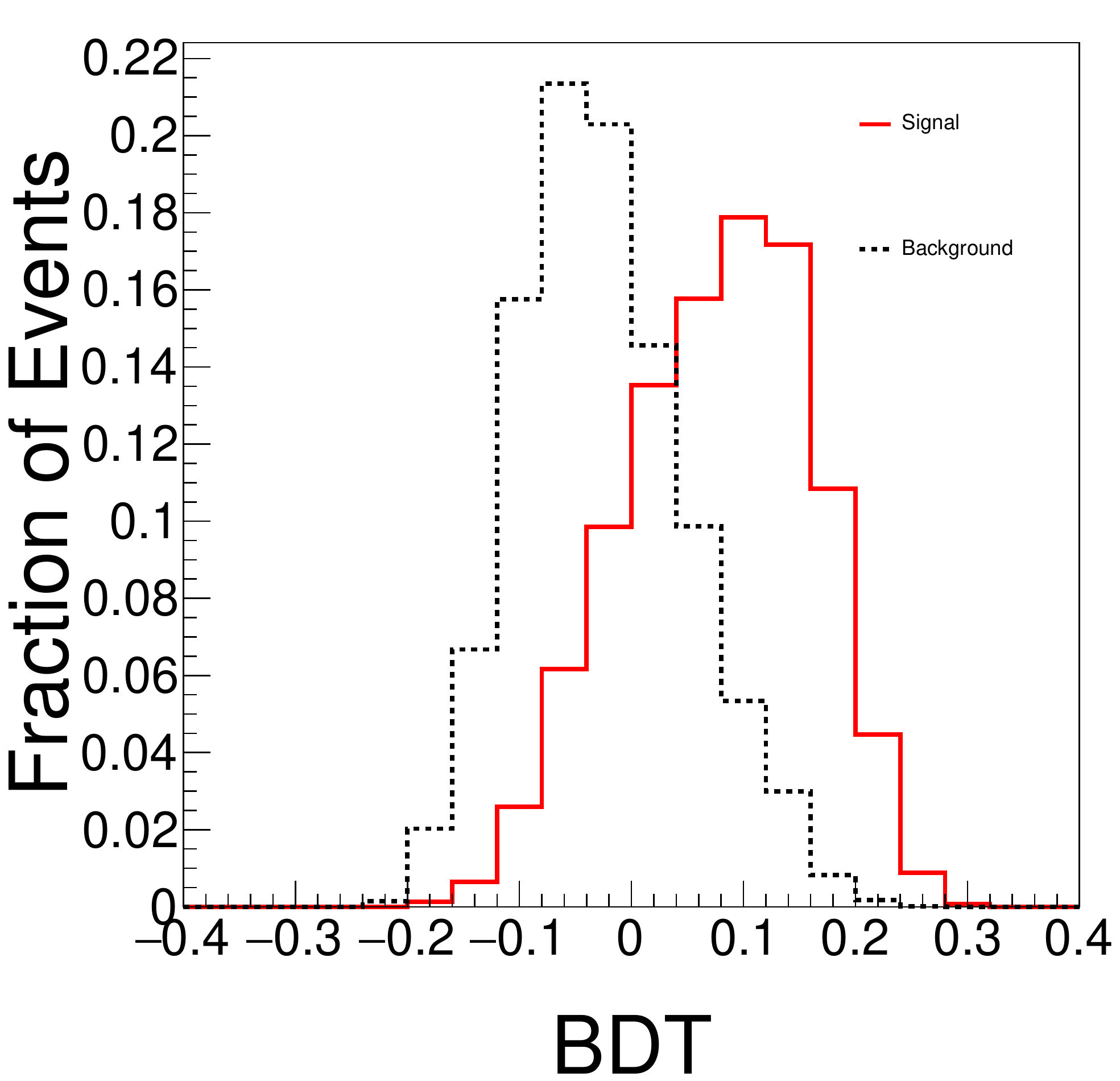}}
  \subfigure[0BH]{
  \label{bdt_resolved}
  \includegraphics[width=0.3\textwidth]{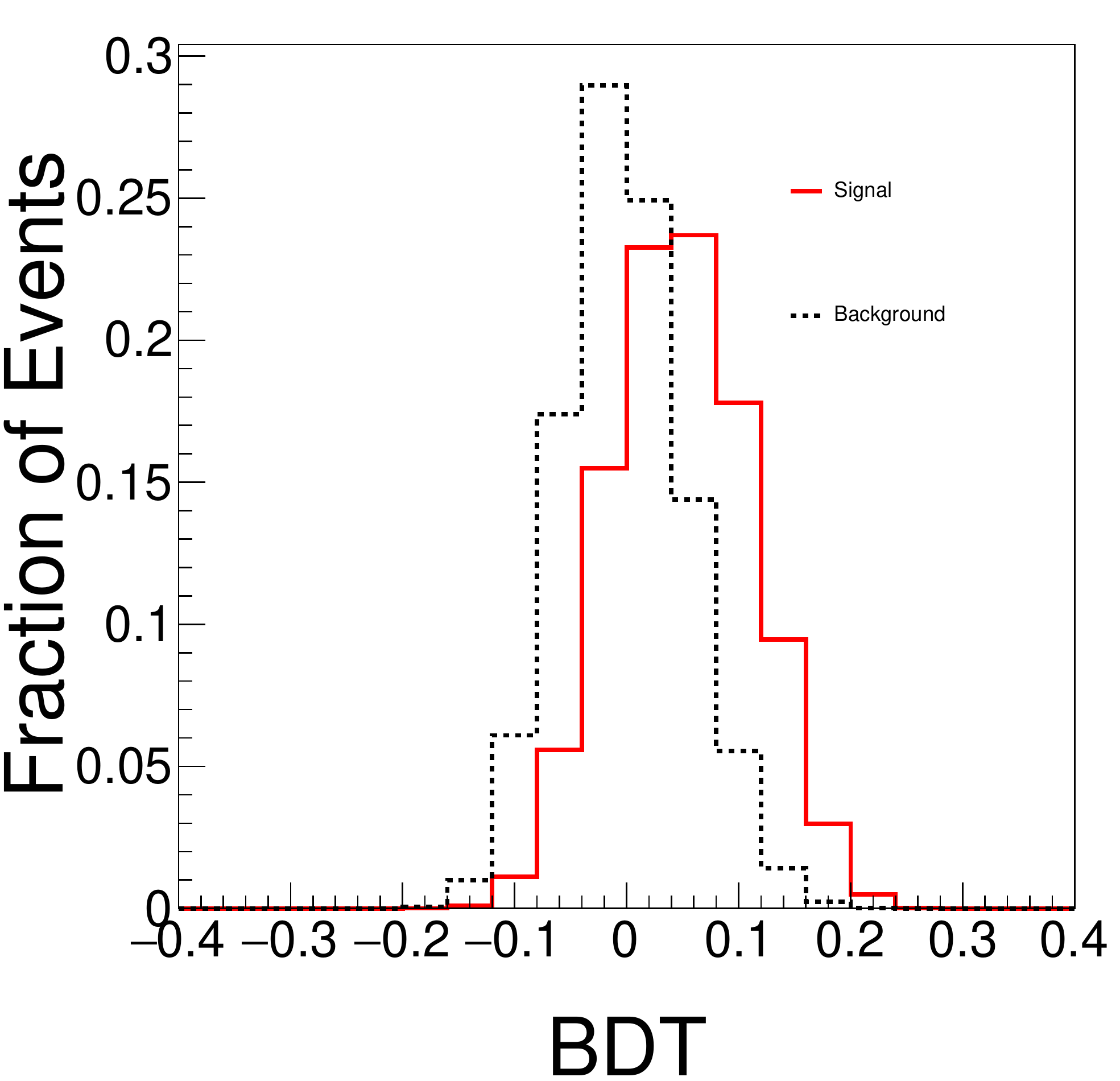}}
  \caption{The response of the discriminants to the SM signal and background at 100 TeV collider with (a) 2BH case, (b) 1BH case, and (c) 0BH case.} \label{fig:SMBDT100}
\end{figure}

The number of events for the 2BH, 1BH and 0BH cases are listed in Table
\ref{number:SM100}. 
As one would expect, more events with highly boosted Higgs bosons can be
observed at the 100 TeV collider than at the LHC.  This is illustrated by the
SM curves of
Fig. (\ref{fig:gb2}).  
To enable signal/background discrimination in these three cases, 
we apply the BDT method as described above.
The results are shown in Fig.~\ref{fig:SMBDT100} and Table \ref{significance:SM100}. 
We can set the BDT cut in a region where the residual background is
effectively zero.  

These results show that it is possible in principle to discover the SM signal
at a 100 TeV collider.  The large rate of the VBF process at high
collision energy leaves enough signal events after all measures for
background reduction have been applied.  We note, however, that pileup
effects in a high luminosity run might pose a challenge for analyses such as
this one.  For a final verdict on the detectability of the SM signal, a more
sophisticated full-simulation study would be necessary which is beyond the
scope of this paper.

\begin{center}
\begin{table}
  \begin{center}
  \begin{tabular}{|c|ccc|}
  \hline
  Process           & $\sigma\times\mathcal{L}$    &  $n_b=4$             &  VBF    \\ 
  \hline
  SM signal         & $4.28\times 10^5$            &  $1.03\times 10^5$   &  $8.96\times 10^4$          \\ 
  \hline
  $pp\to 4b2j$      & $5.02\times 10^{10}$         &  $1.21\times 10^{10}$&  $8.51\times 10^9$\\ 
  $pp\to 2b4j$      & $5.04\times 10^{12}$         &  $2.47\times 10^{6}$ &  $1.83\times 10^6$ \\ 
  $pp\to t\bar{t}\to 2b4j$  & $3.93\times 10^{11}$ &  $1.93\times 10^5$   &  $6.20\times 10^4$\\
  \hline
  \end{tabular}
  \end{center}
  \caption{The cut efficiencies of b-tagging and VBF at 100 TeV collider are demonstrated. Here, the total integrated luminosity is assumed to be $\mathcal{L}=30$ ab$^{-1}$. b-tagging efficiency is $\epsilon_b=0.7$, and miss tagging rate is $\epsilon_{miss}=0.001$. \label{cut:SM100}} 
\end{table}
\end{center}

\begin{center}
\begin{table}
  \begin{center}
  \begin{tabular}{|c|ccc|}
  \hline
                    &  2-boosted  Higgs    &  1-boosted Higgs  &  0-boosted Higgs          \\ 
                    & (2BH) & (1BH) & (0BH) \\
  \hline
  SM Signal         &  $4265$        &  $1.76\times 10^4$       &  $6.77\times 10^4$       \\ 
  \hline
  $pp\to 4b2j$      &  $3.65\times 10^8$          &  $2.01\times 10^9$   &  $6.13\times 10^9$\\ 
  $pp\to 2b4j$      &  $8.35\times 10^4$          &  $4.40\times 10^5$   &  $1.31\times 10^6$ \\ 
  $pp\to t\bar{t}\to 2b4j$  & $8244$    &  $2.20\times 10^4$       &  $3.18\times 10^4$\\
  \hline
  \end{tabular}
  \end{center}
  \caption{The numbers of events for 2BH, 1BH and 0BH cases at 100 TeV collider are tabulated. \label{number:SM100}} 
\end{table}
\end{center}

\begin{center}
\begin{table}
  \begin{center}
  \begin{tabular}{|c|ccc|}
  \hline
                    &  2-boosted  Higgs    &  1-boosted Higgs  &  0-boosted Higgs          \\ 
                    & (2BH) & (1BH) & (0BH) \\
  \hline
  SM Signal with BDT cut     &  $298$         &  $12$       &  $90$              \\ 
  \hline
  Background with BDT cut    &  $0$           &  $0$        &  $78$ \\
  \hline
  $S/B$                      &  -      &  -   &  $1.15$  \\
  \hline
  $S/\sqrt{S+B}$             &  $17.27$       &  $3.43$     &  $6.923$   \\
  \hline
  \end{tabular}
  \end{center}
  \caption{The significances of the BDT at 100 TeV collider are demonstrated. \label{significance:SM100}}
\end{table}
\end{center}

\section{Multi-Higgs production in VBF processes with dimension-6 operators}
\label{Sec:hhh-EFT}

In this section, we extend our study of the $pp\to hhjj$ process
to contributions beyond the SM.  Given
the phenomenological Lagrangian~\ref{lvh}, such effects are parameterized in
terms of the coefficients $g_i$, where we focus in particular on the $\gvb2$
coupling that describes a non-SM double-Higgs interaction with the transverse
polarization components of the vector bosons.

\subsection{Effective-Theory description}
If we introduce the gauge symmetry of the SM in the phenomenological
description, any anomalous effects can be 
re-expressed in terms of higher-dimensional gauge-invariant operators.  To avoid
redundancy, it is convenient to choose a particular operator basis such as the
SILH basis~\cite{Giudice:2007fh} that we adopt for this work.  As usual, we truncate the
power-series expansion of the gauge-invariant effective theory at dimension
six.  This truncation allows us to express the phenomenological couplings
in~\ref{lvh} in terms of a small number of SILH operator coefficients. Such a
simplification allows for relating quantitative results for the VBF Higgs-pair
process to the analysis of existing data, as well as to studies of different
processes and interations.

We list the relevant terms in the SILH basis,
\begin{align}
	\begin{split}
{\cal L}_\text{SILH} \supset
&\frac{ic_Wg}{2m_\rho^2}\left( H^\dagger  \sigma^i \overleftrightarrow {D^\mu} H \right )( D^\nu  W_{\mu \nu})^i
+\frac{ic_Bg'}{2m_\rho^2}\left( H^\dagger  \overleftrightarrow {D^\mu} H \right )( \partial^\nu  B_{\mu \nu})  \\
& +\frac{ic_{HW} g}{16\pi^2f^2}
(D^\mu H)^\dagger \sigma^i(D^\nu H)W_{\mu \nu}^i
+\frac{ic_{HB}g^\prime}{16\pi^2f^2}
(D^\mu H)^\dagger (D^\nu H)B_{\mu \nu}
 \\
&+\frac{c_\gamma {g'}^2}{16\pi^2f^2}\frac{g^2}{g_\rho^2}H^\dagger H B_{\mu\nu}B^{\mu\nu}.
\label{lsilh}
\end{split}
\end{align}
In the following, we make use of the result from \cite{Kilian:2017nio,
  Kilian:2018bhs} in Table \ref{relation} which relates the phenomenological
coefficients such as $g_{V, b}$ to the parameters of~\ref{lsilh}.   In
particular, the couplings $g_{V, b}$ that we are interested in, only depend on
$c_{HW},c_{HB}$ and $c_\gamma$, but not on $c_W$ and $c_B$ which determine
$g_{V, a}$.   The SILH parameterization has been evaluated as
the low-energy limit of
various ultra-violet theories, such as models where the Higgs becomes a pseudo
Nambu-Goldstone boson theory or holographic completions with extra
dimensions; cf.\ \cite{Qi:2019ocx, Agrawal:2019bpm, Xu:2019xuo,
  Li:2019ghf}.   We note the correlations between $g_{V, a1}$ and $g_{V, a2}$,
$g_{V, b1}$ and $g_{V, b2}$ which follow from the dimension-six truncation,
and should be tested against real data if actual deviations from the SM show
up. 

In the SILH effective Lagrangian, Higgs interactions include further
dimension-six
operators such as  
\begin{eqnarray}
  \frac{c_H}{2f^2}\partial^\mu \left( H^\dagger H \right) \partial_\mu \left( H^\dagger H \right)
+ \frac{c_T}{2f^2}\left (H^\dagger {\overleftrightarrow { D^\mu}} H \right)  \left(   H^\dagger{\overleftrightarrow D}_\mu H\right).
\end{eqnarray}
The coefficients of these operators are switched off in our analysis because
they are to be measured in different processes.  In particular, the first one
entails a global shift to all Higgs interactions which is equivalent to a
modified Higgs total width, while the latter violates the custodial symmetry
of weak interactions and globally modifies $ZZ$ vs.\ $WW$ Higgs couplings.  In
our current work we assume that no custodial-symmetry violation beyond the SM
is present, as detailed below.

In the third column of Table \ref{relation}, we present some typical numerical
values for these parameters. 
Considering $g \sim 0.654 , g' \sim 0.350, v \sim 246 \text{GeV}, \text{tan}
\theta = g'/g= 0.535 $, $\alpha = \frac{g^2v^2}{32\pi^2f^2} = 8.2 \times
10^{-5}$ with $f = 1$ TeV and $g_\rho \sim \frac{4 \pi}{\sqrt{3}}$, $m_\rho =
g_\rho f =7.3$ TeV, we can simplify the expressions in this table with 
\begin{align}
&\zeta_h \sim 1, \quad 
 \zeta_A \sim 1, \quad 
 y_{ZA} \sim \alpha \frac{g'}{g} (c_{HW} - c_{HB}) \quad \\
& \zeta_Z\sim  1- \frac{1}{8} \alpha^2 (\frac{g'}{g})^2 (c_{HW} - c_{HB})^2, \quad 
 \zeta_{AZ} \sim  \alpha \frac{g'}{4g} (c_{HW} - c_{HB}),  \quad 
   \zeta_W  \sim 1- \frac{\alpha}{2} c_{HW}
\end{align}
(Our notation slightly differs from the definitions for $\bar{c}_i$ used in Ref.~\cite{ Ellis:2018gqa}.)
To simplify our discussion on the future collider sensitivity on $g_{V,b2}$
and $g_{V,a2}$, we assume custodial-symmetry relations, namely that
$g_{W,b2}=g_{Z,b2}$ and $g_{X,b2}=g_{A,b2}=0$ and 
$g_{W,a2}=g_{Z,a2}$.  This roughly renders $ c_\gamma, c_{HB} , c_B \sim 0$,
so we can perform a two-parameter study on $c_W$ and $c_{HW}$.   
The interactions proportional to 
$g_{W,b2}$ and $g_{W, a2}$ 
($g_{Z,b2}$ and $g_{Z, a2}$) account for dominant contributions to the
cross section of $pp\to hh jj$
process, up to $70\%$ (30$\%$), respectively.
\begin{center}
\begin{table}
	\footnotesize 
  \begin{center}
  \begin{tabular}{|c|c|c|c|}
  \hline
           &  SILH         & numerics  with assumptions below             \\
  \hline
   $\gwb1$            &  $c_{HW}\frac{g^2v^2}{32\pi^2f^2}\zeta_h\zeta^2_W$  & $10^{-4} c_{HW}$    \\
  \hline
  $\gwb2$            &   $\gwb1 \zeta_h$  & $10^{-4} c_{HW}$     \\
  \hline
  $\gab1$           &  $-c_\gamma\frac{g^2v^2}{8\pi^2f^2}\frac{{g^\prime}^2}{g^2_\rho}\cos^2{\theta}\zeta_h\zeta_A^2$       & -$10^{-6} c_\gamma$    \\
  \hline
  $\gab2$           &    $\gab1 \zeta_h$           & -$10^{-6} c_\gamma$         \\
  \hline
  \multirow{2}*{$\gxb1$}     & $\frac{g{g^\prime}v^2}{64\pi^2f^2}\left[(c_{HW}-c_{HB})+8c_\gamma\frac{g^2}{g^2_\rho}\sin^2\theta\right]\zeta_h\zeta_{A}\zeta_{Z}$  & \\
                             & $+ c_\gamma\frac{g^2v^2}{4\pi^2f^2}\frac{{g^\prime}^2}{g^2_\rho}\cos^2{\theta}\zeta_h\zeta_{AZ}^2$    & $10^{-5} (c_{HW}-c_{HB})$    \\
  \hline
$\gxb2$    & $\gxb1 \zeta_h$   & $10^{-5} (c_{HW}-c_{HB})$   \\
  \hline
$\gzb1$     &  $\frac{g^2v^2}{32\pi^2f^2}(c_{HW}+c_{HB}\tan^2{\theta})\zeta_h\zeta^2_Z - c_\gamma\frac{g^2v^2}{8\pi^2f^2}\frac{{g^\prime}^2}{g^2_\rho}\cos^2{\theta}\zeta_h\zeta_{AZ}^2$   &     \\
  &  $-\frac{g{g^\prime}v^2}{64\pi^2f^2}\left[(c_{HW}-c_{HB})+8c_\gamma\frac{g^2}{g^2_\rho}\sin^2\theta\right]\zeta_h\zeta_{AZ}\zeta_{Z}$    & $10^{-4} (c_{HW}+0.29c_{HB})$     \\
  \hline
$\gzb2$     &  $\gzb1 \zeta_h$    & $10^{-4} (c_{HW}+0.29 c_{HB})$      \\ 
\hline
  $\gwa1$ & $\left[1-\left(c_{W}\frac{g^2v^2}{m^2_{\rho}}+c_{HW}\frac{g^2v^2}{16\pi^2f^2}\right)\right]\zeta_h\zeta^2_W$.  & $1- 2\times10^{-4} (3 c_W + c_{HW}$ ) \\
   \hline
  $\gwa2$ &$\left[1-3\left(c_{W}\frac{g^2v^2}{m^2_{\rho}}+c_{HW}\frac{g^2v^2}{16\pi^2f^2}\right)\right]\zeta^2_h\zeta^2_W$   & $1- 6\times 10^{-4} (3 c_W + c_{HW}$ )  \\
  \hline
  $\gza1$ &$ \left[1-\left(c_{W}\frac{g^2v^2}{m^2_{\rho}}+c_{B}\frac{{g^\prime}^2v^2}{m^2_{\rho}}+c_{HW}\frac{g^2v^2}{16\pi^2f^2}+c_{HB}\frac{{g^\prime}^2v^2}{16\pi^2f^2}\right)\right]\zeta_h\zeta^2_Z$  & $1- 2\times10^{-4} [3 (c_{W}+0.29c_{B})+c_{HW}+0.29 c_{HB}]$    \\
   \hline
  $\gza2$ & $\left[1-3\left(c_{W}\frac{g^2v^2}{m^2_{\rho}}+c_{B}\frac{{g^\prime}^2v^2}{m^2_{\rho}}+c_{HW}\frac{g^2v^2}{16\pi^2f^2}+c_{HB}\frac{{g^\prime}^2v^2}{16\pi^2f^2}\right)\right]\zeta^2_h\zeta^2_Z$.   & $1- 6\times10^{-4}  [3 (c_{W}+0.29c_{B})+c_{HW}+0.29 c_{HB}]$    \\
  \hline
\end{tabular}
  \end{center}
  \caption{\label{table:parameter}  Relations between the phenomenological
    Lagrangian parameters in~(\ref{eft}-\ref{lvh}) (first column), the SILH
    effective Lagrangian~\ref{lsilh} (second column),  The extra parameters $\zeta^n_h$, $\zeta^n_W$, $\zeta^n_Z$, $\zeta^n_A$,
    $\zeta^n_{AZ}$ (defined in
  \cite{Kilian:2017nio, Kilian:2018bhs})  are induced by the Higgs and gauge-boson wave-function
    normalization, respectively.
} 
\label{relation}
\end{table}
\end{center}

\subsection{Higgs-pair couplings to transverse vector polarizations in VBF} 

\begin{figure}[bthp]
  \centering
  \subfigure{
\label{pth_2h014b2}   \thesubfigure
  \includegraphics[width=0.40\textwidth]{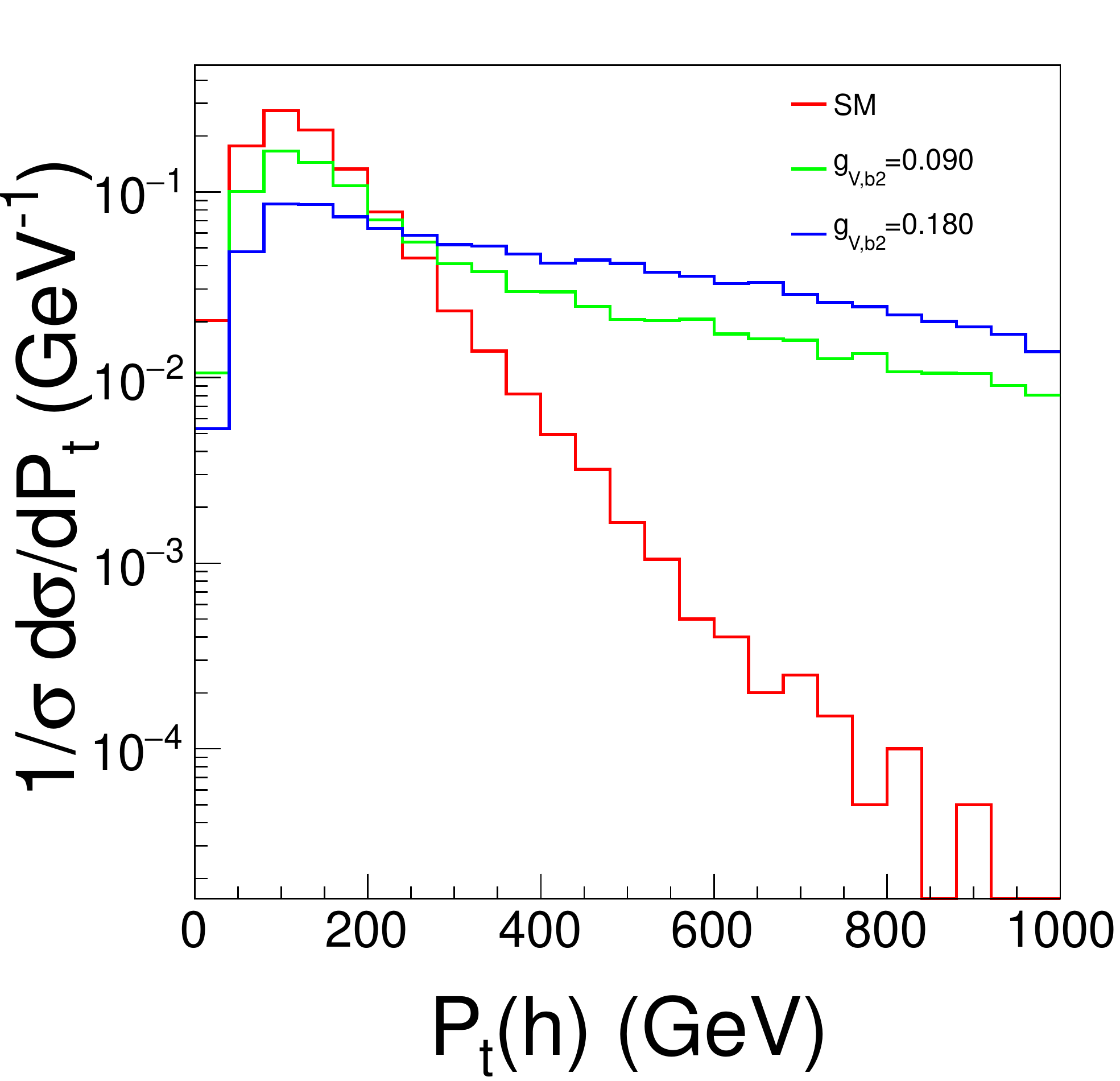}}
  \subfigure{
\label{mhh_2h014b2}   \thesubfigure
  \includegraphics[width=0.40\textwidth]{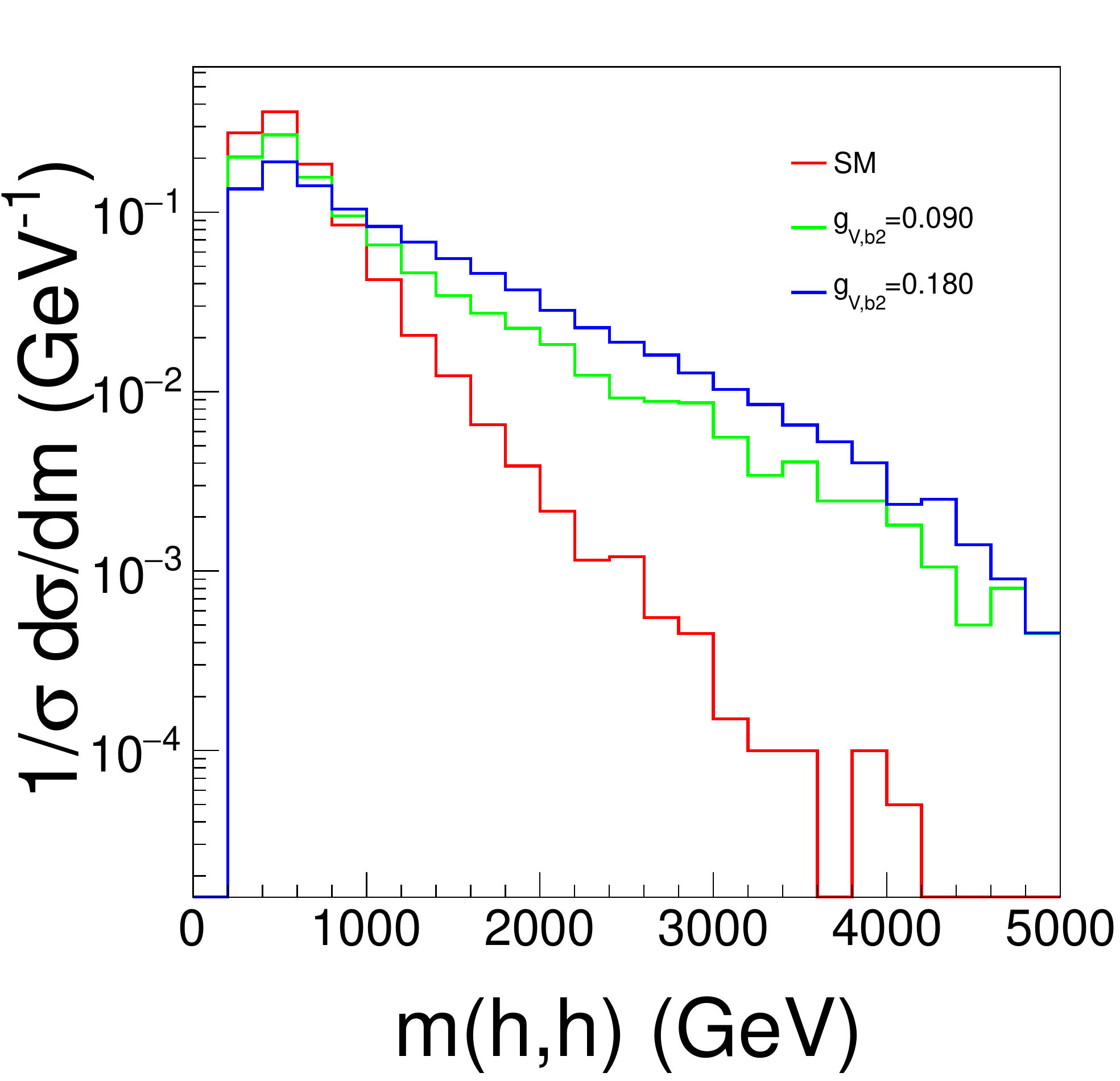}}
  \subfigure{
\label{pth_2h100b2}   \thesubfigure
  \includegraphics[width=0.40\textwidth]{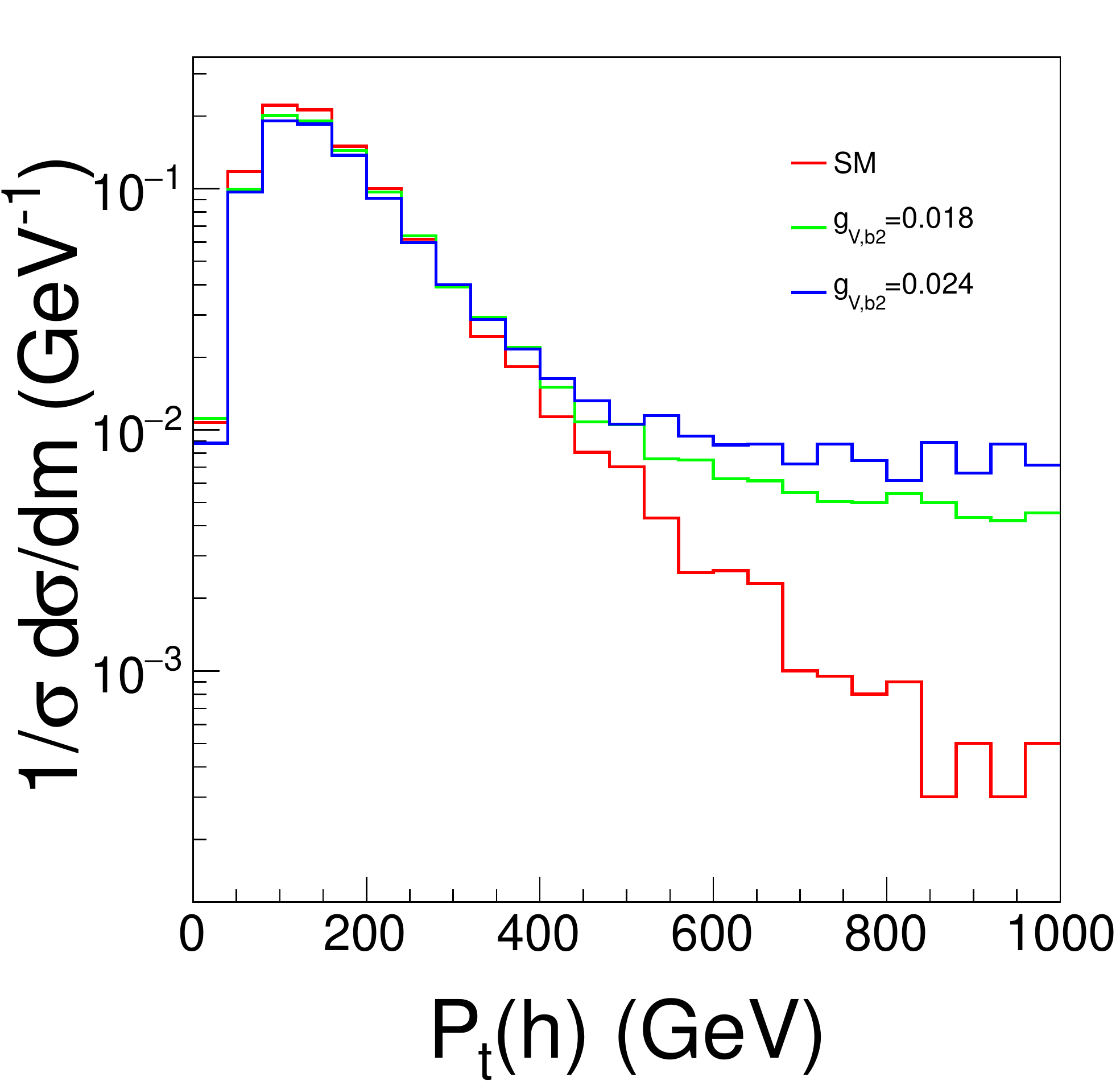}}
  \subfigure{
\label{mhh_2h100b2}   \thesubfigure
  \includegraphics[width=0.40\textwidth]{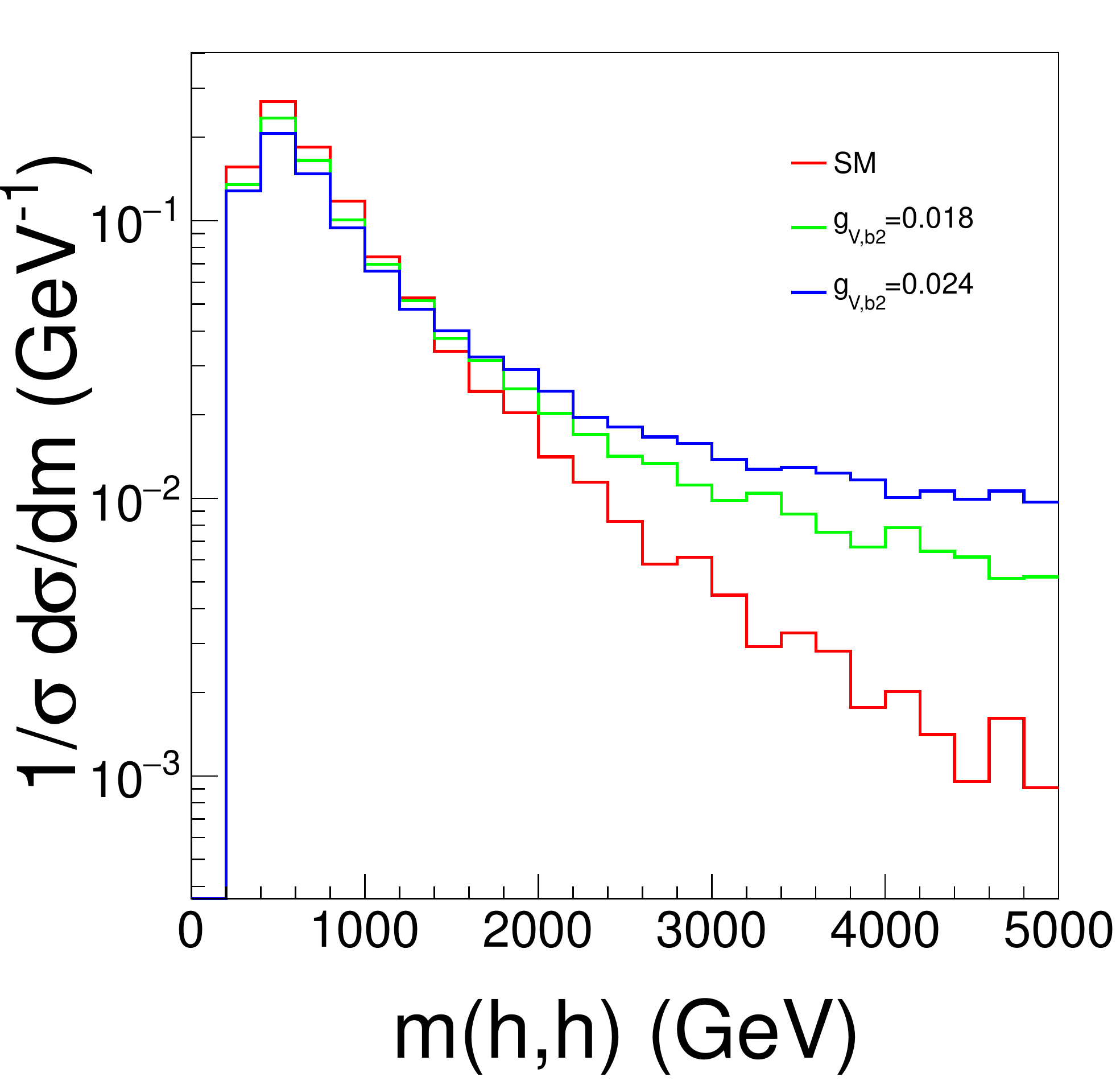}}
  \caption{Distributions of (a) the $P_t$ of the leading Higgs and (b) the
    invariant mass of the Higgs pair at the 14 TeV LHC, with $\gvb2=0.09$ and
    $\gvb2=0.18$. The corresponding distributions for a 100 TeV collider are
    shown in (c) and (d), with $\gvb2=0.018$ and
    $\gvb2=0.024$.} \label{fig:gb2} 
\end{figure}

Introducing non-SM effects proportional to the $\gvb2$ couplings, we consider
kinematical distributions and their discriminating power. 
In Fig.~\ref{fig:gb2}, we display the distributions of the $P_t$ of the
leading Higgs boson and the invariant mass the Higgs-bosons pair, respectively.
The subfigures (a) and (b) show the distributions at the LHC with collision
energy 14 TeV.  The new-physics effect is illustrated by the green and blue
curves which correspond to two different values of $\gvb2$, namely
$\gvb2=0.09$ and $\gvb2=0.18$, repectively.  We observe a huge enhancement at
high $P_t$ in the curves which include the new
interaction of Higgs bosons with transversal gauge bosons.  For the selected
parameter values, 
the fraction of events in the region with $P_t > 200$ GeV increases to $50\%$
and $70\%$, respectively, while in the SM this fraction is just $18\%$.

The corresponding distributions at a 100 TeV hadron collider are shown in the
Fig.~\ref{fig:gb2} (c) and (d), where we select $\gvb2 = 0.018$ and $0.024$.
In this case, the fraction of events in the region with $P_t > 200$ GeV
increases to $50\%$ and $60\%$, respectively, while in the SM, it is $25\%$.

Subfigures (b) and (d) contain the Higgs-pair invariant-mass
distributions for the LHC and for the 100 TeV collider, which are
likewise enhanced in the high-mass region if anomalous effects are included.


As a concrete numerical example for the analysis in presence of new-physics
effects, in Table \ref{cut:gb2014} we demonstrate the cut flow for the value
$\gvb2=0.18$ at the 14 TeV LHC. 
For this parameter value, the total signal cross section
($\sigma\times\mathcal{L}$) is enhanced by a factor of 4 over the SM value.
As Figure (\ref{fig:gb2}a) demonstrates, most of enhancement occurs in the
boosted region. After b-tagging and VBF cuts have been applied, there are more than $600$
signal events left that can be observed. In Table.~\ref{number:gb2014} we
present the resulting numbers for the three classes of events.
In this scenario, the signal accounts for $25\%$ and $35\%$ of the total
events in the 2BH and 1BH classes, respectively. Compared with the results
given in Table \ref{significance:SM014}, the increase in signal events in the
2BH (1BH) case amounts to a factor 35 (10), respectively.

\begin{figure}
  \setcounter{subfigure}{0}
  \centering
  \subfigure[2BH]{
  \label{bdt_boost}
  \includegraphics[width=0.3\textwidth]{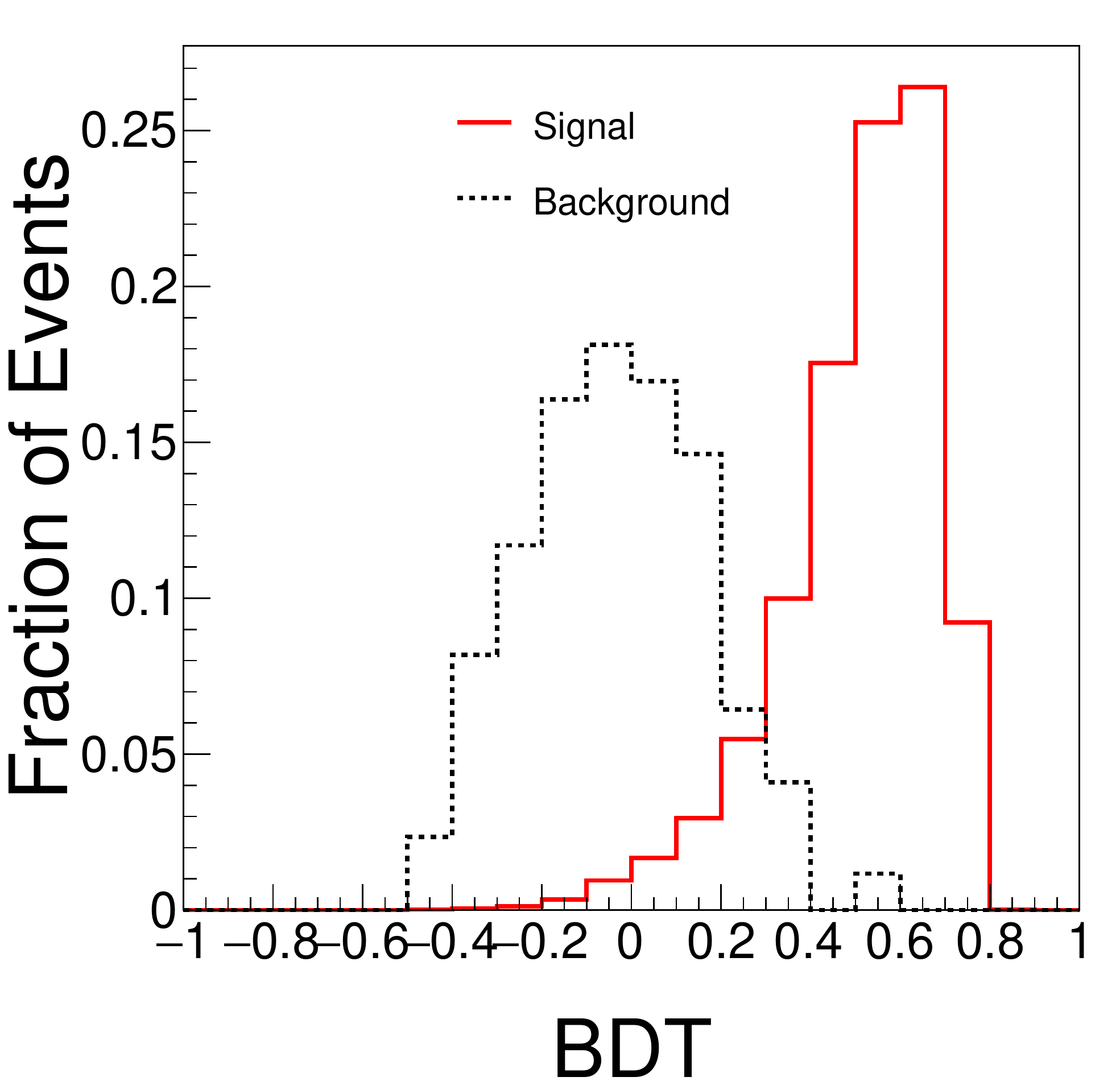}}
  \subfigure[1BH]{
  \label{bdt_inter}
  \includegraphics[width=0.3\textwidth]{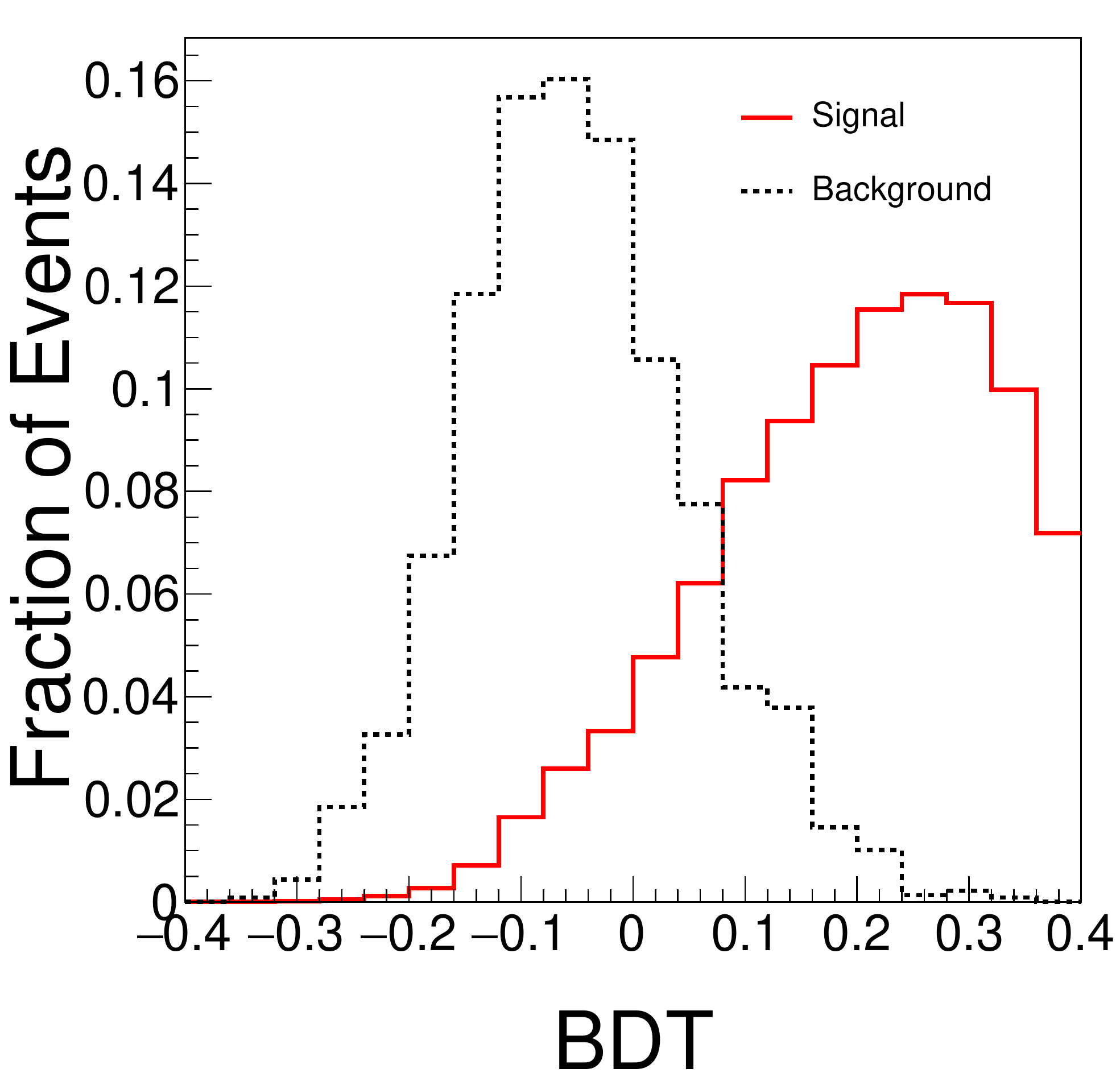}}
  \subfigure[0BH]{
  \label{bdt_resolved}
  \includegraphics[width=0.3\textwidth]{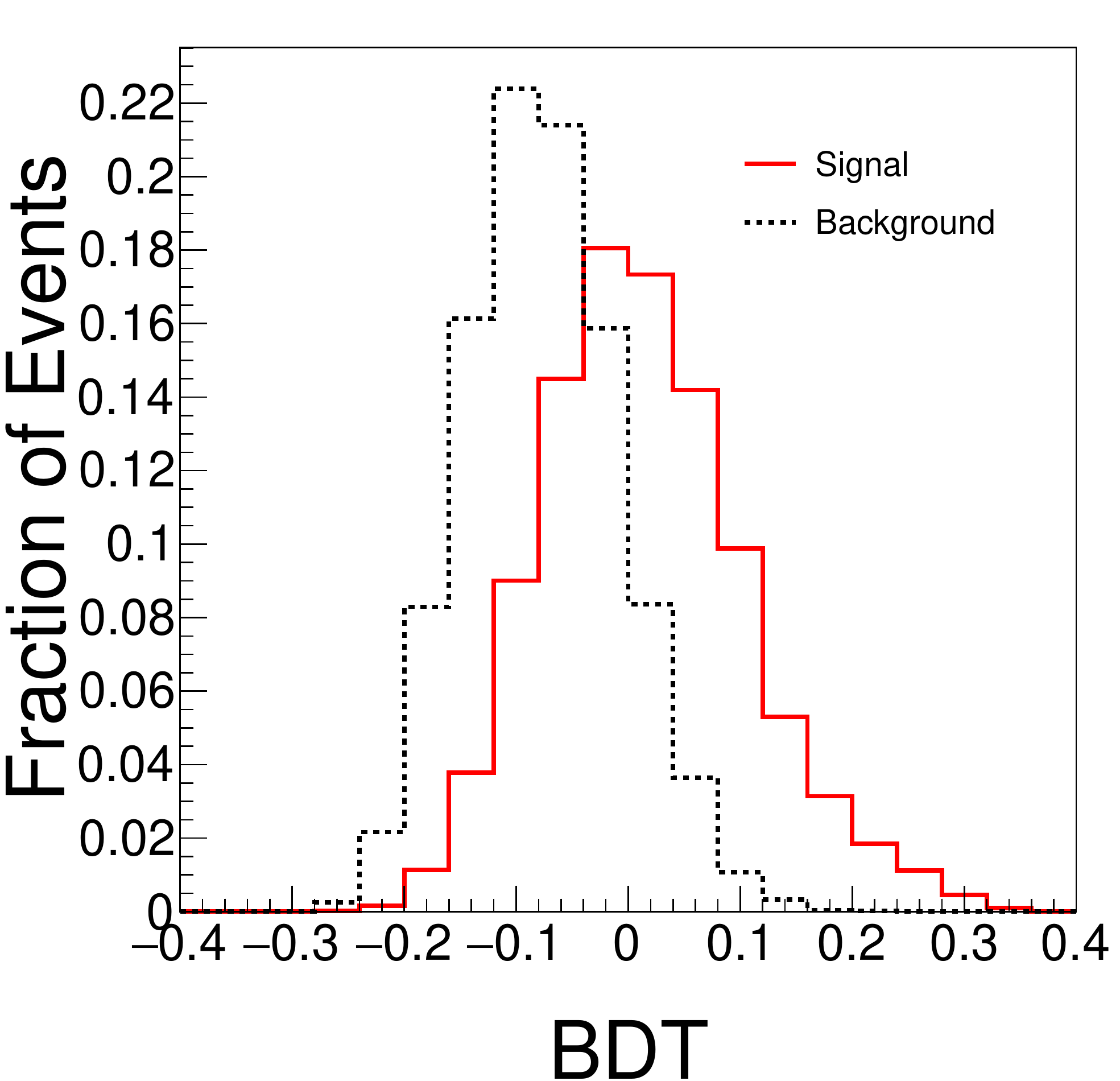}}
  \caption{The response of the discriminants to the $\gvb2$ signal and background at 14 TeV LHC with (a) 2BH, (b) 1BH, and (c) 0BH.} \label{fig:gb2BDT014}
\end{figure}

When the BDT method is applied, this result is further improved, as shown in
Fig.~\ref{fig:gb2BDT014} and Table.~\ref{significance:gb2014}.  By comparing
this with the results of Table \ref{significance:SM014}, we conclude that
isolating the boosted-Higgs region significantly enhances the discovery
potential of this process, both for the 2BH and 1BH cases.

\begin{table}
  \begin{center}
  \begin{tabular}{|c|ccc|}
  \hline
  Process           & $\sigma\times\mathcal{L}$   &  $n_b=4$             &  VBF    \\ 
  \hline
  $g_{V,b2}=0.18$ signal & $4243$                 &  $1019$              &  $617$          \\ 
  \hline
  $pp\to 4b2j$      & $2.28\times 10^8$           &  $5.47\times 10^7$   &  $1.86\times 10^7$\\ 
  $pp\to 2b4j$      & $2.38\times 10^{10}$        &  $1.14\times 10^{4}$ &  $3.85\times 10^4$ \\ 
  $pp\to t\bar{t}\to 2b4j$  & $7.89\times 10^8$   &  $387$               &  $58$\\
  \hline
  \end{tabular}
  \end{center}
  \caption{The cut efficiencies of b-tagging and VBF at 14 TeV LHC are demonstrated. Here, the total integrated luminosity is assumed to be $\mathcal{L}=3$ ab$^{-1}$. b-tagging efficiency is $\epsilon_b=0.7$, and mis-tagging rate is $\epsilon_{miss}=0.001$. \label{cut:gb2014}} 
\end{table}

\begin{table}
  \begin{center}
  \begin{tabular}{|c|ccc|}
  \hline
                    &  2-boosted  Higgs    &  1-boosted Higgs  &  0-boosted Higgs          \\ 
                    & (2BH) & (1BH) & (0BH) \\
  \hline
  $g_{V,b2}=0.18$ Signal         &  $153$          &  $217$       &  $247$             \\ 
  \hline
  $pp\to 4b2j$      &  $1.17\times 10^5$           &  $1.56\times 10^6$   &  $1.69\times 10^7$\\ 
  $pp\to 2b4j$      &  $28$          &  $349$      &  $3.81\times 10^4$ \\ 
  $pp\to t\bar{t}\to 2b4j$  & $3$    &  $13$       &  $42$\\
  \hline
  \end{tabular}
  \end{center}
  \caption{The numbers of 2BH, 1BH and 0BH cases at 14 TeV LHC are tabulated. \label{number:gb2014}} 
\end{table}


By contrast, in the 0BH case the signal significance is too low for the chosen
collider parameters and benchmark values of $\gvb2$.

\begin{center}
\begin{table}
  \begin{center}
  \begin{tabular}{|c|ccc|}
  \hline
                    &  2-boosted  Higgs    &  1-boosted Higgs  &  0-boosted Higgs          \\ 
                    & (2BH) & (1BH) & (0BH) \\
  \hline
  $g_{V,b2}=0.18$ Signal with BDT cut     &  $54$    &  $30$       &  $10$              \\ 
  \hline
  Background with BDT cut    &  $0$       & $0$      & $3430$ \\
  \hline
  $S/B$                      &  -  & - &  $2.99\times 10^{-3}$  \\
  \hline
  $S/\sqrt{S+B}$             &  $7.348$       &  $5.477$    &  $0.175$   \\
  \hline
  \end{tabular}
  \end{center}
  \caption{The significances of the BDT at 14 TeV collider are demonstrated. \label{significance:gb2014}}
\end{table}
\end{center}



\subsection{Unitarity limits and discovery reach}

The interactions of Higgs bosons with transverse gauge bosons in the
Lagrangian~(\ref{lvh}) involve derivative couplings, and
therefore are enhanced over the SM interactions for high values of the
four-momenta.  This is clearly visible in Fig.~\ref{fig:gb2}, where the
contribution of the new interactions dominates for sufficiently large values of the
transverse momentum $P_t$ or the Higgs-pair invariant mass~$m(h,h)$.

If this behavior is naively extrapolated, the computed amplitudes will violate
unitarity constraints.  In Ref.~\cite{Kilian:2018bhs}, we have derived a generic
unitarity bound for the $pp\to hhjj$ process, which relates the value of
$\gvb2$ to a UV cutoff $\Lambda_{UV}$.
\begin{eqnarray}
  \frac{\Lambda^4_{UV}}{2^9\pi^2v^4}|\gvb2|^2\leq\frac{1}{4},
  \label{unitarityb2}
\end{eqnarray}
For energy-momentum values beyond $\Lambda_{UV}$, the EFT
expansion breaks down.  We expect higher-order contributions and,
eventually, a new structure of underlying interactions to dampen the rise of
the amplitudes, e.g., a resonance.  Conversely, for the EFT description to
remain useful, the value of $\gvb2$ must be such that the cutoff implied
by~(\ref{unitarityb2}) is outside the accessible kinematic range, or we would
have to apply a cutoff or form factor on the calculated distributions.

\begin{figure}[bthp]
  \centering
  \subfigure{
  \label{cutoff014_boost}   \thesubfigure
  \includegraphics[width=0.40\textwidth]{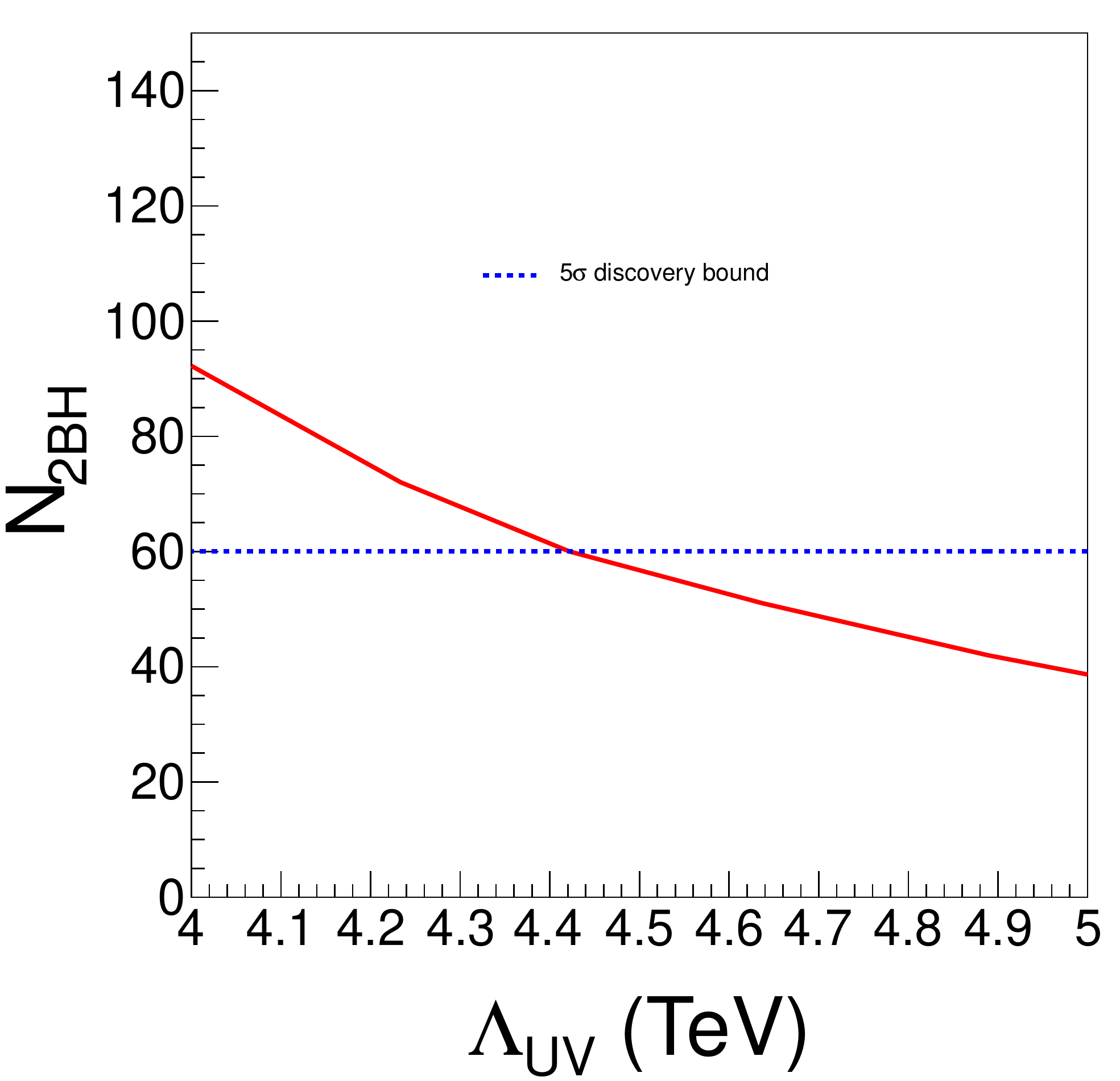}}
  \subfigure{
  \label{cutoff014_inter}   \thesubfigure
  \includegraphics[width=0.40\textwidth]{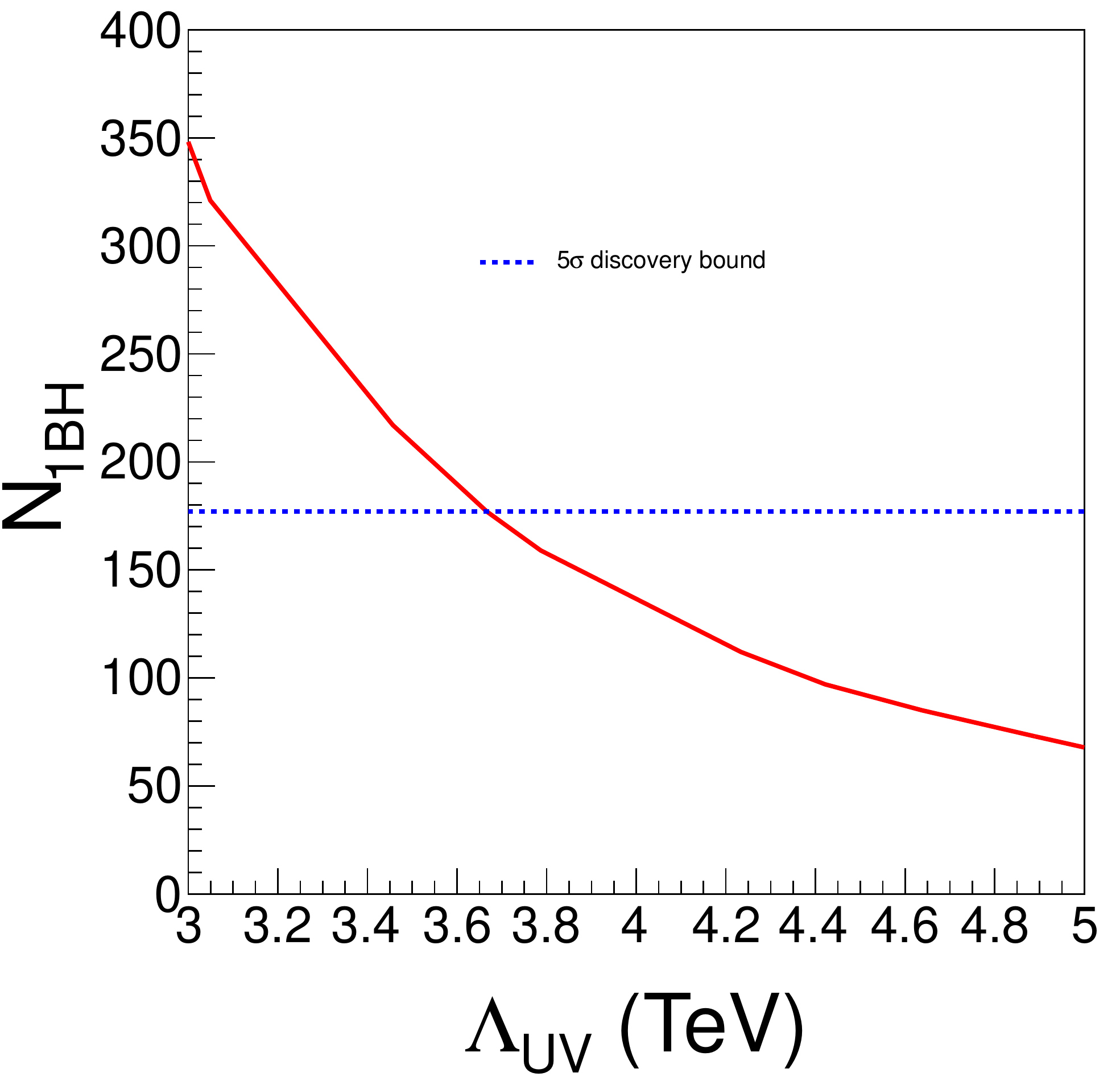}}
  \subfigure{
  \label{cutoff014_boost}   \thesubfigure
  \includegraphics[width=0.40\textwidth]{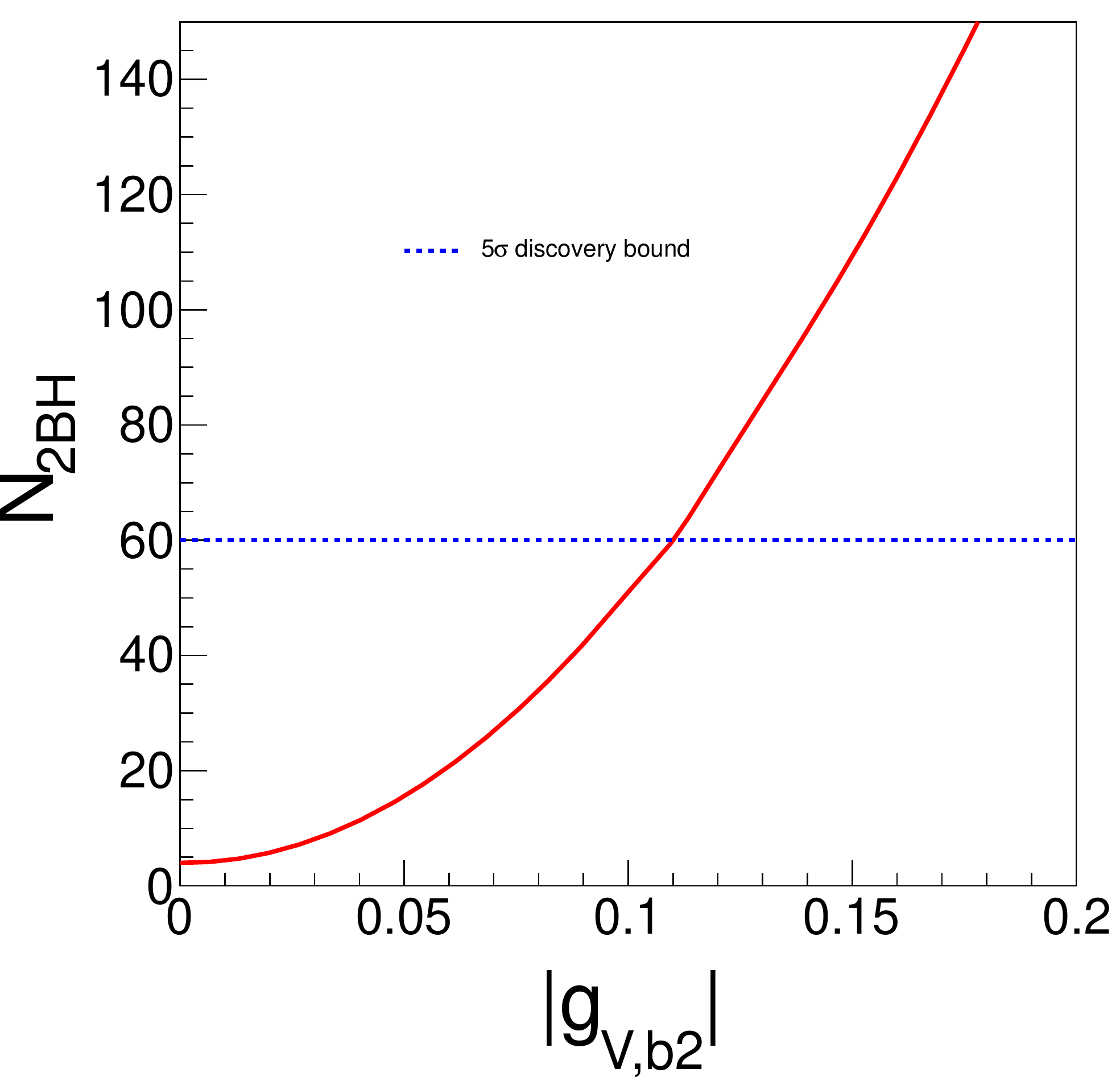}}
  \subfigure{
  \label{cutoff014_inter}   \thesubfigure
  \includegraphics[width=0.40\textwidth]{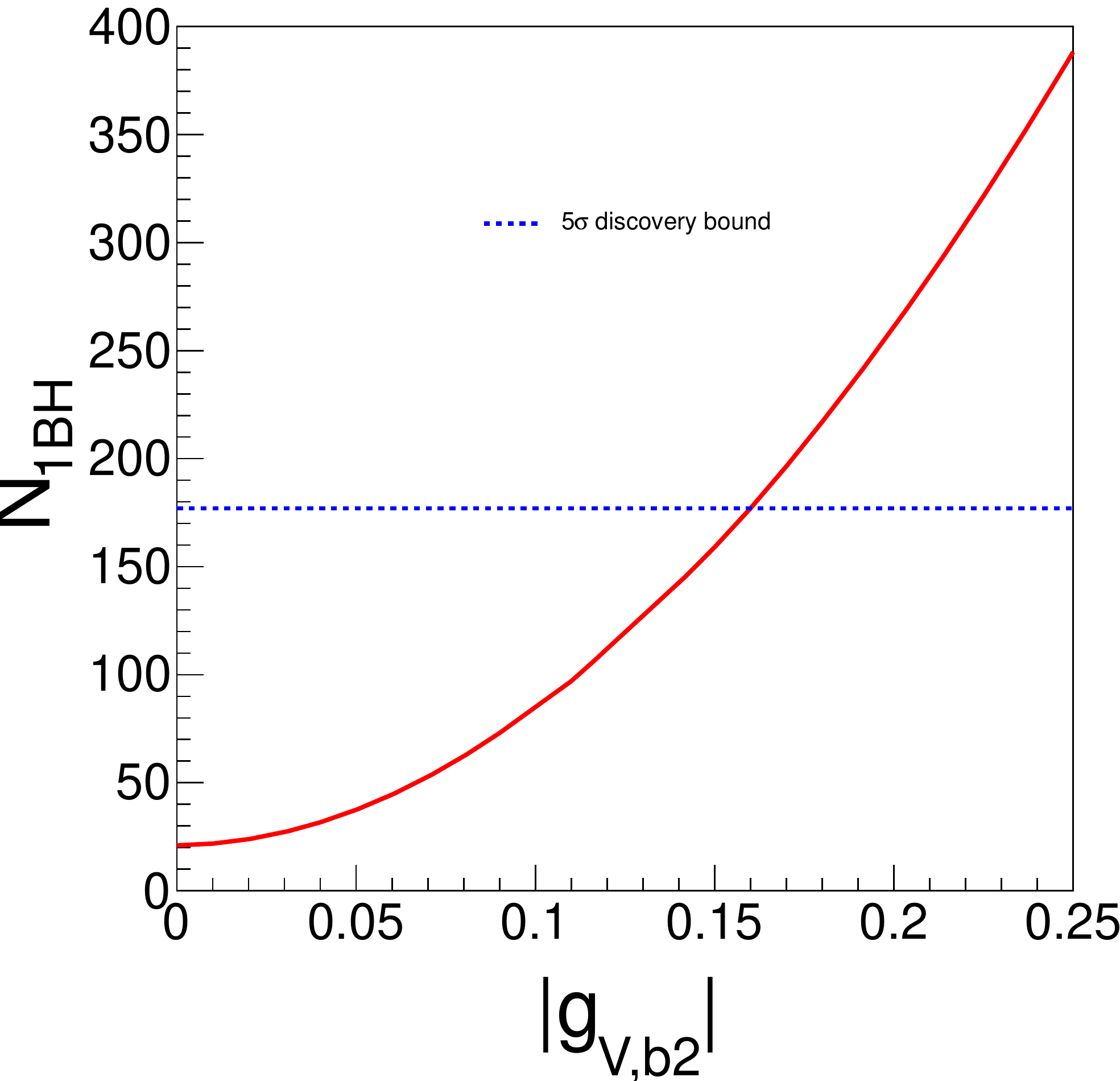}}
  \caption{The $5\sigma$ discovery constraints on new physics cutoff at 14 TeV LHC are demonstrated. (a) is the constraint obtained from 2BH case, and (b) is the constraint obtained from 1BH case. (c) and (d) are the corresponding constraints on the $\gvb2$ in the 2BH and 1BH case, respectively.} \label{fig:cons014}
\end{figure}

In Fig.~\ref{fig:cons014}, we display the sensitivity on $\Lambda_{UV}$ by
using the (a) 2BH case and (b) 1BH case at the 14 TeV LHC, respectively.  To this
end, we determine the maximally allowed $\gvb2$ value as a function of
$\Lambda_{UV}$ using~(\ref{unitarityb2}).  The y-axis indicates the number of
signal events that result for corresponding values of the cutoff
$\Lambda_{UV}$ and the maximal $\gvb2$, respectively.  The
dashed blue line, for each plot, marks the $5\sigma$ discovery threshold as it
follows from the signal and background event rates derived above.

The crossing points of the signal curves and the discovery thresholds in
Fig.~\ref{fig:cons014} determine the sensitivity of the analysis to heavy new
physics, assuming that the bound~(\ref{unitarityb2}) for $\gvb2$ is saturated.
We conclude that the effect of a heavy resonance, for instance, can be
accessible up to $4.4$ TeV in the 2BH case and up to $3.6$ TeV in the 1BH
case.  The cleaner environment of the 2BH case is clearly preferred.  By
contrast, the discovery reach of the 0BH case is limited to about $2.4$
TeV. 

We can apply the analogous analysis to a 100 TeV hadron collider.
Based on the results obtained for the LHC with $\sqrt{s}= 14$ TeV where the
2BH case has been most useful, below we only present the results
for this analysis which requires two highly boosted Higgs bosons.

As shown above, at a 100 TeV machine the signal can be discovered already in
the SM case where no enhancement is present.  Therefore, we define the
observability of new physics in terms of the deviation in rate,
$N_{2BH}-N^{SM}_{2BH}$,  
where $N^{SM}_{2BH}$ is the number of events in the 2BH category in the SM case. 

\begin{figure}[bthp]
  \centering
  \subfigure{
  \label{cutoff100_boost} \thesubfigure
  \includegraphics[width=0.40\textwidth]{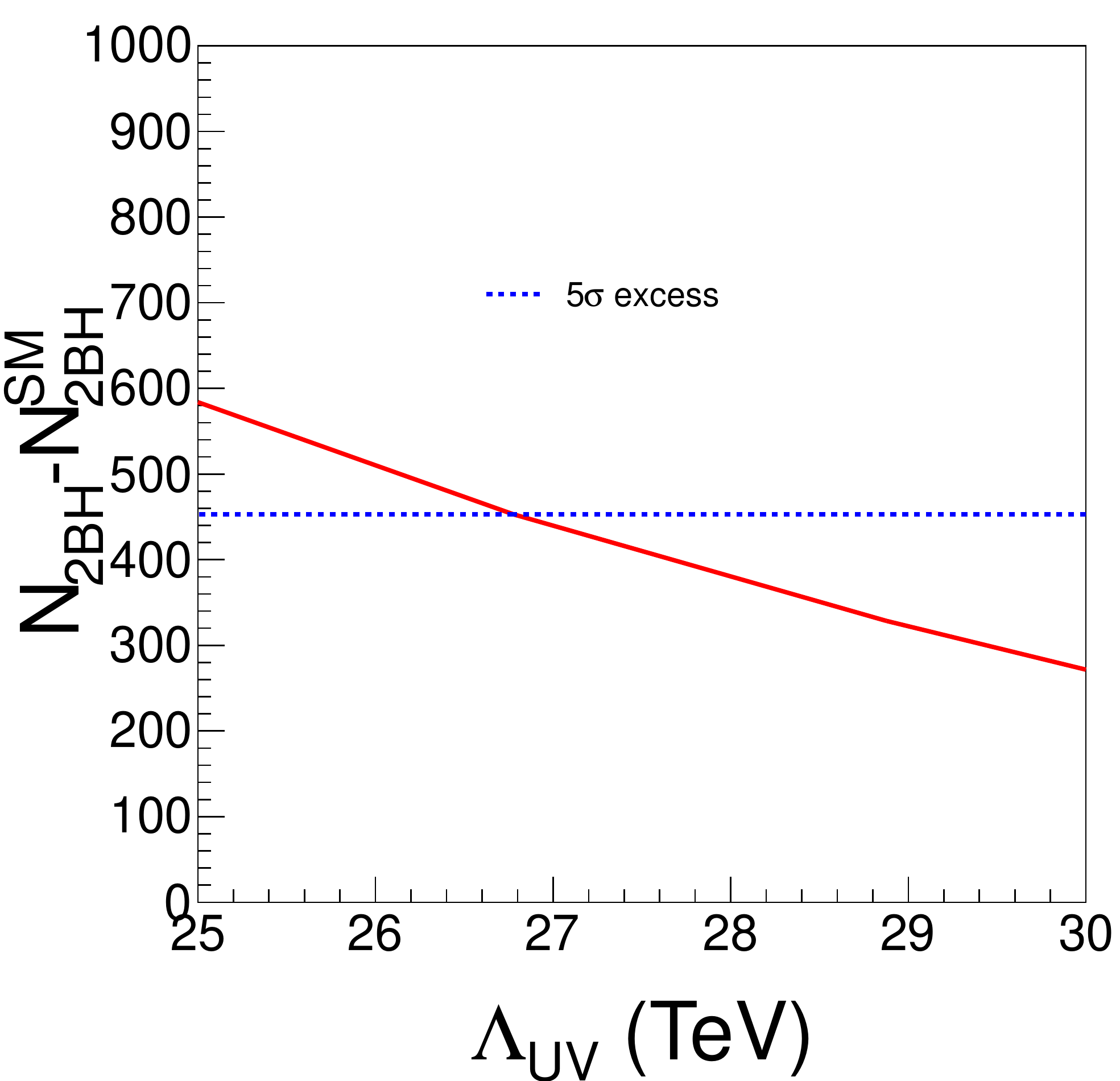}}
  \subfigure{
  \label{cutoff100_boost} \thesubfigure
  \includegraphics[width=0.40\textwidth]{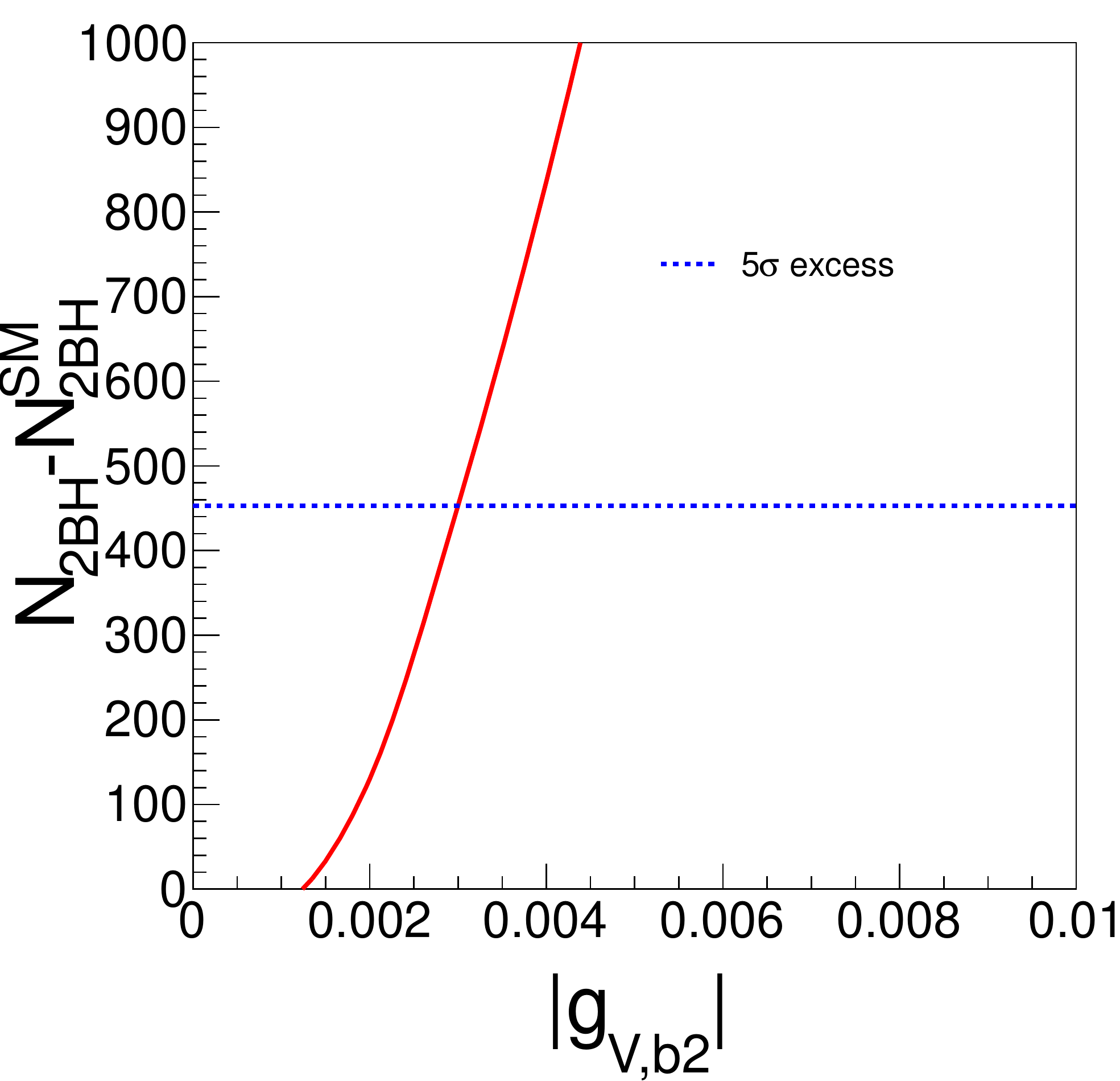}}
  \caption{The $5\sigma$ excess constraint obtained from 2BH events on new physics cutoff at 100 TeV hadron collider is demonstrated in (a), while (b) is the corresponding $\gvb2$.} \label{fig:cons100}
\end{figure}

By using the BDT analysis as before, we derive the $5\sigma$ excess bound
for new physics at the 100 TeV collider,  
as marked in Fig.~\ref{fig:cons100} by the blue dashed line.
We read off this figure that this collider can yield a
meaningful constraint on $\gvb2$ at a value which corresponds to a
sensitivity in the new-physics scale of $\Lambda_{UV}=27$ TeV.

\section{Two-parameter bounds}
\label{projgvb2}

So far, we have considered only the dependence of the Higgs-pair VBF process
on the couplings $\gvb2$ to the transverse polarization of vector bosons,
which dominate the distribution in the highly boosted region.  In this
section, we complement the discussion by taking into account also the $\gva2$
couplings which describe the interaction of a Higgs pair with longitudinally
polarized vector bosons.  This interaction exists in the SM but can receive a
correction if dimension-6 operators are included.

In order to express the cross sections of $pp \to hhjj$ in terms of the
parameters given in Eq. (\ref{eft}), we impose some simplifying assumptions on the phenomenological
information from expected data on single-Higgs processes.  For
instance, the $WWh$ vertex should be strongly constrained by data from the
Higgs decay to $WW$ as well as from VBF single-Higgs production.  As stated
before, we ignore the universal ambiguity in the Higgs-coupling determination
due to undetected Higgs decays at the LHC, which can be lifted by
model-independent measurements at
Higgs factories such as the CEPC, the ILC, or the CLIC collider.  Accordingly,
we fix $\gwa1$ and $\gwb1$, the single-Higgs
couplings to longitudinal and transverse $W$ polarizations, to their SM
values.  As a further simplification, we 
assume the custodial-symmetry 
relations $g_{W,a i}=g_{Z,a i}$ and $g_{W,b i}=g_{Z,b i}$ ($i=1,2$) whenever
contributions of Z bosons are considered.  

Since the amplitudes are at most linear in the parameters, we can parameterize the cross section of $pp\to hhjj$ in terms of $g_{V,a2}$ as given below
\bea
\label{pphha2}
\sigma(pp \to hhjj) &=& \sigma^0_{a_2} + \sigma^1_{a_2} g_{V,a2} + \sigma^2_{a_2}  g_{V,a2}^2\,.
\eea
We compute the coefficients $\sigma^0_{a_2}$,
$\sigma^1_{a_2}$, and $\sigma^2_{a_2}$ numerically using Monte-Carlo methods, evaluating the
total cross section for a sufficient number of different coupling values.
The results 
are given in Table \ref{vbf2hefta2}. 

\begin{table}
  \begin{center}
  \begin{tabular}{|c|c|c|c|}
  \hline
& $\sigma^{0}_{a_2}$ (fb) & $\sigma^1_{a_2}$(fb) & $\sigma^2_{a_2}$(fb)  \\
\hline
14 TeV & $17.71$ & -$29.33$& $12.68$ \\ 
\hline
27 TeV & $88.3$ & - $152.2$ & $68.1$ \\  
\hline
100 TeV & $1601.2$ &  -$2963.8$ & $1401$  \\
\hline
 \hline
  \end{tabular}
  \end{center}
      \caption{Coefficients $\sigma^0_{a_2}$, $\sigma^1_{a_2}$, and $\sigma^2_{a_2}$ in the expression~\eqref{pphha2} for 
        $pp\to hhjj$ at three different collider energies. \label{vbf2hefta2}}  
\end{table}

Analogously, we can parameterize the cross section as a function on $g_{V,
  b2}$ as follows 
\bea
\label{pphhb2}
\sigma(pp \to hhjj) &=& \sigma^0_{b_2} + \sigma^1_{b_2} g_{V,b2} + \sigma^2_{b_2}  g_{V,b2}^2 \,,
\eea
The numerical results for $\sigma^0_{b_2}$, $\sigma^1_{b_2}$, and
$\sigma^2_{b_2}$ are given in Table \ref{vbf2heftb2}.

When both $g_{V,b2}$ and $g_{V,a2}$ are turned on, we have to include a mixed
coefficient proportional to $\gvb2\gva2$.  The corresponding results can be
found in 
\cite{Kilian:2018bhs}; the values are $1.7$ fb for 14 TeV, $9.6$ fb for 27
TeV, and 
$95$ fb for 100 TeV, respectively.

\begin{table}
  \begin{center}
  \begin{tabular}{|c|c|c|c|}
  \hline
& $\sigma^{0}_{b_2}$(fb) & $\sigma^1_{b_2}$(fb) & $\sigma^2_{b_2}$(fb)  \\
\hline
14 TeV & $1.06$ & $1.52$& $106.8$ \\ 
\hline
27 TeV & $4.2$ & $6.97$ & $1135.2$ \\  
\hline
100 TeV & $38.4$ &  $83.08$ & $54070$  \\
\hline
 \hline
  \end{tabular}
  \end{center}
      \caption{Coefficients $\sigma^0_{b_2}$, $\sigma^1_{b_2}$, and $\sigma^2_{b_2}$ in the expression~\eqref{pphhb2} for 
        $pp\to hhjj$ at three different collider energies. \label{vbf2heftb2}}  
\end{table}

The numerical results given in Table \ref{vbf2hefta2} and Table
\ref{vbf2heftb2} have been obtained with Madgraph5 and Whizard by independent
calculations, with very good
numerical agreement. It should be pointed out that the results
in Table \ref{vbf2hefta2} clearly reflect the strong gauge cancellation
which occurs between individual terms of~\eqref{pphha2} in the SM limit. Some
of the coefficients $\sigma^0_{a_2}$, $\sigma^1_{a_2}$, and $\sigma^2_{a_2}$
are one order of magnitude larger than that of the cross section in the SM, as
given as $\sigma^0_{b_2}$ in Table \ref{vbf2heftb2}.  Using this
parameterization and applying the results of our study, we derive the parameter
ranges tabulated in Table \ref{bound:gva2gvb2}.  Outside the given limits, the
deviation from the SM can be detectable as a $5\sigma$ discovery.

\begin{table}
  \begin{center}
  \begin{tabular}{|c|ccc|}
  \hline
                    &  14 TeV (3 ab$^{-1}$)   &  27 TeV (3 ab$^{-1}$)  &  100 TeV  (30 ab$^{-1}$)       \\ 
  \hline
  $\delta\gva2$    &  $(-0.31, 0.39)$           &  $(-0.11,0.13)$       &  $(-0.013, 0.047)$              \\ 
  \hline
  $\gvb2$     &  $(-0.10, 0.11)$           &  $(-0.03, 0.02)$   &  $(-0.003, 0.003)$\\ 
  \hline
  \end{tabular}
  \end{center}
  \caption{$5\sigma$ discovery (excess) bounds of $\gva2$ and $\gvb2$ at a 14
    TeV, 27 TeV and 100 TeV hadron collider, assuming the respective total
    integrated luminosity in brackets. \label{bound:gva2gvb2}} 
\end{table}

\begin{figure}
  \centering
  \subfigure[14 TeV]{
  \label{ga214}
  \includegraphics[width=0.3\textwidth]{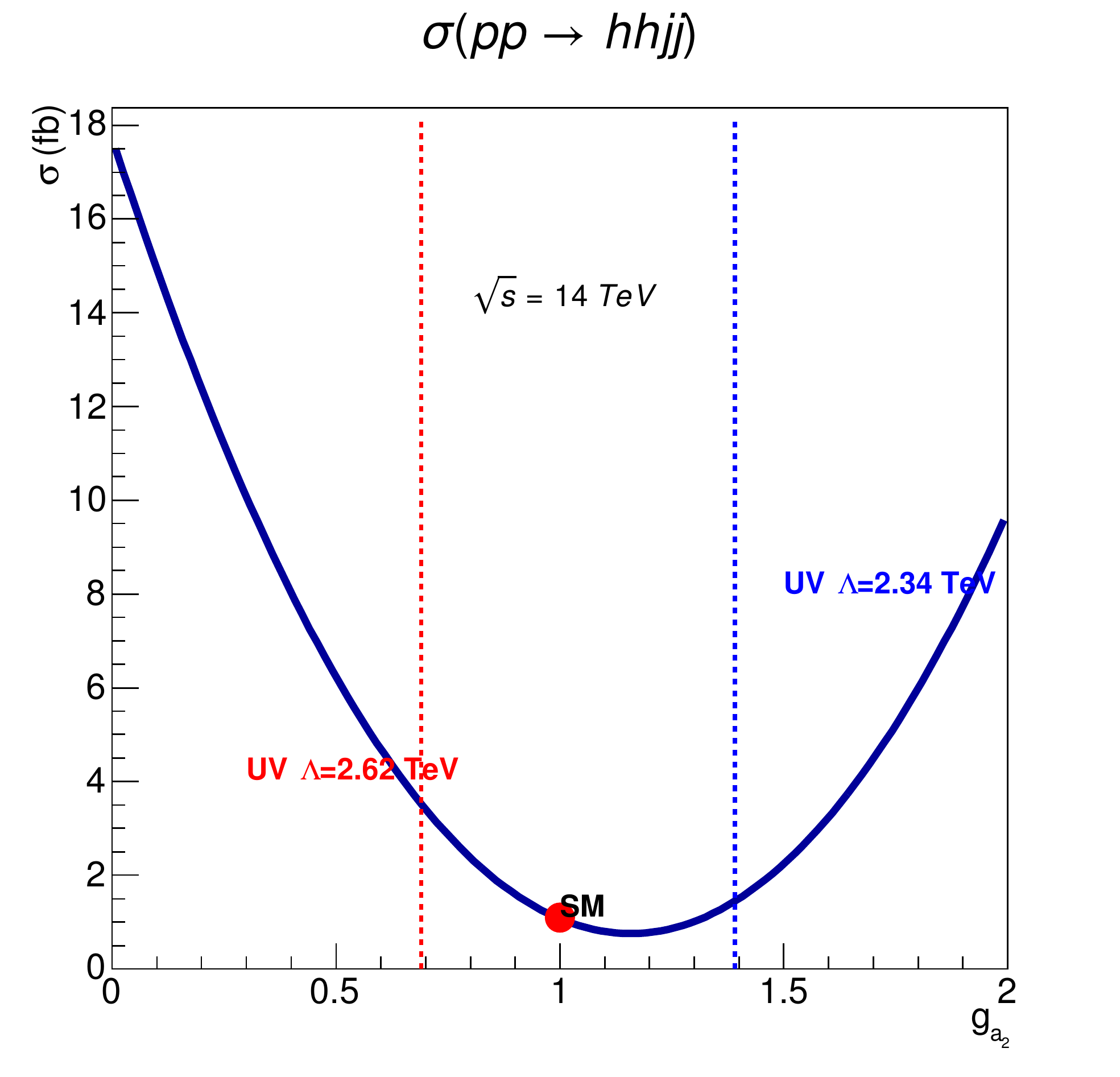}}
  \subfigure[27 TeV]{
  \label{ga227}
  \includegraphics[width=0.3\textwidth]{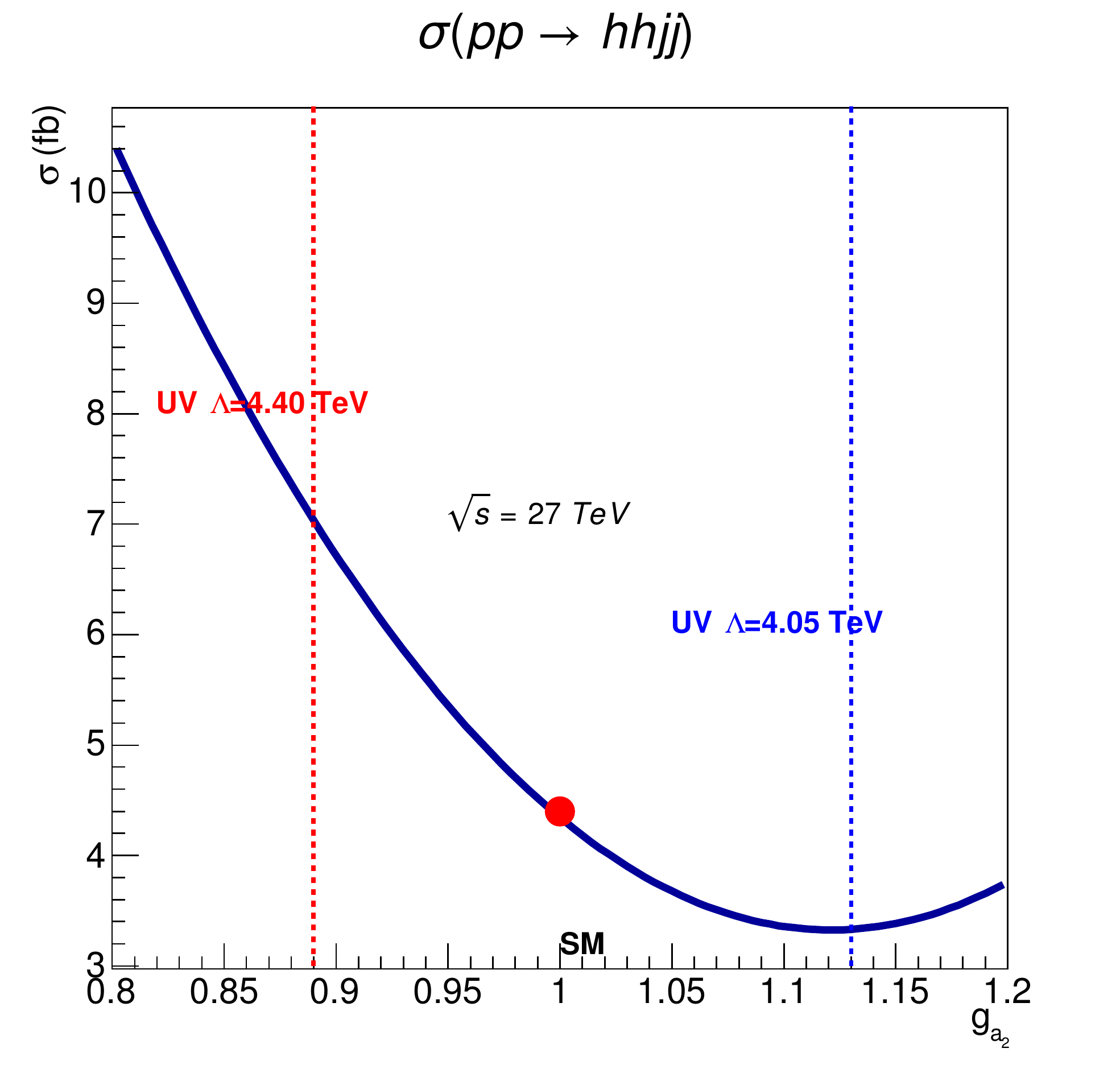}}
  \subfigure[100 TeV]{
  \label{ga2100}
  \includegraphics[width=0.3\textwidth]{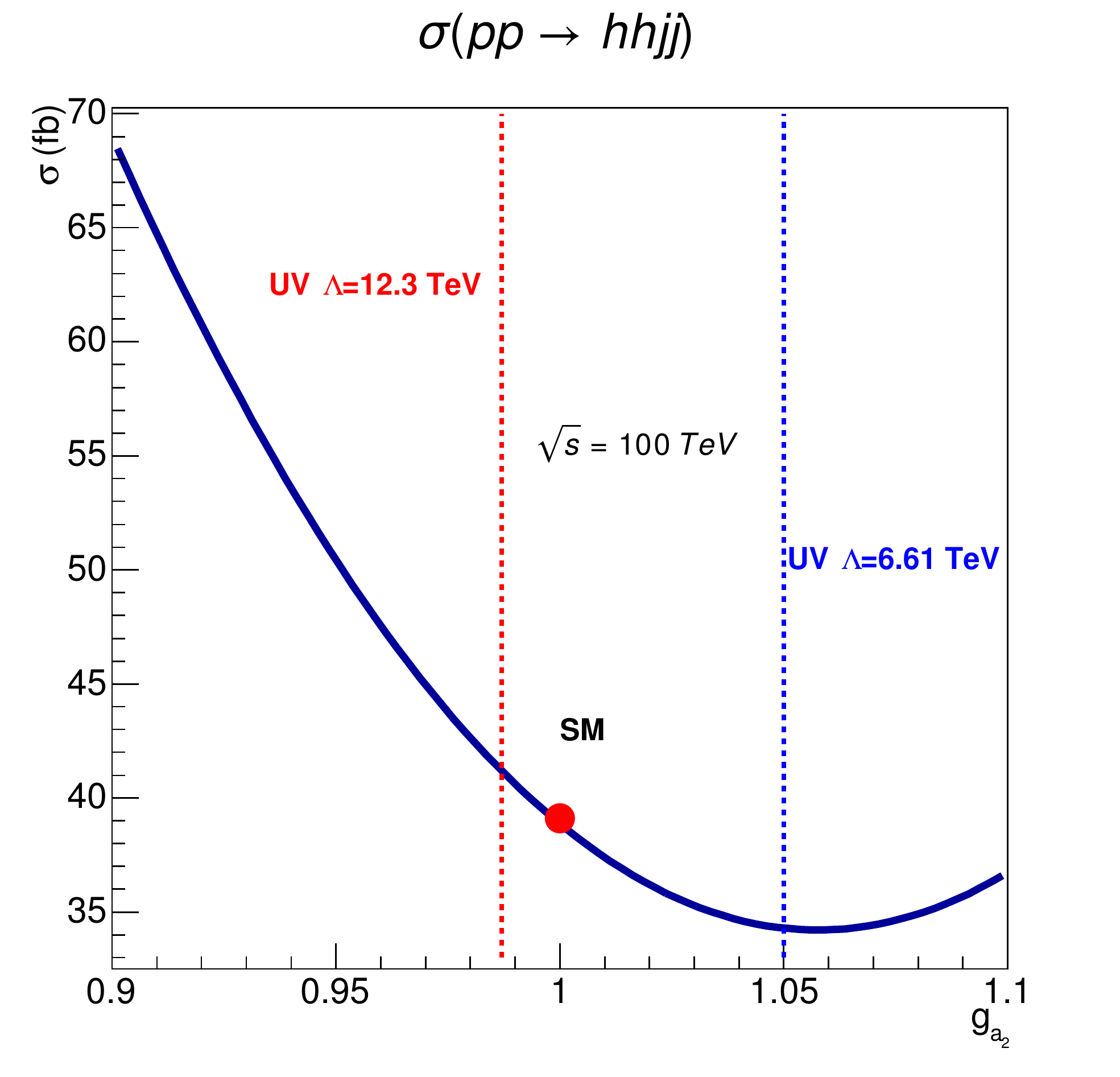}}
  \subfigure[14 TeV]{
  \label{gb214}
  \includegraphics[width=0.3\textwidth]{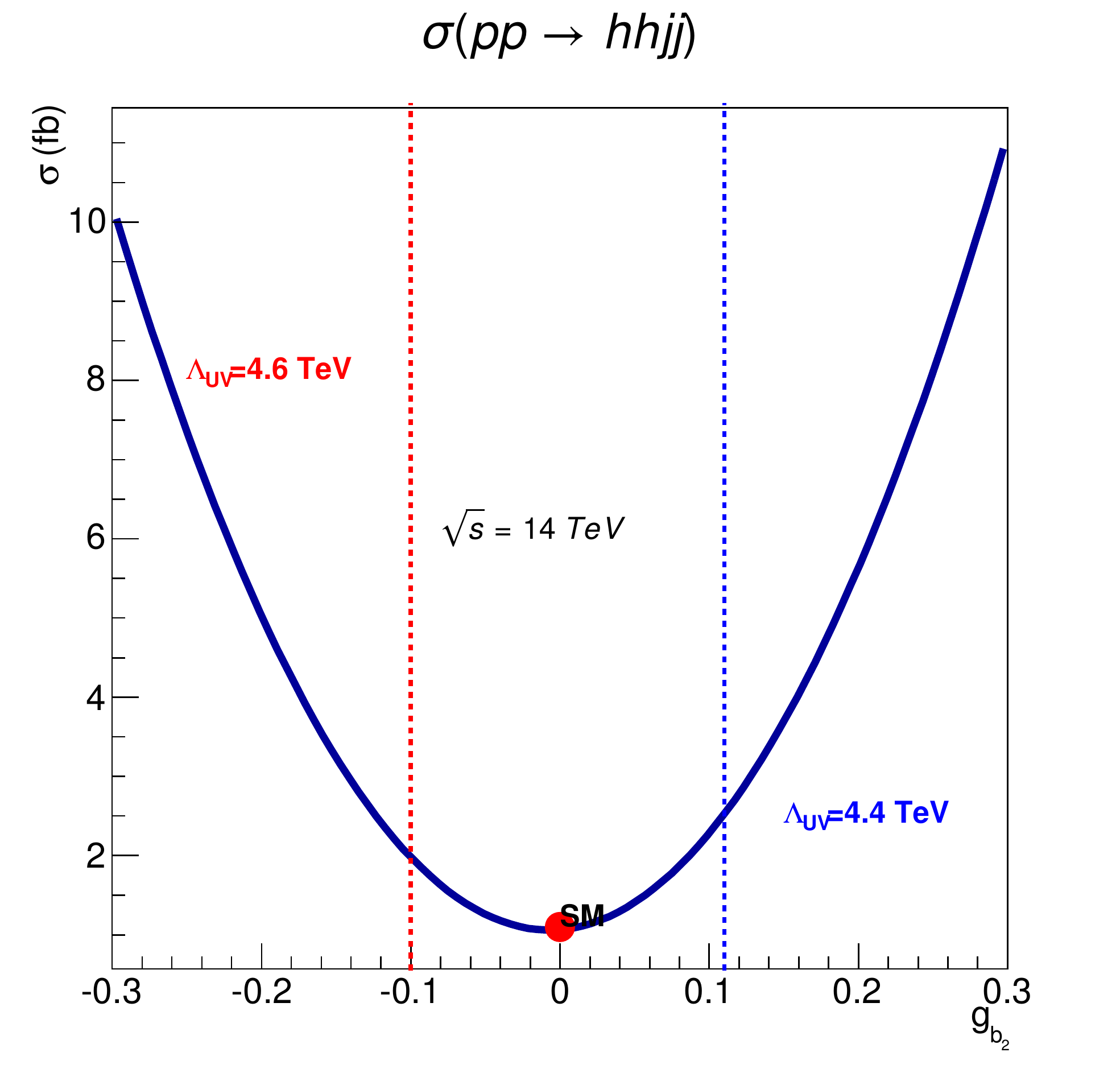}}
  \subfigure[27 TeV]{
  \label{gb227}
  \includegraphics[width=0.3\textwidth]{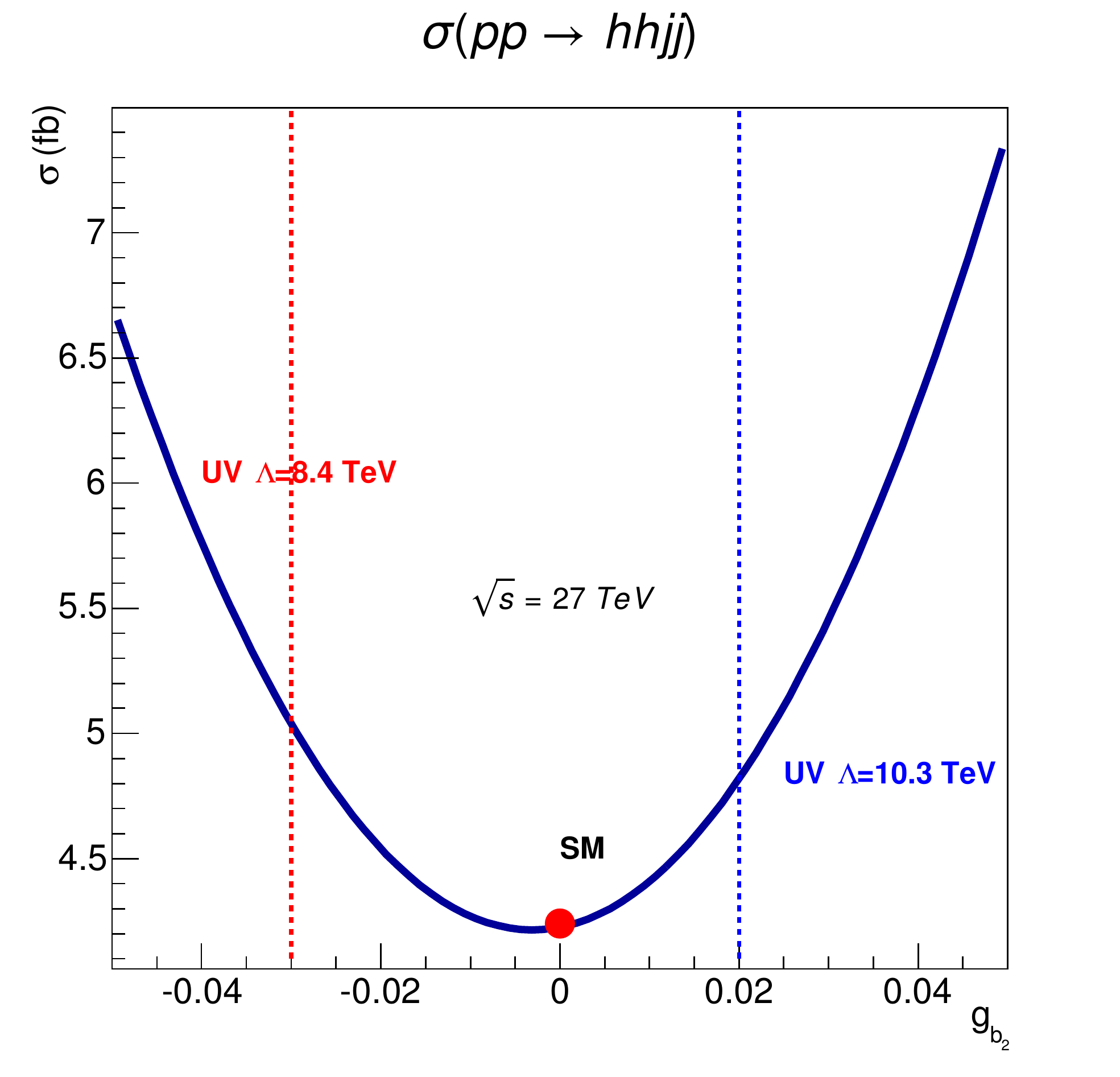}}
  \subfigure[100 TeV]{
  \label{gb2100}
  \includegraphics[width=0.3\textwidth]{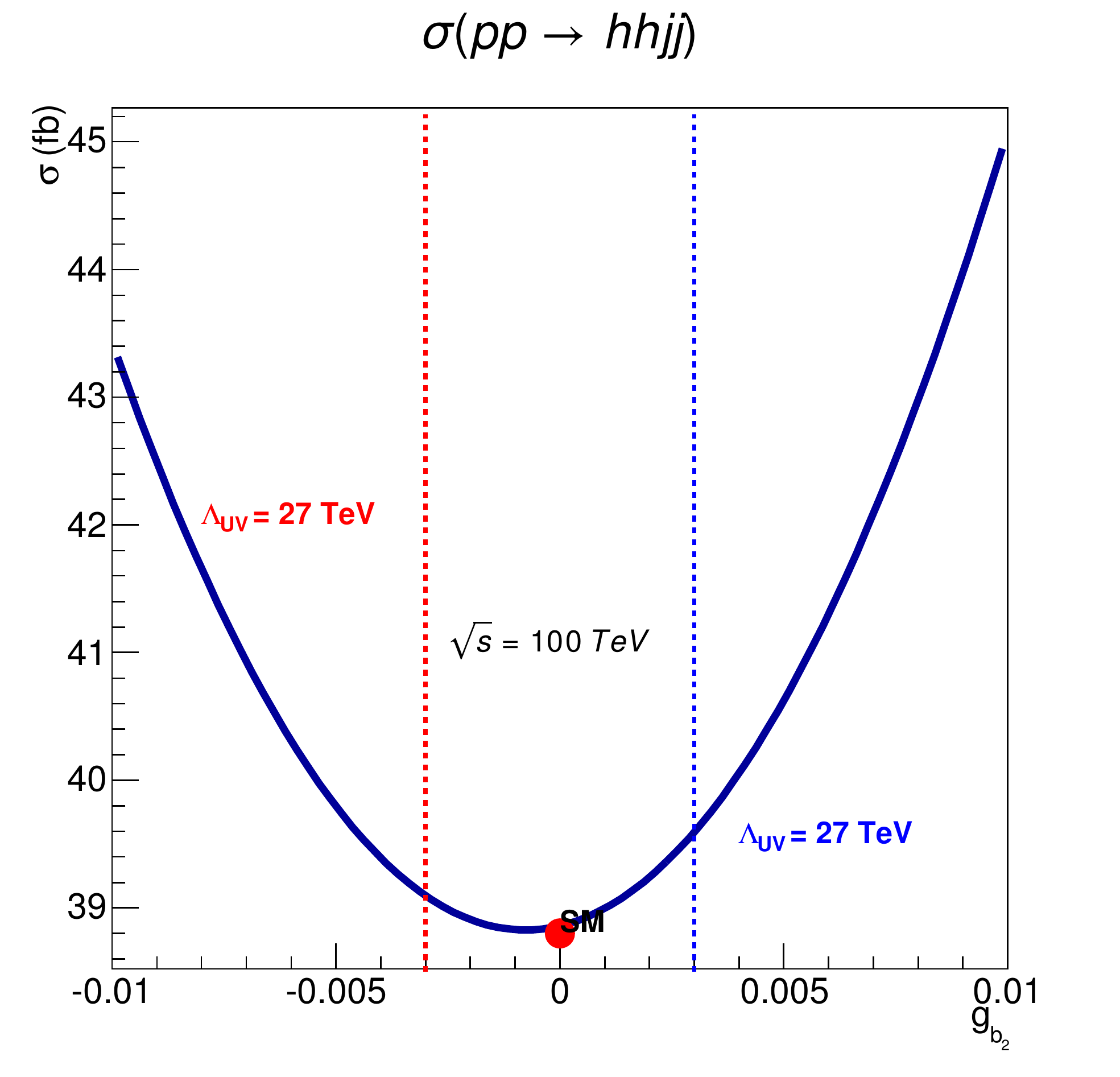}}
  \caption{Total cross section after VBF cuts for the process $pp\to hhjj$ as
    a function of the $WWhh$ couplings $g_{a_2}$ (upper row) and $g_{b_2}$
    (lower row), for three different collider energies. The vertical lines are
    unitarity bounds, 
    which are derived from Eq. (\ref{unitaritya2}) and Eq. \ref{unitarityb2}).\label{fig_hhjj1d}}
\end{figure}

We show the projected bounds on $\gva2$ and $\gvb2$ in
Fig. (\ref{fig_hhjj1d}). We have compared our results for $\gva2$ with those
given in Ref. \cite{Bishara:2016kjn} and found good agreement.  Regarding the
unitarity bounds shown in the plots, the bounds on $\gvb2$ correspond to
Eq. (\ref{unitarityb2}), while for $\gva2$ we make use of the
result~\cite{Kilian:2018bhs} 
\begin{eqnarray}
  \frac{\Lambda^4_{UV}}{2^9\pi^2v^4}|\gva2-\gva1^2|^2\leq\frac{1}{4}\,.
  \label{unitaritya2}
\end{eqnarray}

It is interesting to explore whether it is possible to disentangle the effects of the operator $h^2 W^a_{\mu\nu} W^{a,\mu\nu}$ from those of the operator $h^2 W_\mu^a W^{a,\mu}$. As mentioned in our previous work \cite{Kilian:2018bhs}, the azimuthal angle between two forward jets can be a useful observable for this purpose. 

\begin{figure}
  \centering
  \subfigure[14 TeV]{
  \label{ga214}
  \includegraphics[width=0.3\textwidth]{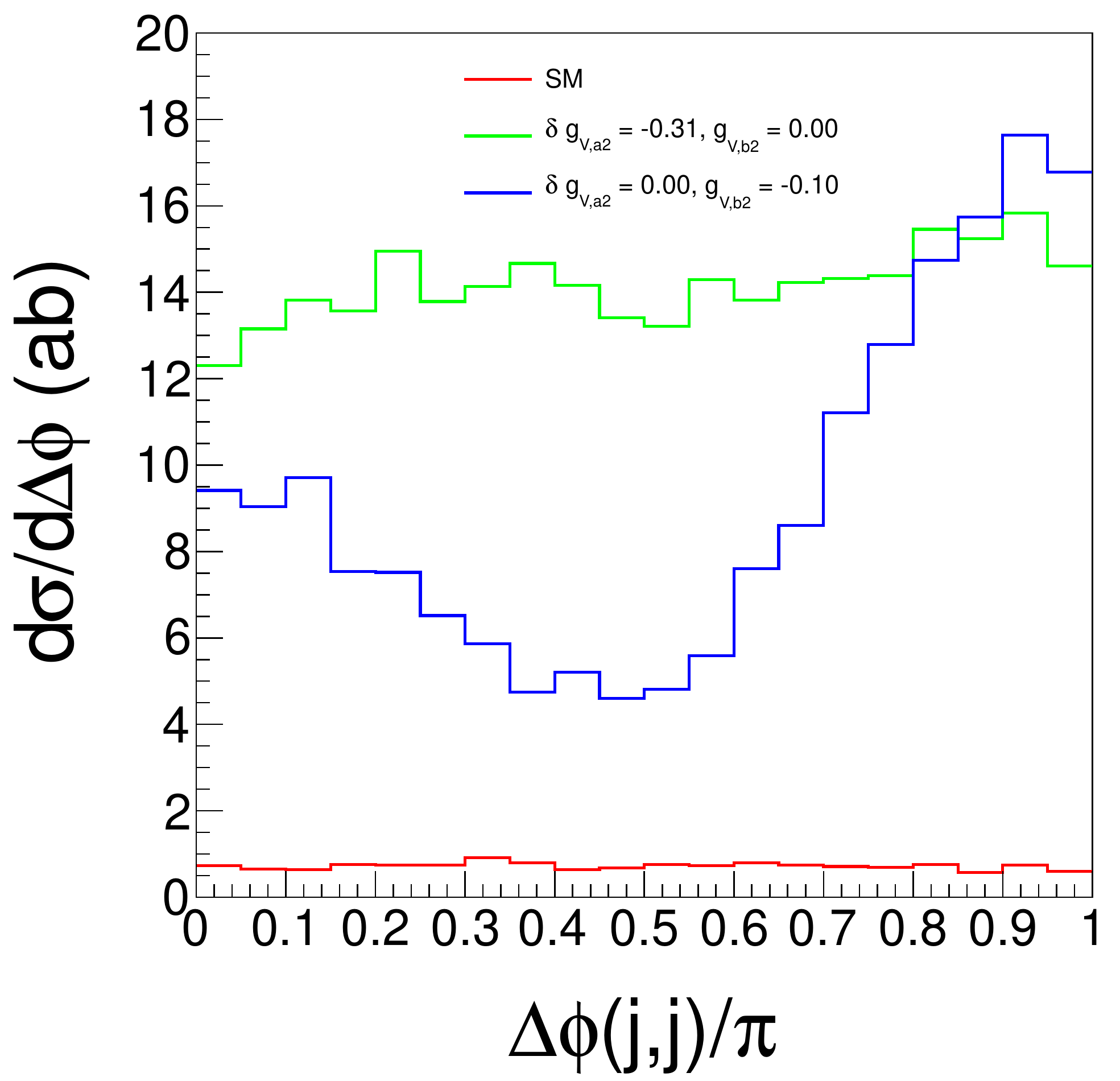}}
  \subfigure[27 TeV]{
  \label{ga227}
  \includegraphics[width=0.3\textwidth]{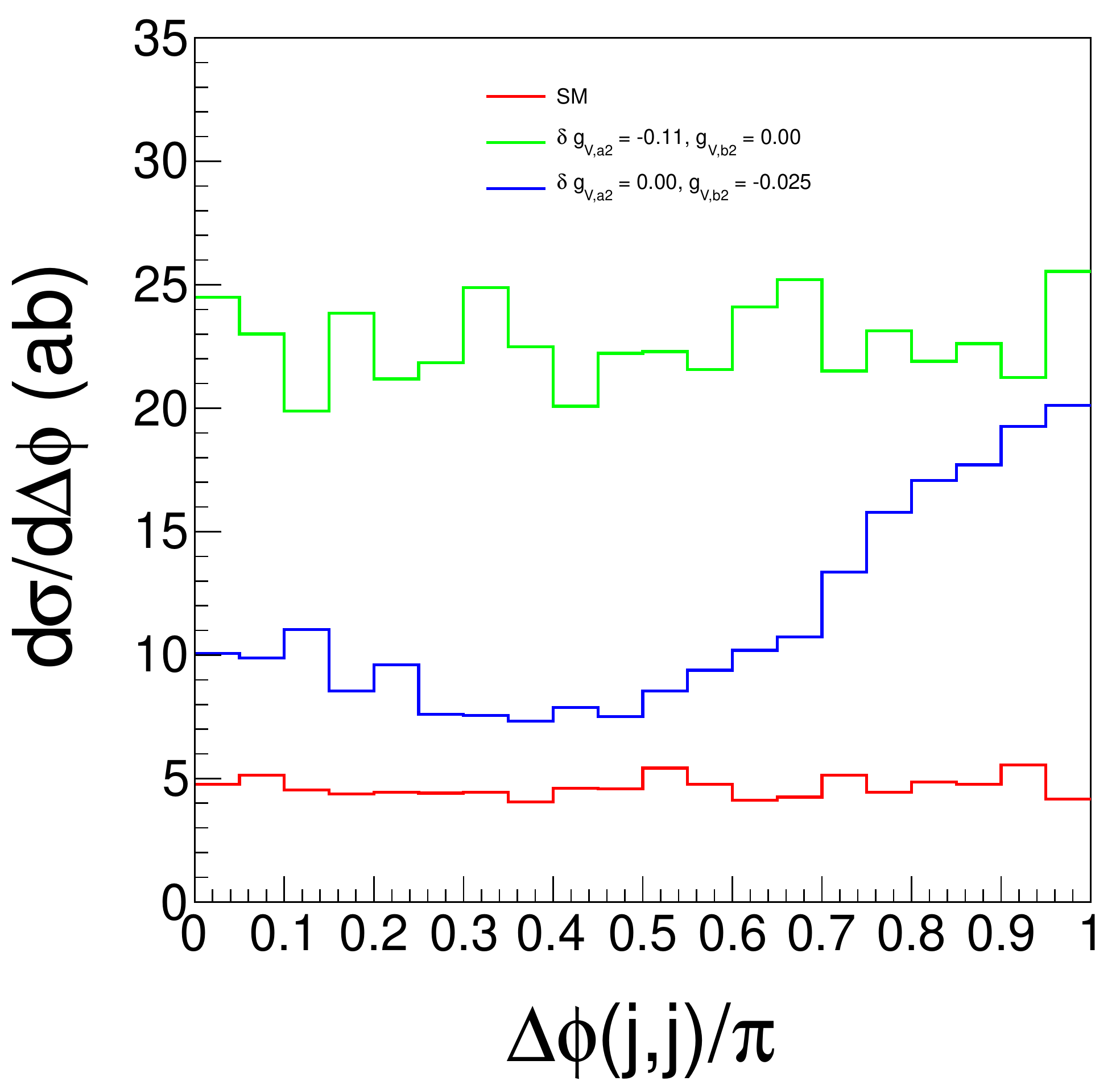}}
  \subfigure[100 TeV]{
  \label{ga2100}
  \includegraphics[width=0.3\textwidth]{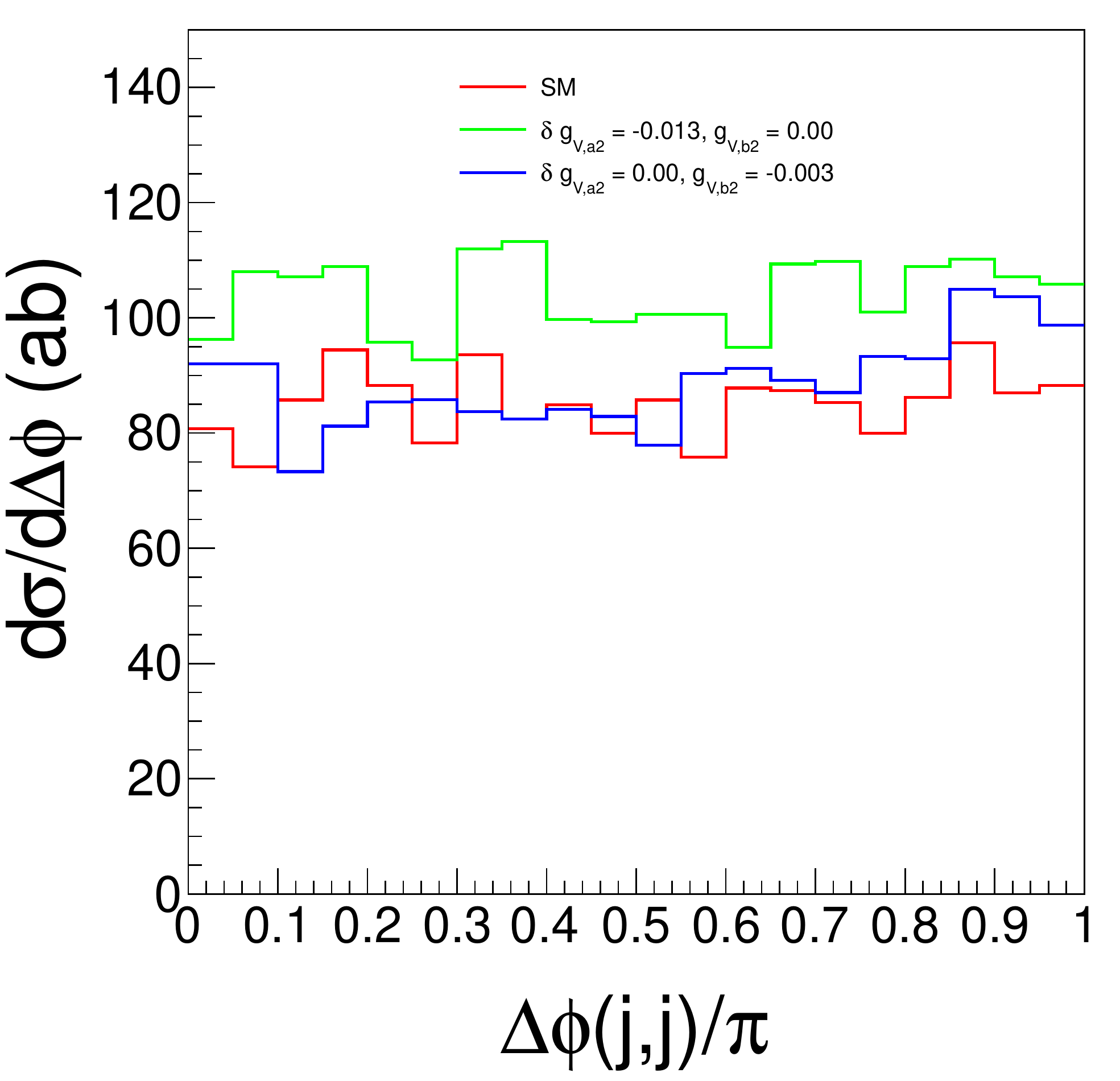}}
  \caption{The relative azimuthal angle distributions of two forward jets for 14 TeV, 27 TeV, and 100 TeV .\label{fig_hhjj1phi}}
\end{figure}

In Fig. (\ref{fig_hhjj1phi}), we display the azimuthal angle distributions 
for the 14 TeV, 27 TeV and 100 TeV cases.  We take into account only events in
category 2BH and apply no further cuts beyond the VBF cuts. The relative azimuthal angle is defined as $\Delta \phi = |\phi(j_1) - \phi(j_2)|$, where $\phi(j_1)$ denotes the azimuthal angle of the leading forward jet, and $ \phi(j_2)$ denotes that of the second leading forward jet. 

We note the main features of the
distributions given in Fig. (\ref{fig_hhjj1phi}):
\begin{itemize}
\item In the SM, the azimuthal-angle distribution of the SM is flat due to the
  dominance of longitudinal polarized vector bosons in the process $pp \to
  hhjj$, as shown by the red curve in each of plot. Similarly, the
  distribution is also flat for the case where the term $h^2 W_{\mu}^a
  W^{a,\mu}$ is turned on, as shown by the green curve in each of plot where
  both the contrbutions of the SM and new physics have been included.  Although new physics represented by the operator $h^2 W_{\mu}^a W^{a,\mu}$ enhances the total cross section, the azimuthal angle distribution remains similar to that of the SM.

\item If the operator $h^2 W_{\mu\nu}^a W^{a,\mu\nu}$ is turned on, the Higgs
  pair being coupled to transversely polarized vector bosons leads to
  distributions that significantly differ from that of the SM.  As the blue
  curves indicate, this type of new interaction causes more events with
  back-to-back outgoing jets. 

\item
  At the 100 TeV collider, the difference in shape of the angular distribution
  is much less pronounced.  In fact, the chosen benchmark values of the
  operator coefficients cause just a small disturbance of the SM signal in
  this plot, which
  is still detectable due to the much larger event rate for the high energy and
  high luminosity of the machine.  To distinguish the two kinds of
  contributions, a more detailed analysis of this and other distributions
  becomes necessary. 

\end{itemize}

\begin{figure}
	\centering
	\includegraphics[width=0.33\textwidth]{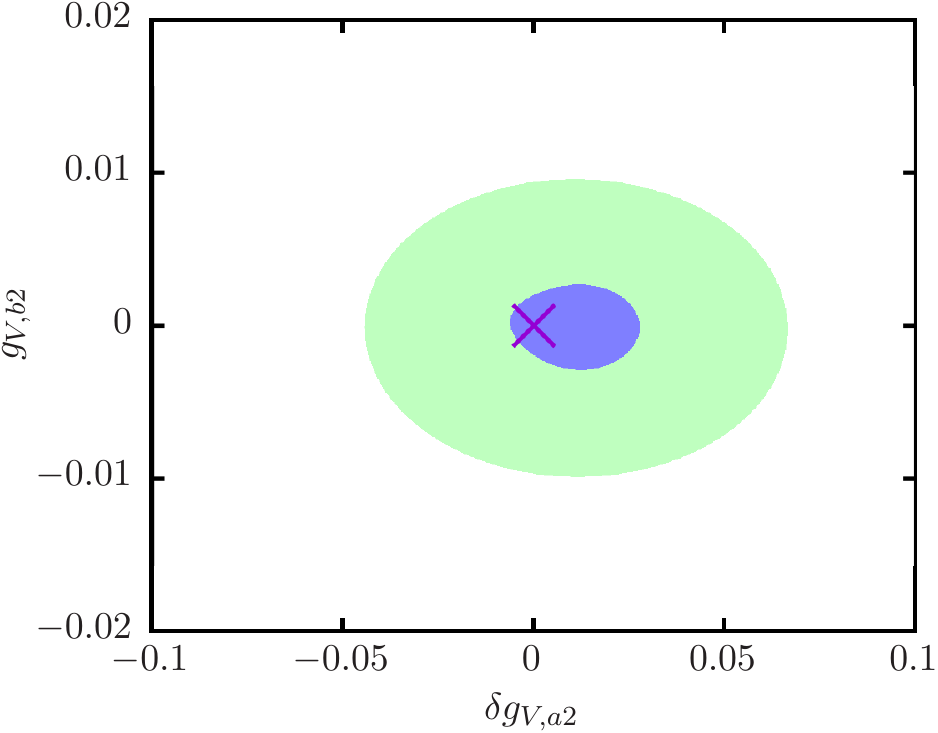}
	\includegraphics[width=0.33\textwidth]{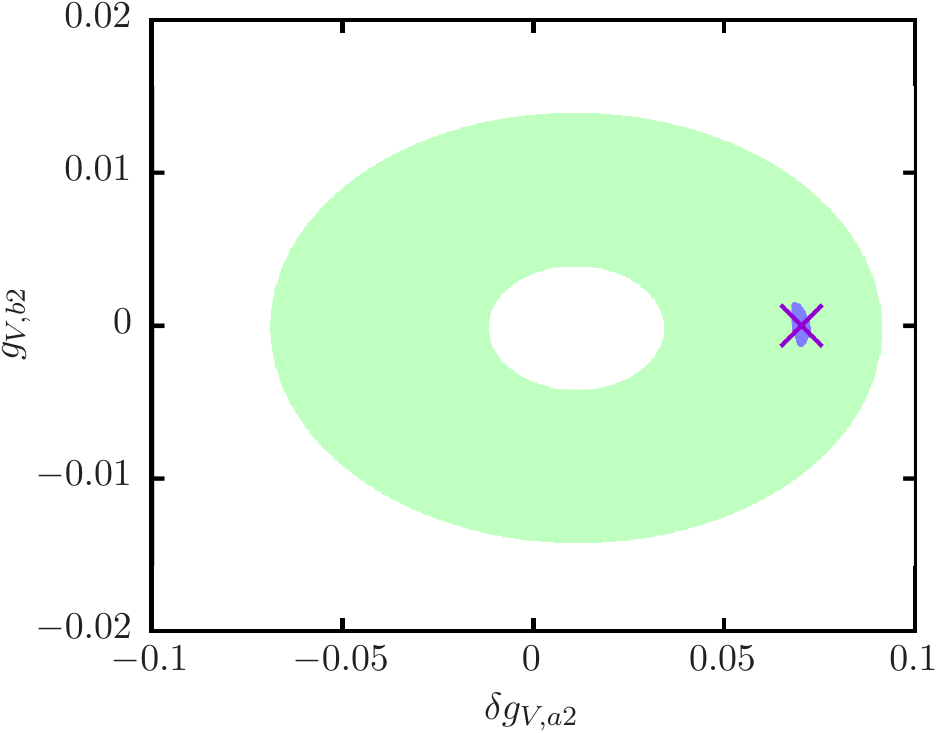}
	\includegraphics[width=0.33\textwidth]{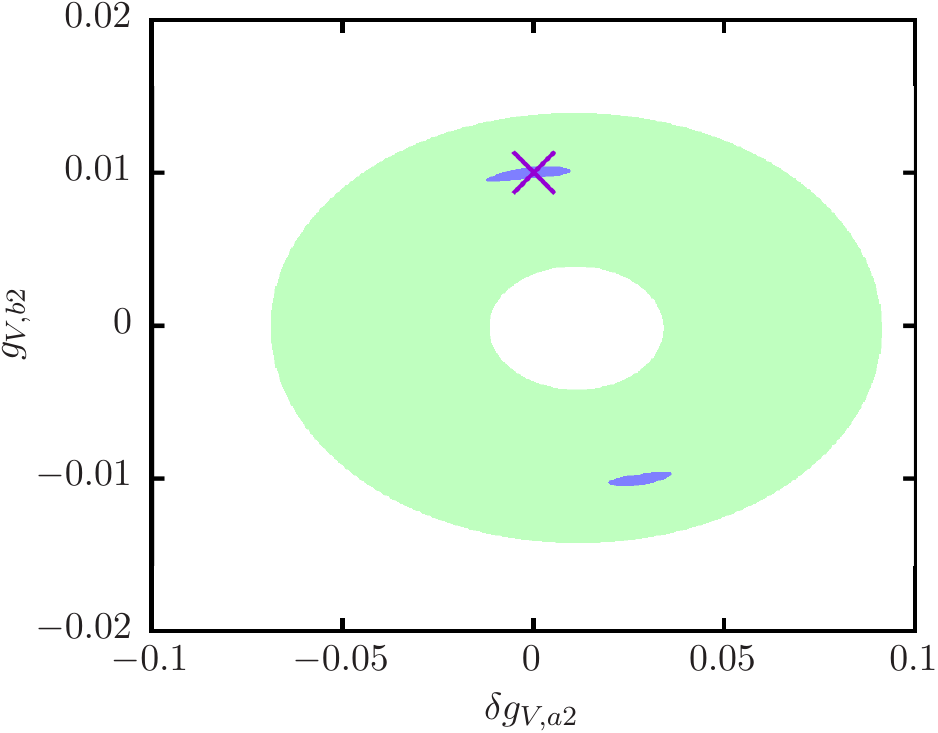}
	\caption{The fit results with inclusive cross section(green) and
          differential cross section(blue) with three different assumptions on
          the central values:
          $(\delta\gva2^{\textrm{true}},\gvb2^{\textrm{true}})=(0,0)$,
          $(0.07,0)$, $(0,0.01)$ with $\sqrt{s}=100$ TeV and an integrated luminosity 30 ab$^{-1}$.
}
\label{fig:fit-smab}
\end{figure}

To fully utilise the differential information, we perform a $\chi^2$-fit on the distribution of $\Delta\phi(j,j)$, using the 2BH events after the BDT cuts with the collision energy $\sqrt{s}=100$ TeV and integrated luminosity 30 ab$^{-1}$. In Fig. \ref{fig:fit-smab},
we show the $2\sigma$ allowed region obtained with differential information
(blue) or just from the total cross section (green).
For the leftmost plot, we assume the SM for the true values of the
coefficients.  It is evident that the information from the differential
distribution significantly improves the precisions on both $\delta\gva2$ and
$\gvb2$. 
The middle plot assumes $(\delta\gva2^{\rm true},\gvb2^{\rm true})=(0.07,0)$
for the true value, i.e., only $\gva2$ receives a new-physics contribution.
The inclusive cross section confines the region allowed by a measurement to a
ring, due to the quadratic dependence on both $\delta\gva2$ and $\gvb2$, while
the differential distribution singles out a small area around the point
$(\delta\gva2,\gvb2)=(0.07,0)$.  
Analogously, for the right plot we assume the true values $(\delta\gva2^{\rm
  true},\gvb2^{\rm true})=(0,0.07)$.  Again, the differential distribution
selects a small region, but in this case there is a two-fold sign ambiguity
left for $\gvb2$. 
This reflects the fact that the effects of $\gvb2$ is dominated by the squared
term, and thus the sign of $\gvb2$ cannot be determined.

Finally, using the relations given in Table \ref{table:parameter}, we can map
this contour onto the plane spanned by $c_W$ and $c_{HW}$.  The assumption of
a linearly realized symmetry and truncation at the dimension-6 order enforces
the constraint $\delta\gva2 
\sim 6\gvb2$, if both couplings depend only on a single parameter
$c_{HW}$.  The two-parameter analysis that we describe in this paper, would allow us
to search for the relation of $c_W$ and $c_{HW}$, which would point to the presence
of contributions that do not follow the simple power-counting assumption
underlying the dimension-six truncation of the SILH basis.  Currently, the experimental data constraints on the parameter set are rather weak, to the level of  $c_V
\sim O(10^3)$~\cite{ Ellis:2018gqa}. In this study, we conclude that we
can reach a sensitivity of up to $c_V \sim O(10)$ at future colliders.

\section{Discussions and Conclusions}
\label{Sec:conc} 

We have studied double-Higgs production in vector-boson at a proton-proton collider,
$pp\to hhjj\to 4b2j$.  We compare the 14 TeV LHC with future high-energy and
high-luminosity colliders of 27 TeV and 100 TeV.  The
analysis of this process benefits greatly from identifying highly-boosted
Higgs bosons.  To this end, we use
the mass-drop method to analyse the jet substructure, and we optimize the
significance by the boosted decision-tree method.  Our results show that the
2BH case where two highly-boosted Higgs are tagged can provide 
the cleanest  experimental environment to discover new physics in the the VBF
signal.  

At the LHC, the number of 2BH events is too small to discover this channel if
the SM is valid without new contributions.  Conversely, at a 100 TeV collider
with a high luminosity of the order of 30 ab$^{-1}$, the number of 2BH events
is large enough to discover the SM signal.

To study the effects of new physics, we use a phenomenological effective
Lagrangian~(\ref{lvh}).  We explore the effect of the interactions of
type $h^2 V_{\mu}^a V^{a,\mu}$ and $h^2 V_{\mu\nu}^a V^{a,\mu\nu}$ and extract
bounds for the associated coefficients $\gva2$ and $\gvb2$. Our
results demonstrate that new-physics scales up to $4.4$ TeV
at the 14 TeV LHC, and $27$ 
TeV at a 100 TeV hadron collider, are within reach of discovery. 

\begin{figure}
  \centering
  \includegraphics[width=.8\linewidth]{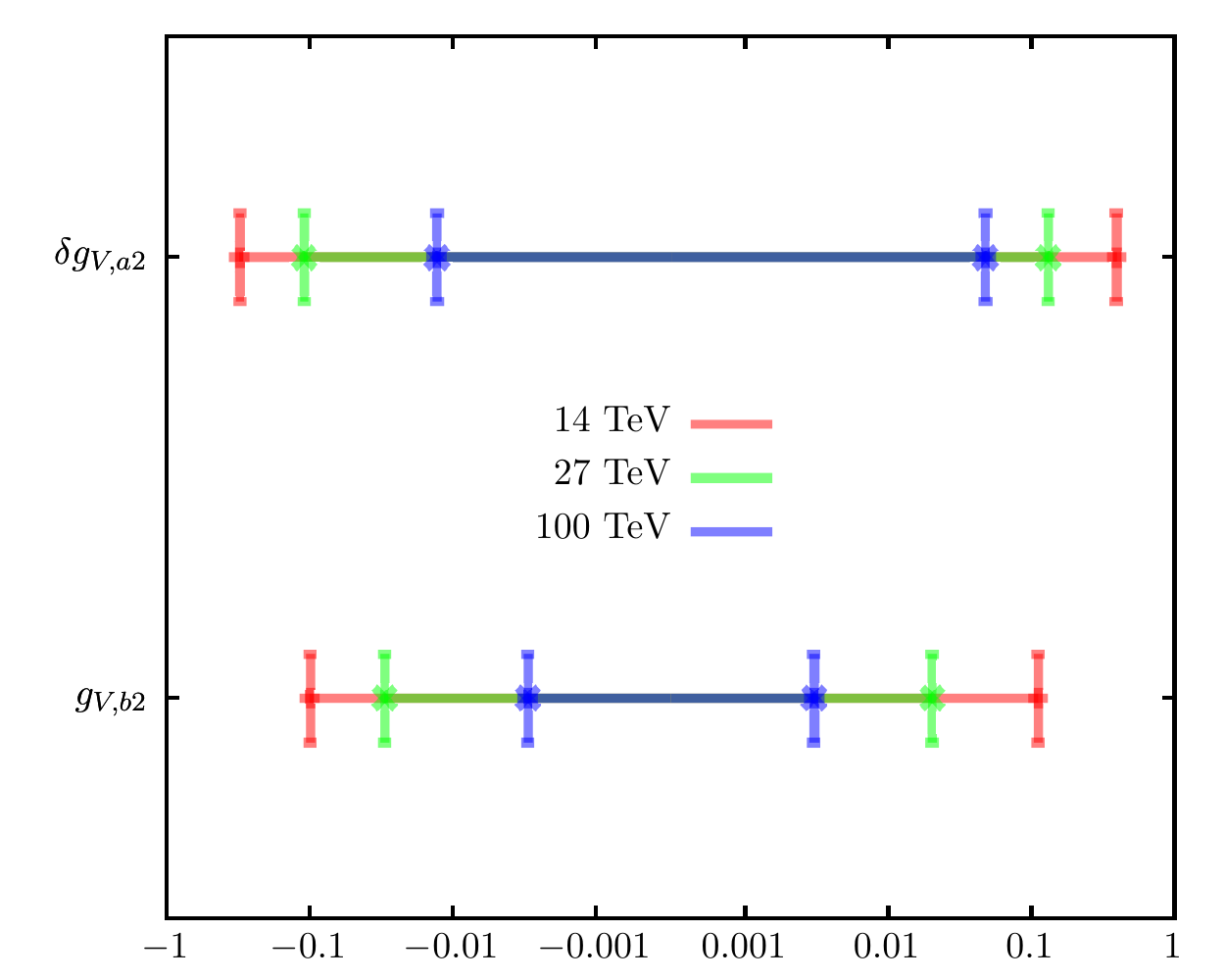}
  \caption{The summarized bounds of $\gva2$ and $\gvb2$ at 14 TeV, 27 TeV and 100 TeV hadron collider, which is the same as Table~\ref{bound:gva2gvb2}.\label{logbounds}}
\end{figure}

Fig.~\ref{logbounds} collects the projected bounds on $\gva2$ and $\gvb2$ that
we have determined
in this work, cf.\ also Table~\ref{bound:gva2gvb2}.
The bounds that can be obtained at 27 TeV, are one order of magnitude stronger
than for the 14 TeV LHC.
The 100 TeV machine can further constrain these parameters by another order of
magnitude.  At both 
the 27 TeV and 100 TeV machines, this offers a significant indirect
sensitivity to new physics in this sector.

Our study does not account for underlying-event or pile-up
effects. These produce additional soft radiation and may render the
extraction of a hard-process signal more difficult. In fact, the mass-drop
method selects two-pronged jets in the final state.  This reduces the impact of QCD jets from
underlying events or pile-up, which are one-pronged. We also use filtering to 
remove soft radiation after reconstructing two sub-jets in the final state,
which is helpful to reject jets from extra radiation produced by the parton
shower.  Recent studies indicate that modern pile-up mitigation
techniques~\cite{Cacciari:2014gra} can minimize the pile-up contamination
efficiently for the 4b final state~\cite{Behr:2015oqq}.  A detailed pile-up analysis can be done but is beyond the scope of the current paper.

For a further improvement of our result, we may consider color-flow
properties~\cite{Maltoni:2002mq} as a tool to further discriminate
$h\to b\bar b$ decays from b jets in the QCD background.  Color-connection
information can be quantified by observables such as the pull
vector~\cite{Gallicchio:2010sw}. In a recent study of double Higgs production
at LHC~\cite{Kim:2019wns}, it was argued that while the color flow is very
different between the double-Higgs signal and the QCD background, this
information may be diluted after applying kinematics cuts. The authors of
\cite{Kim:2019wns} proposed to use jet-image and a Deep Neutral Network
analysis methods to discriminate the signal from background, rather than
constructing the pull vector.   We defer these refinements of our study to
future work.

In summary, the analysis of highly boosted Higgs-boson pairs in vector-boson
fusion is promising and
can be used to significantly improve our knowledge about Higgs-sector
interactions.  In particular, the method should become important for a future
high-energy proton collider where the 
sensitivity is sufficient to extract a signal down to
the SM rate.

\begin{acknowledgments}
W.K. is supported by the Deutsche Forschungsgemeinschaft
(DFG, German Research Foundation) under grant 396021762 -- TRR 257.
S.\, Sun is supported by MIUR in Italy under Contract(No. PRIN 2015P5SBHT) and ERC Ideas Advanced Grant (No. 267985) \textquotedblleft DaMeSyFla";
 Q.S. Yan is supported by the Natural Science Foundation of China
under the grant NO. 11475180 and NO. 11875260. 
X.R. Zhao is supported by the ``Fonds Sp\'ecial de Recherche'' (FSR)
of the UCLouvain.

\end{acknowledgments}

\clearpage

\bibliography{mybib.bib}

\providecommand{\href}[2]{#2}\begingroup\raggedright\begin{thebibliography}{10}

\bibitem{Aad:2012tfa}
{\scshape ATLAS} collaboration, \emph{Observation of a new particle in the
  search for the standard model higgs boson with the atlas detector at the
  lhc},
  \href{https://doi.org/10.1016/j.physletb.2012.08.020}{\emph{Phys.Lett.B}
  {\bfseries 716} (2012) 1} [\href{https://arxiv.org/abs/1207.7214}{{\ttfamily
  1207.7214}}].

\bibitem{Chatrchyan:2012ufa}
{\scshape CMS} collaboration, \emph{Observation of a new boson at a mass of 125
  gev with the cms experiment at the lhc},
  \href{https://doi.org/10.1016/j.physletb.2012.08.021}{\emph{Phys.Lett.B}
  {\bfseries 716} (2012) 30} [\href{https://arxiv.org/abs/1207.7235}{{\ttfamily
  1207.7235}}].

\bibitem{Jones:1979bq}
D.~Jones and S.~Petcov, \emph{Heavy higgs bosons at lep},
  \href{https://doi.org/10.1016/0370-2693(79)91234-6}{\emph{Phys.Lett.B}
  {\bfseries 84} (1979) 440}.

\bibitem{Kilian:2018bhs}
W.~Kilian, S.~Sun, Q.-S. Yan, X.~Zhao and Z.~Zhao, \emph{{Multi-Higgs
  Production and Unitarity in Vector-Boson Fusion at Future Hadron Colliders}},
   \href{https://arxiv.org/abs/1808.05534}{{\ttfamily 1808.05534}}.

\bibitem{Qi:2019ocx}
Y.-H. Qi, J.-H. Yu and S.-H. Zhu, \emph{{Effective Field Theory Perspective on
  Next-to-Minimal Composite Higgs}},
  \href{https://arxiv.org/abs/1912.13058}{{\ttfamily 1912.13058}}.

\bibitem{Agrawal:2019bpm}
P.~Agrawal, D.~Saha, L.-X. Xu, J.-H. Yu and C.~Yuan, \emph{{Shape of Higgs
  Potential at Future Colliders}},
  \href{https://arxiv.org/abs/1907.02078}{{\ttfamily 1907.02078}}.

\bibitem{Xu:2019xuo}
L.-X. Xu, J.-H. Yu and S.-H. Zhu, \emph{{Holographic Completion of Minimal
  Neutral Naturalness Model and Deconstruction}},
  \href{https://arxiv.org/abs/1905.12796}{{\ttfamily 1905.12796}}.

\bibitem{Li:2019ghf}
H.-L. Li, L.-X. Xu, J.-H. Yu and S.-H. Zhu, \emph{{EFTs meet Higgs
  Nonlinearity, Compositeness and (Neutral) Naturalness}},
  \href{https://doi.org/10.1007/JHEP09(2019)010}{\emph{JHEP} {\bfseries 09}
  (2019) 010} [\href{https://arxiv.org/abs/1904.05359}{{\ttfamily
  1904.05359}}].

\bibitem{Aaboud:2017vzb}
{\scshape ATLAS} collaboration, \emph{{Measurement of the Higgs boson coupling
  properties in the $H\rightarrow ZZ^{*} \rightarrow 4\ell$ decay channel at
  $\sqrt{s}$ = 13 TeV with the ATLAS detector}},
  \href{https://doi.org/10.1007/JHEP03(2018)095}{\emph{JHEP} {\bfseries 03}
  (2018) 095} [\href{https://arxiv.org/abs/1712.02304}{{\ttfamily
  1712.02304}}].

\bibitem{Sirunyan:2017exp}
{\scshape CMS} collaboration, \emph{{Measurements of properties of the Higgs
  boson decaying into the four-lepton final state in pp collisions at $
  \sqrt{s}=13 $ TeV}},
  \href{https://doi.org/10.1007/JHEP11(2017)047}{\emph{JHEP} {\bfseries 11}
  (2017) 047} [\href{https://arxiv.org/abs/1706.09936}{{\ttfamily
  1706.09936}}].

\bibitem{Aaboud:2018jqu}
{\scshape ATLAS} collaboration, \emph{{Measurements of gluon-gluon fusion and
  vector-boson fusion Higgs boson production cross-sections in the $H \to
  WW^{\ast} \to e\nu\mu\nu$ decay channel in $pp$ collisions at $\sqrt{s}=13$
  TeV with the ATLAS detector}},
  \href{https://doi.org/10.1016/j.physletb.2018.11.064}{\emph{Phys. Lett.}
  {\bfseries B789} (2019) 508}
  [\href{https://arxiv.org/abs/1808.09054}{{\ttfamily 1808.09054}}].

\bibitem{Aad:2019mbh}
{\scshape ATLAS} collaboration, \emph{{Combined measurements of Higgs boson
  production and decay using up to $80$ fb$^{-1}$ of proton-proton collision
  data at $\sqrt{s}=$ 13 TeV collected with the ATLAS experiment}},
  \href{https://doi.org/10.1103/PhysRevD.101.012002}{\emph{Phys.\ Rev.\ D}
  {\bfseries 101} (2020) 012002}
  [\href{https://arxiv.org/abs/1909.02845}{{\ttfamily 1909.02845}}].

\bibitem{Aad:2020kub}
{\scshape ATLAS} collaboration, \emph{{Search for the $HH \rightarrow b \bar{b}
  b \bar{b}$ process via vector-boson fusion production using proton-proton
  collisions at $\sqrt{s} = 13$ TeV with the ATLAS detector}},
  \href{https://arxiv.org/abs/2001.05178}{{\ttfamily 2001.05178}}.

\bibitem{Aaboud:2018knk}
{\scshape ATLAS} collaboration, \emph{{Search for pair production of Higgs
  bosons in the $b\bar{b}b\bar{b}$ final state using proton-proton collisions
  at $\sqrt{s} = 13$ TeV with the ATLAS detector}},
  \href{https://doi.org/10.1007/JHEP01(2019)030}{\emph{JHEP} {\bfseries 01}
  (2019) 030} [\href{https://arxiv.org/abs/1804.06174}{{\ttfamily
  1804.06174}}].

\bibitem{Sirunyan:2018tki}
{\scshape CMS} collaboration, \emph{{Search for nonresonant Higgs boson pair
  production in the $\mathrm{b\overline{b}b\overline{b}}$ final state at
  $\sqrt{s} =$ 13 TeV}},
  \href{https://doi.org/10.1007/JHEP04(2019)112}{\emph{JHEP} {\bfseries 04}
  (2019) 112} [\href{https://arxiv.org/abs/1810.11854}{{\ttfamily
  1810.11854}}].

\bibitem{Aaboud:2018ftw}
{\scshape ATLAS} collaboration, \emph{{Search for Higgs boson pair production
  in the $\gamma\gamma b\bar{b}$ final state with 13 TeV $pp$ collision data
  collected by the ATLAS experiment}},
  \href{https://doi.org/10.1007/JHEP11(2018)040}{\emph{JHEP} {\bfseries 11}
  (2018) 040} [\href{https://arxiv.org/abs/1807.04873}{{\ttfamily
  1807.04873}}].

\bibitem{Sirunyan:2018iwt}
{\scshape CMS} collaboration, \emph{{Search for Higgs boson pair production in
  the $\gamma\gamma\mathrm{b\overline{b}}$ final state in pp collisions at
  $\sqrt{s}=$ 13 TeV}},
  \href{https://doi.org/10.1016/j.physletb.2018.10.056}{\emph{Phys.\ Lett.\ B}
  {\bfseries 788} (2019) 7} [\href{https://arxiv.org/abs/1806.00408}{{\ttfamily
  1806.00408}}].

\bibitem{Aaboud:2018sfw}
{\scshape ATLAS} collaboration, \emph{{Search for resonant and non-resonant
  Higgs boson pair production in the ${b\bar{b}\tau^+\tau^-}$ decay channel in
  $pp$ collisions at $\sqrt{s}=13$ TeV with the ATLAS detector}},
  \href{https://doi.org/10.1103/PhysRevLett.121.191801}{\emph{Phys.\ Rev.\
  Lett.} {\bfseries 121} (2018) 191801}
  [\href{https://arxiv.org/abs/1808.00336}{{\ttfamily 1808.00336}}].

\bibitem{Sirunyan:2017djm}
{\scshape CMS} collaboration, \emph{{Search for Higgs boson pair production in
  events with two bottom quarks and two tau leptons in proton–proton
  collisions at $\sqrt s$ =13TeV}},
  \href{https://doi.org/10.1016/j.physletb.2018.01.001}{\emph{Phys.\ Lett.\ B}
  {\bfseries 778} (2018) 101}
  [\href{https://arxiv.org/abs/1707.02909}{{\ttfamily 1707.02909}}].

\bibitem{Aaboud:2018zhh}
{\scshape ATLAS} collaboration, \emph{{Search for Higgs boson pair production
  in the $b\bar{b}WW^{*}$ decay mode at $\sqrt{s}=13$ TeV with the ATLAS
  detector}}, \href{https://doi.org/10.1007/JHEP04(2019)092}{\emph{JHEP}
  {\bfseries 04} (2019) 092}
  [\href{https://arxiv.org/abs/1811.04671}{{\ttfamily 1811.04671}}].

\bibitem{Sirunyan:2017guj}
{\scshape CMS} collaboration, \emph{{Search for resonant and nonresonant Higgs
  boson pair production in the $ \mathrm{b}\overline{\mathrm{b}}\mathit{\ell
  \nu \ell \nu } $ final state in proton-proton collisions at $ \sqrt{s}=13 $
  TeV}}, \href{https://doi.org/10.1007/JHEP01(2018)054}{\emph{JHEP} {\bfseries
  01} (2018) 054} [\href{https://arxiv.org/abs/1708.04188}{{\ttfamily
  1708.04188}}].

\bibitem{Aaboud:2018ewm}
{\scshape ATLAS} collaboration, \emph{{Search for Higgs boson pair production
  in the $\gamma\gamma WW^{*}$ channel using $pp$ collision data recorded at
  $\sqrt{s} = 13$ TeV with the ATLAS detector}},
  \href{https://doi.org/10.1140/epjc/s10052-018-6457-x}{\emph{Eur.\ Phys.\ J.\
  C} {\bfseries 78} (2018) 1007}
  [\href{https://arxiv.org/abs/1807.08567}{{\ttfamily 1807.08567}}].

\bibitem{Aaboud:2018ksn}
{\scshape ATLAS} collaboration, \emph{{Search for Higgs boson pair production
  in the $WW^{(*)}WW^{(*)}$ decay channel using ATLAS data recorded at
  $\sqrt{s}=13$ TeV}},
  \href{https://doi.org/10.1007/JHEP05(2019)124}{\emph{JHEP} {\bfseries 05}
  (2019) 124} [\href{https://arxiv.org/abs/1811.11028}{{\ttfamily
  1811.11028}}].

\bibitem{Sirunyan:2018ayu}
{\scshape CMS} collaboration, \emph{{Combination of searches for Higgs boson
  pair production in proton-proton collisions at $\sqrt{s} = $ 13 TeV}},
  \href{https://doi.org/10.1103/PhysRevLett.122.121803}{\emph{Phys.\ Rev.\
  Lett.} {\bfseries 122} (2019) 121803}
  [\href{https://arxiv.org/abs/1811.09689}{{\ttfamily 1811.09689}}].

\bibitem{Aad:2019uzh}
{\scshape ATLAS} collaboration, \emph{{Combination of searches for Higgs boson
  pairs in $pp$ collisions at $\sqrt{s} = $13 TeV with the ATLAS detector}},
  \href{https://doi.org/10.1016/j.physletb.2019.135103}{\emph{Phys.\ Lett.\ B}
  {\bfseries 800} (2020) 135103}
  [\href{https://arxiv.org/abs/1906.02025}{{\ttfamily 1906.02025}}].

\bibitem{Dolan:2013rja}
M.~J. Dolan, C.~Englert, N.~Greiner and M.~Spannowsky, \emph{{Further on up the
  road: $hhjj$ production at the LHC}},
  \href{https://doi.org/10.1103/PhysRevLett.112.101802}{\emph{Phys. Rev. Lett.}
  {\bfseries 112} (2014) 101802}
  [\href{https://arxiv.org/abs/1310.1084}{{\ttfamily 1310.1084}}].

\bibitem{Liu-Sheng:2014gxa}
L.-S. Ling, R.-Y. Zhang, W.-G. Ma, L.~Guo, W.-H. Li and X.-Z. Li, \emph{{NNLO
  QCD corrections to Higgs pair production via vector boson fusion at hadron
  colliders}}, \href{https://doi.org/10.1103/PhysRevD.89.073001}{\emph{Phys.
  Rev.} {\bfseries D89} (2014) 073001}
  [\href{https://arxiv.org/abs/1401.7754}{{\ttfamily 1401.7754}}].

\bibitem{Dolan:2015zja}
M.~J. Dolan, C.~Englert, N.~Greiner, K.~Nordstrom and M.~Spannowsky,
  \emph{{$hhjj$ production at the LHC}},
  \href{https://doi.org/10.1140/epjc/s10052-015-3622-3}{\emph{Eur. Phys. J.}
  {\bfseries C75} (2015) 387}
  [\href{https://arxiv.org/abs/1506.08008}{{\ttfamily 1506.08008}}].

\bibitem{Bishara:2016kjn}
F.~Bishara, R.~Contino and J.~Rojo, \emph{{Higgs pair production in
  vector-boson fusion at the LHC and beyond}},
  \href{https://doi.org/10.1140/epjc/s10052-017-5037-9}{\emph{Eur. Phys. J.}
  {\bfseries C77} (2017) 481}
  [\href{https://arxiv.org/abs/1611.03860}{{\ttfamily 1611.03860}}].

\bibitem{Arganda:2018ftn}
E.~Arganda, C.~Garcia-Garcia and M.~J. Herrero, \emph{{Probing the Higgs
  self-coupling through double Higgs production in vector boson scattering at
  the LHC}}, \href{https://doi.org/10.1016/j.nuclphysb.2019.114687}{\emph{Nucl.
  Phys.} {\bfseries B945} (2019) 114687}
  [\href{https://arxiv.org/abs/1807.09736}{{\ttfamily 1807.09736}}].

\bibitem{Baglio:2012np}
J.~Baglio, A.~Djouadi, R.~Gröber, M.~M. Mühlleitner, J.~Quevillon and
  M.~Spira, \emph{{The measurement of the Higgs self-coupling at the LHC:
  theoretical status}},
  \href{https://doi.org/10.1007/JHEP04(2013)151}{\emph{JHEP} {\bfseries 04}
  (2013) 151} [\href{https://arxiv.org/abs/1212.5581}{{\ttfamily 1212.5581}}].

\bibitem{Frederix:2014hta}
R.~Frederix, S.~Frixione, V.~Hirschi, F.~Maltoni, O.~Mattelaer, P.~Torrielli
  et~al., \emph{{Higgs pair production at the LHC with NLO and parton-shower
  effects}}, \href{https://doi.org/10.1016/j.physletb.2014.03.026}{\emph{Phys.
  Lett.} {\bfseries B732} (2014) 142}
  [\href{https://arxiv.org/abs/1401.7340}{{\ttfamily 1401.7340}}].

\bibitem{Dreyer:2018qbw}
F.~A. Dreyer and A.~Karlberg, \emph{{Vector-Boson Fusion Higgs Pair Production
  at N$^3$LO}}, \href{https://doi.org/10.1103/PhysRevD.98.114016}{\emph{Phys.
  Rev. D} {\bfseries 98} (2018) 114016}
  [\href{https://arxiv.org/abs/1811.07906}{{\ttfamily 1811.07906}}].

\bibitem{Dreyer:2018rfu}
F.~A. Dreyer and A.~Karlberg, \emph{{Fully differential Vector-Boson Fusion
  Higgs Pair Production at Next-to-Next-to-Leading Order}},
  \href{https://doi.org/10.1103/PhysRevD.99.074028}{\emph{Phys. Rev. D}
  {\bfseries 99} (2019) 074028}
  [\href{https://arxiv.org/abs/1811.07918}{{\ttfamily 1811.07918}}].

\bibitem{Dreyer:2020urf}
F.~A. Dreyer, A.~Karlberg and L.~Tancredi, \emph{{On the impact of
  non-factorisable corrections in VBF single and double Higgs production}},
  \href{https://doi.org/10.1007/JHEP10(2020)131}{\emph{JHEP} {\bfseries 10}
  (2020) 131} [\href{https://arxiv.org/abs/2005.11334}{{\ttfamily
  2005.11334}}].

\bibitem{Dreyer:2020xaj}
F.~A. Dreyer, A.~Karlberg, J.-N. Lang and M.~Pellen, \emph{{Precise predictions
  for double-Higgs production via vector-boson fusion}},
  \href{https://doi.org/10.1140/epjc/s10052-020-08610-7}{\emph{Eur. Phys. J. C}
  {\bfseries 80} (2020) 1037}
  [\href{https://arxiv.org/abs/2005.13341}{{\ttfamily 2005.13341}}].

\bibitem{Englert:2017gdy}
C.~Englert, Q.~Li, M.~Spannowsky, M.~Wang and L.~Wang, \emph{{VBS ${\rm W}^\pm
  {\rm W}^\pm {\rm H}$ production at the HL-LHC and a 100 TeV $pp$-collider}},
  \href{https://doi.org/10.1142/S0217751X17501068}{\emph{Int. J. Mod. Phys.}
  {\bfseries A32} (2017) 1750106}
  [\href{https://arxiv.org/abs/1702.01930}{{\ttfamily 1702.01930}}].

\bibitem{Nordstrom:2018ceg}
K.~Nordström and A.~Papaefstathiou, \emph{{$VHH$ production at the
  High-Luminosity LHC}},
  \href{https://doi.org/10.1140/epjp/i2019-12614-2}{\emph{Eur. Phys. J. Plus}
  {\bfseries 134} (2019) 288}
  [\href{https://arxiv.org/abs/1807.01571}{{\ttfamily 1807.01571}}].

\bibitem{Plehn:1996wb}
T.~Plehn, M.~Spira and P.~M. Zerwas, \emph{{Pair production of neutral Higgs
  particles in gluon-gluon collisions}},
  \href{https://doi.org/10.1016/0550-3213(96)00418-X,
  10.1016/S0550-3213(98)00406-4}{\emph{Nucl. Phys.} {\bfseries B479} (1996) 46}
  [\href{https://arxiv.org/abs/hep-ph/9603205}{{\ttfamily hep-ph/9603205}}].

\bibitem{Baur:2002rb}
U.~Baur, T.~Plehn and D.~L. Rainwater, \emph{{Measuring the Higgs Boson Self
  Coupling at the LHC and Finite Top Mass Matrix Elements}},
  \href{https://doi.org/10.1103/PhysRevLett.89.151801}{\emph{Phys. Rev. Lett.}
  {\bfseries 89} (2002) 151801}
  [\href{https://arxiv.org/abs/hep-ph/0206024}{{\ttfamily hep-ph/0206024}}].

\bibitem{Li:2013flc}
Q.~Li, Q.-S. Yan and X.~Zhao, \emph{{Higgs Pair Production: Improved
  Description by Matrix Element Matching}},
  \href{https://doi.org/10.1103/PhysRevD.89.033015}{\emph{Phys. Rev.}
  {\bfseries D89} (2014) 033015}
  [\href{https://arxiv.org/abs/1312.3830}{{\ttfamily 1312.3830}}].

\bibitem{Cao:2015oaa}
Q.-H. Cao, B.~Yan, D.-M. Zhang and H.~Zhang, \emph{{Resolving the Degeneracy in
  Single Higgs Production with Higgs Pair Production}},
  \href{https://doi.org/10.1016/j.physletb.2015.11.045}{\emph{Phys. Lett.}
  {\bfseries B752} (2016) 285}
  [\href{https://arxiv.org/abs/1508.06512}{{\ttfamily 1508.06512}}].

\bibitem{Cao:2016zob}
Q.-H. Cao, G.~Li, B.~Yan, D.-M. Zhang and H.~Zhang, \emph{{Double Higgs
  production at the 14 TeV LHC and a 100 TeV $pp$ collider}},
  \href{https://doi.org/10.1103/PhysRevD.96.095031}{\emph{Phys. Rev.}
  {\bfseries D96} (2017) 095031}
  [\href{https://arxiv.org/abs/1611.09336}{{\ttfamily 1611.09336}}].

\bibitem{Baur:2002qd}
U.~Baur, T.~Plehn and D.~L. Rainwater, \emph{{Determining the Higgs Boson
  Selfcoupling at Hadron Colliders}},
  \href{https://doi.org/10.1103/PhysRevD.67.033003}{\emph{Phys. Rev.}
  {\bfseries D67} (2003) 033003}
  [\href{https://arxiv.org/abs/hep-ph/0211224}{{\ttfamily hep-ph/0211224}}].

\bibitem{Ren:2017jbg}
J.~Ren, R.-Q. Xiao, M.~Zhou, Y.~Fang, H.-J. He and W.~Yao, \emph{{LHC Search of
  New Higgs Boson via Resonant Di-Higgs Production with Decays into 4W}},
  \href{https://doi.org/10.1007/JHEP06(2018)090}{\emph{JHEP} {\bfseries 06}
  (2018) 090} [\href{https://arxiv.org/abs/1706.05980}{{\ttfamily
  1706.05980}}].

\bibitem{Baur:2003gp}
U.~Baur, T.~Plehn and D.~L. Rainwater, \emph{{Probing the Higgs selfcoupling at
  hadron colliders using rare decays}},
  \href{https://doi.org/10.1103/PhysRevD.69.053004}{\emph{Phys. Rev.}
  {\bfseries D69} (2004) 053004}
  [\href{https://arxiv.org/abs/hep-ph/0310056}{{\ttfamily hep-ph/0310056}}].

\bibitem{Yao:2013ika}
W.~Yao, \emph{{Studies of measuring Higgs self-coupling with $HH\rightarrow
  b\bar b \gamma\gamma$ at the future hadron colliders}},  in
  \emph{{Proceedings, 2013 Community Summer Study on the Future of U.S.
  Particle Physics: Snowmass on the Mississippi (CSS2013): Minneapolis, MN,
  USA, July 29-August 6, 2013}}, 2013,
  \href{https://arxiv.org/abs/1308.6302}{{\ttfamily 1308.6302}},
  \href{http://www.slac.stanford.edu/econf/C1307292/docs/submittedArxivFiles/1308.6302.pdf}{http://www.slac.stanford.edu/econf/C1307292/docs/submittedArxivFiles/1308.6302.pdf}.

\bibitem{Kling:2016lay}
F.~Kling, T.~Plehn and P.~Schichtel, \emph{{Maximizing the significance in
  Higgs boson pair analyses}},
  \href{https://doi.org/10.1103/PhysRevD.95.035026}{\emph{Phys. Rev.}
  {\bfseries D95} (2017) 035026}
  [\href{https://arxiv.org/abs/1607.07441}{{\ttfamily 1607.07441}}].

\bibitem{Chang:2018uwu}
J.~Chang, K.~Cheung, J.~S. Lee, C.-T. Lu and J.~Park, \emph{{Higgs-boson-pair
  production $H(\to b\bar{b})H(\to \gamma\gamma)$ from gluon fusion at the
  HL-LHC and HL-100 TeV hadron collider}},
  \href{https://doi.org/10.1103/PhysRevD.100.096001}{\emph{Phys. Rev.}
  {\bfseries D100} (2019) 096001}
  [\href{https://arxiv.org/abs/1804.07130}{{\ttfamily 1804.07130}}].

\bibitem{Kim:2018uty}
J.~H. Kim, Y.~Sakaki and M.~Son, \emph{{Combined analysis of double Higgs
  production via gluon fusion at the HL-LHC in the effective field theory
  approach}}, \href{https://doi.org/10.1103/PhysRevD.98.015016}{\emph{Phys.
  Rev.} {\bfseries D98} (2018) 015016}
  [\href{https://arxiv.org/abs/1801.06093}{{\ttfamily 1801.06093}}].

\bibitem{He:2015spf}
H.-J. He, J.~Ren and W.~Yao, \emph{{Probing new physics of cubic Higgs boson
  interaction via Higgs pair production at hadron colliders}},
  \href{https://doi.org/10.1103/PhysRevD.93.015003}{\emph{Phys. Rev.}
  {\bfseries D93} (2016) 015003}
  [\href{https://arxiv.org/abs/1506.03302}{{\ttfamily 1506.03302}}].

\bibitem{Papaefstathiou:2012qe}
A.~Papaefstathiou, L.~L. Yang and J.~Zurita, \emph{{Higgs boson pair production
  at the LHC in the $b \bar{b} W^+ W^-$ channel}},
  \href{https://doi.org/10.1103/PhysRevD.87.011301}{\emph{Phys. Rev.}
  {\bfseries D87} (2013) 011301}
  [\href{https://arxiv.org/abs/1209.1489}{{\ttfamily 1209.1489}}].

\bibitem{Baur:2003gpa}
U.~Baur, T.~Plehn and D.~L. Rainwater, \emph{{Examining the Higgs boson
  potential at lepton and hadron colliders: A Comparative analysis}},
  \href{https://doi.org/10.1103/PhysRevD.68.033001}{\emph{Phys. Rev.}
  {\bfseries D68} (2003) 033001}
  [\href{https://arxiv.org/abs/hep-ph/0304015}{{\ttfamily hep-ph/0304015}}].

\bibitem{Dolan:2012rv}
M.~J. Dolan, C.~Englert and M.~Spannowsky, \emph{{Higgs self-coupling
  measurements at the LHC}},
  \href{https://doi.org/10.1007/JHEP10(2012)112}{\emph{JHEP} {\bfseries 10}
  (2012) 112} [\href{https://arxiv.org/abs/1206.5001}{{\ttfamily 1206.5001}}].

\bibitem{Barr:2013tda}
A.~J. Barr, M.~J. Dolan, C.~Englert and M.~Spannowsky, \emph{{Di-Higgs final
  states augMT2ed -- selecting $hh$ events at the high luminosity LHC}},
  \href{https://doi.org/10.1016/j.physletb.2013.12.011}{\emph{Phys. Lett.}
  {\bfseries B728} (2014) 308}
  [\href{https://arxiv.org/abs/1309.6318}{{\ttfamily 1309.6318}}].

\bibitem{deLima:2014dta}
D.~E. Ferreira~de Lima, A.~Papaefstathiou and M.~Spannowsky, \emph{{Standard
  model Higgs boson pair production in the $(b\bar{b})(b\bar{b})$ final
  state}}, \href{https://doi.org/10.1007/JHEP08(2014)030}{\emph{JHEP}
  {\bfseries 08} (2014) 030} [\href{https://arxiv.org/abs/1404.7139}{{\ttfamily
  1404.7139}}].

\bibitem{Behr:2015oqq}
J.~K. Behr, D.~Bortoletto, J.~A. Frost, N.~P. Hartland, C.~Issever and J.~Rojo,
  \emph{{Boosting Higgs pair production in the $b\bar{b}b\bar{b}$ final state
  with multivariate techniques}},
  \href{https://doi.org/10.1140/epjc/s10052-016-4215-5}{\emph{Eur. Phys. J.}
  {\bfseries C76} (2016) 386}
  [\href{https://arxiv.org/abs/1512.08928}{{\ttfamily 1512.08928}}].

\bibitem{Barger:2013jfa}
V.~Barger, L.~L. Everett, C.~B. Jackson and G.~Shaughnessy, \emph{{Higgs-Pair
  Production and Measurement of the Triscalar Coupling at LHC(8,14)}},
  \href{https://doi.org/10.1016/j.physletb.2013.12.013}{\emph{Phys. Lett.}
  {\bfseries B728} (2014) 433}
  [\href{https://arxiv.org/abs/1311.2931}{{\ttfamily 1311.2931}}].

\bibitem{Barr:2014sga}
A.~J. Barr, M.~J. Dolan, C.~Englert, D.~E. Ferreira~de Lima and M.~Spannowsky,
  \emph{{Higgs Self-Coupling Measurements at a 100 TeV Hadron Collider}},
  \href{https://doi.org/10.1007/JHEP02(2015)016}{\emph{JHEP} {\bfseries 02}
  (2015) 016} [\href{https://arxiv.org/abs/1412.7154}{{\ttfamily 1412.7154}}].

\bibitem{Papaefstathiou:2015iba}
A.~Papaefstathiou, \emph{{Discovering Higgs boson pair production through rare
  final states at a 100 TeV collider}},
  \href{https://doi.org/10.1103/PhysRevD.91.113016}{\emph{Phys. Rev.}
  {\bfseries D91} (2015) 113016}
  [\href{https://arxiv.org/abs/1504.04621}{{\ttfamily 1504.04621}}].

\bibitem{Li:2015yia}
Q.~Li, Z.~Li, Q.-S. Yan and X.~Zhao, \emph{{Probe Higgs boson pair production
  via the 3$\ell$2j+$E\!\!\!/$ mode}},
  \href{https://doi.org/10.1103/PhysRevD.92.014015}{\emph{Phys. Rev.}
  {\bfseries D92} (2015) 014015}
  [\href{https://arxiv.org/abs/1503.07611}{{\ttfamily 1503.07611}}].

\bibitem{Zhao:2016tai}
X.~Zhao, Q.~Li, Z.~Li and Q.-S. Yan, \emph{{Discovery potential of Higgs boson
  pair production through 4$\ell$+$E\!\!\!/$ final states at a 100 TeV
  collider}}, \href{https://doi.org/10.1088/1674-1137/41/2/023105}{\emph{Chin.
  Phys.} {\bfseries C41} (2017) 023105}
  [\href{https://arxiv.org/abs/1604.04329}{{\ttfamily 1604.04329}}].

\bibitem{Contino:2016spe}
R.~Contino et~al., \emph{{Physics at a 100 TeV pp collider: Higgs and EW
  symmetry breaking studies}},
  \href{https://doi.org/10.23731/CYRM-2017-003.255}{\emph{CERN Yellow Rep.}
  (2017) 255} [\href{https://arxiv.org/abs/1606.09408}{{\ttfamily
  1606.09408}}].

\bibitem{Goncalves:2018yva}
D.~Gonçalves, T.~Han, F.~Kling, T.~Plehn and M.~Takeuchi, \emph{{Higgs boson
  pair production at future hadron colliders: From kinematics to dynamics}},
  \href{https://doi.org/10.1103/PhysRevD.97.113004}{\emph{Phys. Rev.}
  {\bfseries D97} (2018) 113004}
  [\href{https://arxiv.org/abs/1802.04319}{{\ttfamily 1802.04319}}].

\bibitem{Kilian:2007gr}
W.~Kilian, T.~Ohl and J.~Reuter, \emph{{WHIZARD: Simulating Multi-Particle
  Processes at LHC and ILC}},
  \href{https://doi.org/10.1140/epjc/s10052-011-1742-y}{\emph{Eur. Phys. J.}
  {\bfseries C71} (2011) 1742}
  [\href{https://arxiv.org/abs/0708.4233}{{\ttfamily 0708.4233}}].

\bibitem{Pumplin:2002vw}
J.~Pumplin, D.~R. Stump, J.~Huston, H.~L. Lai, P.~M. Nadolsky and W.~K. Tung,
  \emph{{New generation of parton distributions with uncertainties from global
  QCD analysis}},
  \href{https://doi.org/10.1088/1126-6708/2002/07/012}{\emph{JHEP} {\bfseries
  07} (2002) 012} [\href{https://arxiv.org/abs/hep-ph/0201195}{{\ttfamily
  hep-ph/0201195}}].

\bibitem{Alwall:2014hca}
J.~Alwall, R.~Frederix, S.~Frixione, V.~Hirschi, F.~Maltoni, O.~Mattelaer
  et~al., \emph{{The automated computation of tree-level and next-to-leading
  order differential cross sections, and their matching to parton shower
  simulations}}, \href{https://doi.org/10.1007/JHEP07(2014)079}{\emph{JHEP}
  {\bfseries 07} (2014) 079} [\href{https://arxiv.org/abs/1405.0301}{{\ttfamily
  1405.0301}}].

\bibitem{Mangano:2002ea}
M.~L. Mangano, M.~Moretti, F.~Piccinini, R.~Pittau and A.~D. Polosa,
  \emph{{ALPGEN, a generator for hard multiparton processes in hadronic
  collisions}},
  \href{https://doi.org/10.1088/1126-6708/2003/07/001}{\emph{JHEP} {\bfseries
  07} (2003) 001} [\href{https://arxiv.org/abs/hep-ph/0206293}{{\ttfamily
  hep-ph/0206293}}].

\bibitem{Sjostrand:2007gs}
T.~Sjostrand, S.~Mrenna and P.~Z. Skands, \emph{{A Brief Introduction to PYTHIA
  8.1}}, \href{https://doi.org/10.1016/j.cpc.2008.01.036}{\emph{Comput. Phys.
  Commun.} {\bfseries 178} (2008) 852}
  [\href{https://arxiv.org/abs/0710.3820}{{\ttfamily 0710.3820}}].

\bibitem{Cacciari:2011ma}
M.~Cacciari, G.~P. Salam and G.~Soyez, \emph{{FastJet User Manual}},
  \href{https://doi.org/10.1140/epjc/s10052-012-1896-2}{\emph{Eur. Phys. J.}
  {\bfseries C72} (2012) 1896}
  [\href{https://arxiv.org/abs/1111.6097}{{\ttfamily 1111.6097}}].

\bibitem{Cacciari:2008gp}
M.~Cacciari, G.~P. Salam and G.~Soyez, \emph{{The anti-$k_t$ jet clustering
  algorithm}}, \href{https://doi.org/10.1088/1126-6708/2008/04/063}{\emph{JHEP}
  {\bfseries 04} (2008) 063} [\href{https://arxiv.org/abs/0802.1189}{{\ttfamily
  0802.1189}}].

\bibitem{Butterworth:2008iy}
J.~M. Butterworth, A.~R. Davison, M.~Rubin and G.~P. Salam, \emph{{Jet
  substructure as a new Higgs search channel at the LHC}},
  \href{https://doi.org/10.1103/PhysRevLett.100.242001}{\emph{Phys. Rev. Lett.}
  {\bfseries 100} (2008) 242001}
  [\href{https://arxiv.org/abs/0802.2470}{{\ttfamily 0802.2470}}].

\bibitem{Thaler:2010tr}
J.~Thaler and K.~Van~Tilburg, \emph{{Identifying Boosted Objects with
  N-subjettiness}}, \href{https://doi.org/10.1007/JHEP03(2011)015}{\emph{JHEP}
  {\bfseries 03} (2011) 015} [\href{https://arxiv.org/abs/1011.2268}{{\ttfamily
  1011.2268}}].

\bibitem{Marzani:2019hun}
S.~Marzani, G.~Soyez and M.~Spannowsky, \emph{{Looking inside jets: an
  introduction to jet substructure and boosted-object phenomenology}},
  vol.~958. Springer, 2019,
  \href{https://doi.org/10.1007/978-3-030-15709-8}{10.1007/978-3-030-15709-8},
  [\href{https://arxiv.org/abs/1901.10342}{{\ttfamily 1901.10342}}].

\bibitem{Dokshitzer:1997in}
Y.~L. Dokshitzer, G.~D. Leder, S.~Moretti and B.~R. Webber, \emph{{Better jet
  clustering algorithms}},
  \href{https://doi.org/10.1088/1126-6708/1997/08/001}{\emph{JHEP} {\bfseries
  08} (1997) 001} [\href{https://arxiv.org/abs/hep-ph/9707323}{{\ttfamily
  hep-ph/9707323}}].

\bibitem{Wobisch:1998wt}
M.~Wobisch and T.~Wengler, \emph{{Hadronization corrections to jet
  cross-sections in deep inelastic scattering}},  in \emph{{Monte Carlo
  generators for HERA physics. Proceedings, Workshop, Hamburg, Germany,
  1998-1999}}, pp.~270--279, 1998,
  \href{https://arxiv.org/abs/hep-ph/9907280}{{\ttfamily hep-ph/9907280}}.

\bibitem{Giudice:2007fh}
G.~Giudice, C.~Grojean, A.~Pomarol and R.~Rattazzi, \emph{{The
  Strongly-Interacting Light Higgs}},
  \href{https://doi.org/10.1088/1126-6708/2007/06/045}{\emph{JHEP} {\bfseries
  06} (2007) 045} [\href{https://arxiv.org/abs/hep-ph/0703164}{{\ttfamily
  hep-ph/0703164}}].

\bibitem{Kilian:2017nio}
W.~Kilian, S.~Sun, Q.-S. Yan, X.~Zhao and Z.~Zhao, \emph{{New Physics in
  multi-Higgs boson final states}},
  \href{https://doi.org/10.1007/JHEP06(2017)145}{\emph{JHEP} {\bfseries 06}
  (2017) 145} [\href{https://arxiv.org/abs/1702.03554}{{\ttfamily
  1702.03554}}].

\bibitem{Ellis:2018gqa}
J.~Ellis, C.~W. Murphy, V.~Sanz and T.~You, \emph{{Updated Global SMEFT Fit to
  Higgs, Diboson and Electroweak Data}},
  \href{https://doi.org/10.1007/JHEP06(2018)146}{\emph{JHEP} {\bfseries 06}
  (2018) 146} [\href{https://arxiv.org/abs/1803.03252}{{\ttfamily
  1803.03252}}].

\bibitem{Cacciari:2014gra}
M.~Cacciari, G.~P. Salam and G.~Soyez, \emph{{SoftKiller, a particle-level
  pileup removal method}},
  \href{https://doi.org/10.1140/epjc/s10052-015-3267-2}{\emph{Eur. Phys. J. C}
  {\bfseries 75} (2015) 59} [\href{https://arxiv.org/abs/1407.0408}{{\ttfamily
  1407.0408}}].

\bibitem{Maltoni:2002mq}
F.~Maltoni, K.~Paul, T.~Stelzer and S.~Willenbrock, \emph{{Color Flow
  Decomposition of QCD Amplitudes}},
  \href{https://doi.org/10.1103/PhysRevD.67.014026}{\emph{Phys. Rev. D}
  {\bfseries 67} (2003) 014026}
  [\href{https://arxiv.org/abs/hep-ph/0209271}{{\ttfamily hep-ph/0209271}}].

\bibitem{Gallicchio:2010sw}
J.~Gallicchio and M.~D. Schwartz, \emph{{Seeing in Color: Jet Superstructure}},
  \href{https://doi.org/10.1103/PhysRevLett.105.022001}{\emph{Phys. Rev. Lett.}
  {\bfseries 105} (2010) 022001}
  [\href{https://arxiv.org/abs/1001.5027}{{\ttfamily 1001.5027}}].

\bibitem{Kim:2019wns}
J.~H. Kim, M.~Kim, K.~Kong, K.~T. Matchev and M.~Park, \emph{{Portraying Double
  Higgs at the Large Hadron Collider}},
  \href{https://doi.org/10.1007/JHEP09(2019)047}{\emph{JHEP} {\bfseries 09}
  (2019) 047} [\href{https://arxiv.org/abs/1904.08549}{{\ttfamily
  1904.08549}}].

\end{thebibliography}\endgroup
\bibliographystyle{JHEP}

\end{document}